\documentclass[a4paper,11pt]{article}

\usepackage{frcursive}
\usepackage{amsmath,amssymb,amsthm}
\usepackage{cancel}
\usepackage{mathrsfs}
\usepackage[rightcaption]{sidecap}
\usepackage{hyperref}
\usepackage{fancyhdr}
\hypersetup{
    colorlinks=true,
    linkcolor=blue,
    filecolor=magenta,      
    urlcolor=cyan,
    citecolor=blue
}
\usepackage{orcidlink}
\usepackage{faktor}
\usepackage{float}
\usepackage{blkarray}
\usepackage{longtable}
\usepackage{adjustbox}

\usepackage[margin=1in]{geometry}

\usepackage{dynkin-diagrams}
\usepackage{wrapfig}
\usepackage{tikz-cd} 
\usepackage{tikz}
\usetikzlibrary{braids}

\usepackage{quiver}
\usepackage{arydshln}
\usepackage{mathrsfs}
\usepackage[skins,theorems]{tcolorbox}
\usepackage{fdsymbol}
\usepackage{lscape}
\usepackage{float}

\usepackage{comment}
\usepackage{caption}
\usepackage{subcaption}
\usepackage{multicol}
\usepackage{multirow}
\usepackage{booktabs}
\usepackage[bbgreekl]{mathbbol}
\usepackage{authblk} 
\usepackage{orcidlink}
\usepackage[backend=biber, sorting=none, style=numeric, maxbibnames=999]{biblatex}
\definecolor{amber}{rgb}{1.0, 0.49, 0.0}
\definecolor{Green}{rgb}{0.0, 0.5, 0.0}
\definecolor{purple}{rgb}{0.7,0,0.7}

\DeclareSymbolFontAlphabet{\mathbbm}{bbold}
\DeclareSymbolFontAlphabet{\mathbb}{AMSb}%

\renewcommand{\arraystretch}{1.5}

\newcommand*{\Resize}[2]{\resizebox{#1}{!}{$#2$}}%

\newtheorem{thm}{Theorem}[section]

\newtheorem{rmk}[thm]{Remark}

\newtheorem{coj}[thm]{Conjecture}

\newtheorem*{que*}{Question}
\newtheorem*{ans*}{Answer}

\newcommand{\Sk}{\operatorname{Sk}}
\newcommand{\R}{{\mathbb{R}}}

\newcommand{\Q}{{\mathbb{Q}}}

\makeatletter

\newcommand{\smstrut}{\vphantom{q^{-1}x_i x_j}}
\newenvironment{customsmallmatrix}{%
  \null\,\vcenter\bgroup
  \Let@\restore@math@cr\default@tag
  \baselineskip6\ex@
  \lineskip1.5\ex@
  \lineskiplimit\lineskip
  \ialign\bgroup
    \hfil$\m@th\scriptstyle\smstrut##$\hfil
    &&\thickspace\hfil$\m@th\scriptstyle##$\hfil\crcr
}{%
  \crcr\egroup\egroup\,
}
\makeatother

\title{Worldsheet skeins and block decomposition of quivers}

\author[1--3]{Sachin Chauhan \orcidlink{0000-0003-1045-3428}\thanks{sachin.chauhan5921@gmail.com}}
\author[4]{Piotr Kucharski \orcidlink{0000-0002-9599-5658}\thanks{piotr.kucharski@mimuw.edu.pl}}
\author[5]{Lukas Nakamura \orcidlink{0009-0007-9950-4178}\thanks{lukas.nakamura@imj-prg.fr}}
\author[6]{\break Dmitry Noshchenko
\orcidlink{0000-0002-9639-5603}\thanks{dsnoshchenko@stp.dias.ie}}
\author[7]{ Ramadevi Pichai\orcidlink{0000-0003-4331-8222}\thanks{ramadevi@iitb.ac.in}}
\author[8]{\break Vivek Kumar Singh \orcidlink{0000-0001-9141-2331}\thanks{vks2024@nyu.edu}}
\author[9,10]{Marko Sto\v si\'c\orcidlink{0000-0002-4464-396X}\thanks{mstosic@ciencias.ulisboa.pt}
}
\affil[1]{Centre for Geometry and Physics, Uppsala University, Box 524, 751 20 Uppsala, Sweden}
\affil[2]{The Institute of Mathematical Sciences, 
IV Cross Road, C.I.T. Campus, \break Taramani, Chennai, India-600113}
\affil[3]{Homi Bhabha National Institute,
Training School Complex, Anushakti Nagar, Mumbai, India 400094}
\affil[4]{Institute of Mathematics, University of Warsaw, ul. Banacha 2, 02-097 Warsaw, Poland}
\affil[5]{Institut de Mathématiques de Jussieu-Paris Rive Gauche, CNRS, \break Sorbonne Université, Université Paris Cité, Paris, France}
\affil[6]{School of Theoretical Physics, Dublin Institute for Advanced Studies,\break 10 Burlington Road, Dublin 4, D04 C932, Ireland}
\affil[7]{Department of Physics, Indian Institute of Technology Bombay,\break Powai, Mumbai 400076, India}
\affil[8]{Center for Quantum and Topological Systems, NYUAD Research Institute, \break New York University Abu Dhabi, Abu Dhabi, United Arab Emirates}
\affil[9]
{CEMS.UL, Departamento de Ci\^encias Matem\'aticas, Faculdade de Ci\^encias, Universidade de Lisboa, Edif\'icio C6, Campo Grande, 1749-016 Lisbon, Portugal}
\affil[10]
{Mathematical Institute SANU, Knez Mihajlova 36, 11000 Belgrade, Serbia}
\date{ }

\begin{document}

\hfill
\begin{tabular}{r}
UUITP-24/26 \\
DIAS-STP-26-22
\end{tabular}
{\let\newpage\relax\maketitle}

\maketitle
\abstract{
Following the knot-quiver correspondence, we investigate quiver counterparts of the HOMFLYPT skein relations by interpreting link invariants in terms of holomorphic worldsheets and their multicover contributions. Extending this framework to more general worldsheet geometries, we conjecture that, up to quiver equivalence, the adjacency matrix of the quiver associated with any link in a skein triple admits a block decomposition whose diagonal blocks are quiver adjacency matrices associated with the other two links. We verify this conjecture in numerous examples. 
}
\newpage
\tableofcontents
\newpage

\section{Introduction}

The HOMFLYPT polynomial is a link invariant defined by the skein relations that compare links which differ only within a 3-ball containing over-crossing, under-crossing or no-crossing~\cite{HOMFLY,PT}. It is also conjectured that for every link, one can associate a quiver through the knot-quiver correspondence \cite{Kucharski:2017poe,Kucharski:2017ogk,Jankowski:2021flt}. Specifically, the correspondence suggests that the (anti-)symmetrically coloured HOMFLYPT invariants can be identified, in a non-unique way, with the motivic Donaldson-Thomas invariants of a symmetric quiver. This is currently proven for all rational and, more generally, arborescent knots and links \cite{Stosic:2017wno,Stosic:2020xwn}.

Recent developments show that skein relations for the 
HOMFLYPT invariants coloured by one-column Young diagrams, admit a categorification relevant for the triply graded homology theory of a link \cite{hogancamp2026skein,HogancampRoseWedrich2024}.
On the other hand, already in the original formulation of the knot-quiver correspondence, it was observed that symmetric quivers for simple knots contain some information about their coloured HOMFLYPT homologies \cite{Kucharski:2017ogk}. This relation was further explored in~\cite{Ekholm:2021gyu}. Independently, categorifications of quiver invariants and transformations have been developed, whose relation to BPS algebras and coloured link homologies remains to be understood \cite{DotsenkoFeiginReineke2022,DotsenkoFeiginKucharskiReineke2026,VanGarderen2024,DFKLN2026}.

Motivated by the above, we are interested in the following open problem which was first highlighted in~\cite{Chauhan:2025iwt}:

\begin{que*}
    Can the skein relations for HOMFLYPT polynomials of links be lifted to the level of quivers?
\end{que*}

\begin{figure}[h]
\centering
\includegraphics[scale=0.3]{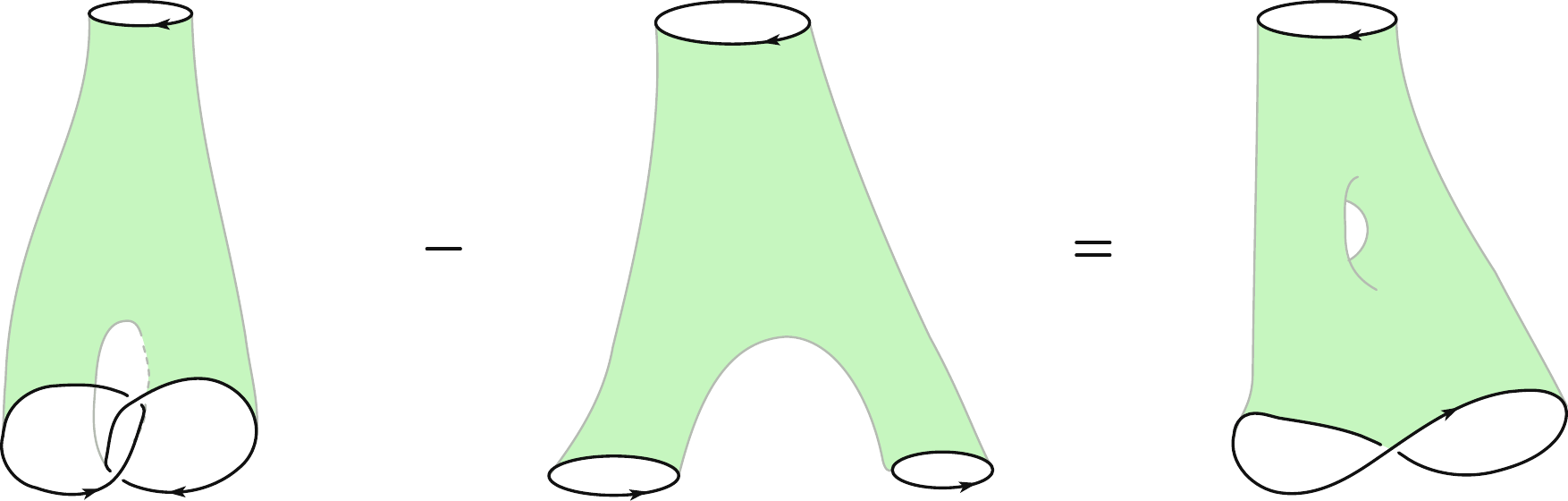}
\caption{An example of a worldsheet skein relation. \label{fig:skein-relation}}
\end{figure}

\noindent To attempt this question, we turn to the completely different perspective: the geometric interpretation of the knot–quiver correspondence via the holomorphic curve count \cite{Ekholm:2018eee,Ekholm:2019lmb,Ekholm:2019yqp,Nak26}. To be more precise, we will reinterpret the HOMFLYPT skein relations by considering the worldsheet skein approach introduced in \cite{Ekholm:2024ceb}. This naturally generalizes the standard knot-quiver correspondence, which is restricted to genus-zero worldsheets. 
In this geometric setting, the skein relation keeps track of the topology of the worldsheets whose boundary links undergo the usual HOMFLYPT skein
(see Figure \ref{fig:skein-relation}). 

The study of worldsheet skein relations allows us formulate the main conjecture:

\begin{ans*}
    Up to equivalence, the quiver adjacency matrix of any single link in the HOMFLYPT skein relation admits a block decomposition into the quiver adjacency matrices of the other two links. 
\end{ans*}

We confirm these decompositions in various examples. The examples also suggest that quiver equivalences play an important role as well. Most often than not, the conjecture holds for non-minimal quivers which are generated by operations of linking, unlinking and removal of redundant pair \cite{Ekholm:2019lmb,Kucharski:2023jds}.

In the original formulation of the knot-quiver correspondence, the correspondence for links is made somewhat artificial: it requires normalisation of the coloured HOMFLYPT polynomials for links so that they become genuinely polynomial. 
In contrast, our new approach provides a natural and geometrically meaningful extension of the knot-quiver correspondence to multicomponent links. Further, in the worldsheet skein formulation we observe that the partition functions of elementary holomorphic curves (higher-genus worldsheets) readily produce the requisite $q$-Pochhammer symbols.

We believe that our results, although formulated using the geometric language of the worldsheet skein, may also be lifted to the level of homology theories associated with a quiver and a link. In particular, the natural question is whether there exists a deeper connection between skein triangles in link homology and block decompositions of quivers we found. Another major challenge is to construct a quiver to a given knot directly from its diagram. We expect that quiver skein relations, which we hope to derive from the present results, are the important step towards this goal.

The organisation of the paper is as follows.
In Section \ref{sec:Prerequisites}, necessary prerequisites are reviewed. These include physical interpretations of coloured HOMFLYPT polynomials, skein-valued Gromov-Witten partition functions, the worldsheet skein module, the knot-quiver correspondence, and operations on quivers. 
In Section \ref{sec:HighGenusHOMFLYPT}, the worldsheet extension of HOMFLYPT generating functions to higher genera is discussed.
Section \ref{sec:Block_decompositions} states a general conjecture that gives a block decomposition of quivers corresponding to links entering the worldsheet skein. Section \ref{sec:knot_operations} presents how the quivers are affected when certain operations on the corresponding knots are performed. 
Section \ref{sec:Examples} contains examples of block decomposition of quivers for knots and links related by the skein relation, including 2-bridge torus knots and links, twist knots, as well as the two more complicated examples of the knots $6_2$ and $8_{13}$. Finally, Section \ref{sec:Future} describes the research directions following from presented results.

\section{Prerequisites}\label{sec:Prerequisites}
In this section, we review the geometric and algebraic framework underlying our later analysis. We begin with the open \(A\)-model on the conifold and its relation, via the Ooguri-Vafa large-\(N\) transition, to Chern-Simons theory and HOMFLYPT invariants, emphasising how Witten’s skein relation is reflected in open string amplitudes. We then introduce skein-valued Gromov-Witten theory and the worldsheet skein module, which provide invariant counts of holomorphic curves with boundary. Finally, we review the knot-quiver correspondence, its interpretation in terms of basic holomorphic disks and BPS states, and the operations -- unlinking, linking, and addition of redundant pairs -- that preserve the motivic generating series of a quiver.
\subsection{Open \texorpdfstring{$A$}{A}-model, Chern-Simons theory, and HOMFLYPT polynomials}
\label{sec:Open_A-model}
We study the open $A$-model on the deformed conifold $T^*S^3$, with holomorphic maps satisfying boundary conditions on the knot conormal Lagrangian $L_K$. The conormal $L_K$ may be displaced from the zero section $S^3$, producing the basic Ooguri-Vafa annulus. We consider the system with $N$ $A$-branes wrapping the zero section $S^3$ and $M$ $A$-branes wrapping $L_K$. By Witten’s interpretation of Chern-Simons theory as open topological string theory on $T^*S^3$ \cite{WitCSasString}, this open $A$-model partition function is equivalent to the Chern-Simons partition function on $S^3$ with a Wilson loop insertion along the knot $K$. In this description, Wilson loop observables are coloured by arbitrary partitions, or equivalently by arbitrary representations.

Ooguri and Vafa showed that, in the large $N$ geometric transition, the zero section $S^3$ can be shrunk to zero size and replaced by the exceptional $\mathbb{C}P^1$ of the small resolution of the resulting conifold singularity \cite{Ooguri:1999bv}. The target geometry after the transition is the resolved conifold
\[
\mathcal{O}(-1)\oplus \mathcal{O}(-1)\longrightarrow \mathbb{C}P^1,
\]
where the exceptional $\mathbb{C}P^1$ corresponds to the $S^2$ linking the original $S^3$ before the transition. Under this transition, the open curve counts with boundary on $L_K$ are identified with the corresponding Wilson loop observables in Chern-Simons theory. 

\begin{figure}[ht]
    \centering
  \begin{tikzpicture}[
    thick, >=stealth, line cap=round, line join=round,
    over/.style={
        preaction={draw=white,line width=5pt,line cap=butt}
    }
]
        \begin{scope}[xshift=1.0cm,yshift=5.5cm]
                    \draw[->] (1,-1)--(-1,1);
        \draw[->,preaction={draw=white,line width=5pt}] (-1,-1)--(1,1);
        \node at (0,-1.35) {$L_+$};
        \end{scope}

        \begin{scope}[xshift=5cm,yshift=5.5cm]
        \draw[->] (-.9,-1) .. controls (-.35,-.4) and (-.35,.4) .. (-.9,1);
        \draw[->] ( .9,-1) .. controls ( .35,-.4) and ( .35,.4) .. ( .9,1);
        \node at (0,-1.35) {$L_0$};
        \end{scope}
        \begin{scope}[xshift=9cm,yshift=5.5cm]
        \draw[->] (-1,-1)--(1,1);
        \draw[->,preaction={draw=white,line width=5pt}] (1,-1)--(-1,1);
        \node at (0,-1.35) {$L_-$};
        \end{scope}
    \node[anchor=east] at (5.75,2.8)
        {$a\;-\;a^{-1}\quad=\quad z$};
    \draw (6.5,2.7) circle[radius=0.7];

    \begin{scope}[xshift=0.55cm]
        \draw[->]
            (-0.52,-1.4)
            .. controls (-0.37,-0.93) and (-0.27,-0.45)
            .. (-0.14,-0.21)
            -- (0.14,0.21)
            .. controls (0.38,0.55) and (0.57,0.52)
            .. (0.76,0.20)
            .. controls (1.10,-0.35) and (0.44,-0.68)
            .. (0.14,-0.21)
            -- (-0.14,0.21)
            .. controls (-0.31,0.49) and (-0.46,1.07)
            .. (-0.52,1.4);

        \draw[over] (-0.14,-0.21) -- (0.14,0.21);

        \node[anchor=base] at (1.75,-0.20) {$=$};
        \node[anchor=base] at (2.70,-0.20) {$a$};
        \draw[->] (3.55,-1.4) -- (3.55,1.4);
    \end{scope}

    \begin{scope}[xshift=6.75cm]
        \draw[->]
            (-0.52,-1.4)
            .. controls (-0.48,-0.93) and (-0.27,-0.45)
            .. (-0.14,-0.21)
            -- (0.14,0.21)
            .. controls (0.32,0.72) and (0.72,0.66)
            .. (0.84,0.25)
            .. controls (1.03,-0.32) and (0.44,-0.50)
            .. (0.14,-0.21)
            -- (-0.14,0.21)
            .. controls (-0.31,0.49) and (-0.46,1.07)
            .. (-0.52,1.4);

        \draw[over] (0.14,-0.21) -- (-0.14,0.21);

        \node[anchor=base] at (1.75,-0.20) {$=$};
        \node[anchor=base] at (2.95,-0.20) {$a^{-1}$};
        \draw[->] (4.15,-1.4) -- (4.15,1.4);
    \end{scope}
\end{tikzpicture}
    \caption{The links $L_+$, $L_0$, and $L_-$ agree outside a small three-ball and differ only by the local tangle inside it. The corresponding framed HOMFLYPT invariants with the diagrams equipped with blackboard framing satisfy $P(L_+)-zP(L_0)-P(L_-)=0$, 
    where $z=q-q^{-1}$. The other two formulas are the remaining two defining relations of the HOMFLYPT skein.} 
    \label{fig:Skein_relation}
\end{figure}
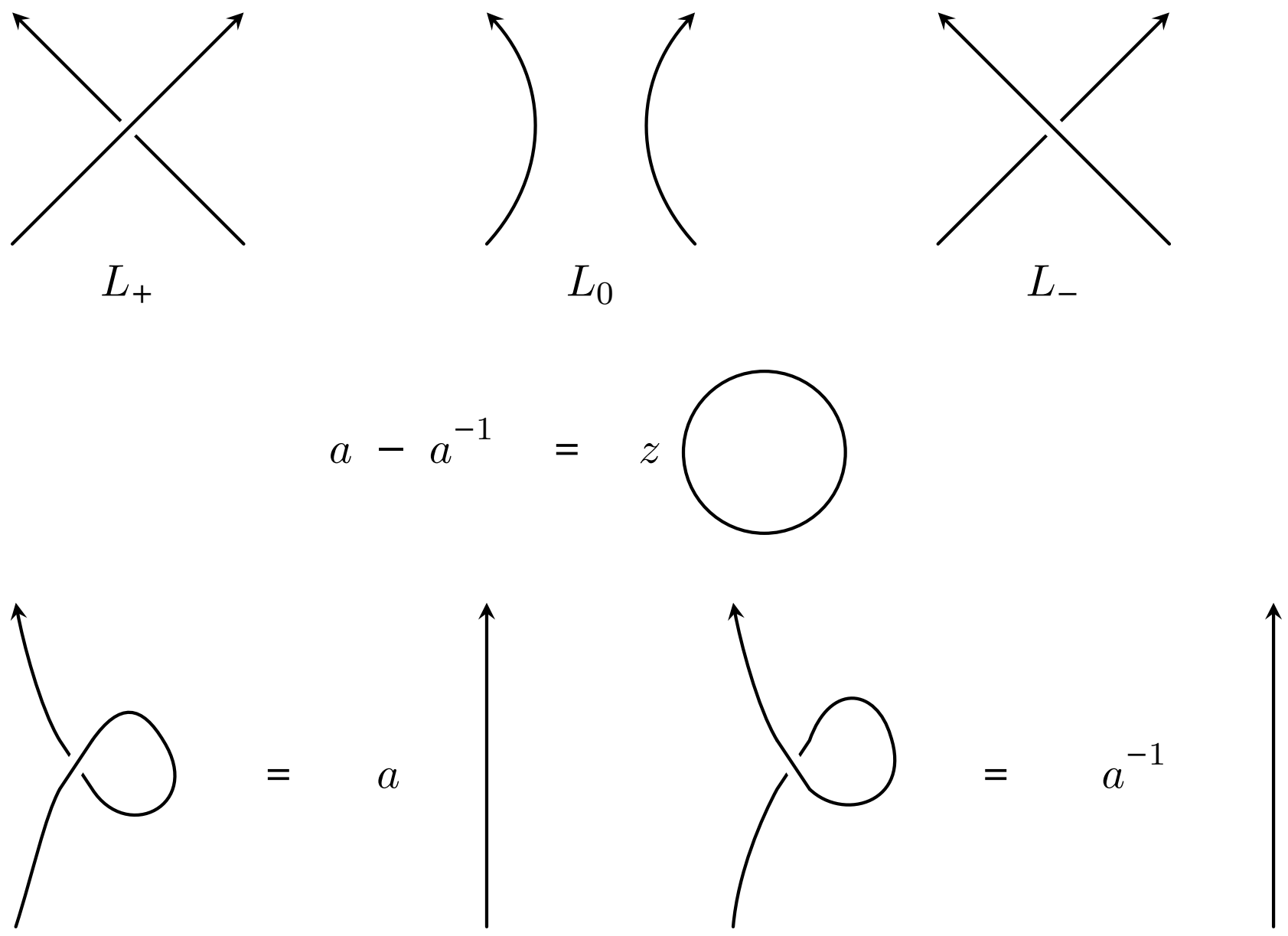

On the other hand, in his famous work \cite{Witten_Jones} Witten showed that one can isolate a crossing of the Wilson loop by cutting out a small three-ball, whose boundary is a four-punctured~\(S^2\). Then the three local configurations: the positive crossing \(L_+\), the smoothing \(L_0\), and the negative crossing \(L_-\), define three states in the associated two-dimensional space of conformal blocks and hence satisfy a linear relation. After appropriate framing corrections, this becomes the skein relation that defines HOMFLYPT polynomials \cite{HOMFLY,PT} (see Figure~\ref{fig:Skein_relation}).\footnote{Another interesting connection to high-energy physics comes from relating tree-level gluon amplitudes to a
state sum of HOMFLYPT coefficients \cite{Guevara:2026zcv}.} 

Coming back to the correspondence between Chern-Simons theory and topological strings, the skein relation can therefore be viewed as a local relation among the open string amplitudes, and ultimately among the corresponding holomorphic curve contributions on the resolved conifold side.

The connection between Wilson loop observables in Chern-Simons theory and HOMFLYPT polynomials of knots can be generalised to different representations (which correspond to coloured HOMFLYPT polynomials).
In particular, when one places a single $A$-brane on $L_K$, the generating function reduces to the series of symmetrically coloured HOMFLYPT polynomials of $K$. Thus the one brane open $A$-model partition function on the knot conormal agrees with the Chern-Simons generating function for Wilson loops coloured by symmetric representations, in accordance with Witten’s gauge theoretic construction of knot invariants.
For symmetric representations relevant for our paper, the coloured skein relations are captured by MOY calculus \cite{MOY1998}. This was generalised further to categorified skein relations for Khovanov-Rozansky homologies, 
as well as for HOMFLYPT homologies where the skein relations are lifted to long exact sequence in homology \cite{hogancamp2026skein,QueffelecRose2016}.

\subsection{The skein-valued Gromov-Witten partition function}

Now we take a closer look at the geometry of open topological strings \cite{Ekholm:2019yqp}.
Let $X$ be a symplectic Calabi-Yau 3-fold and let $L\subset X$ be an oriented and spin Lagrangian with vanishing Maslov class. We assume that $X$ and $L$ are cylindrical near infinity, and we denote their boundaries by $Y=\partial X$ and $\Lambda=\partial L \subseteq Y$. A \emph{($J$-)holomorphic curve} is a map
\begin{equation}
    u:(S,\partial S)\rightarrow (X,L)
\end{equation}
from a  Riemann surface $(S,j)$ to $X$ which maps the boundary of $S$ to $L$ and whose differential is complex linear, i.e.\,it satisfies 
\begin{equation}
    \bar \partial_Ju=\frac{1}{2}(du+J\circ du\circ j)=0.
\end{equation}
for an auxiliary almost complex structure $J$.
 The Calabi-Yau and Maslov zero conditions imply that the formal dimension of the moduli space $\mathcal M$ of holomorphic curves is zero. If $L=\emptyset$ then we can define the Gromov-Witten partition function as
\begin{equation}
    Z_{GW}(X)=\sum_{(u,S)\in\mathcal M}w(u,S)g_s^{-\chi(S)}t^{[u(S)]}
\end{equation}
where $w(u,S)$ is the rational (orbifold) weight of $(u,S)$, $\chi(S)$ is the Euler characteristic of $S$, and $t^{[u(S)]}$ is the element represented by $u(S)$ in the group ring $\mathbb Q[H_2(X)]$. The partition function $Z_{GW}(X)$ is an invariant of the symplectic manifold $X$ and does not depend on the choice of almost complex structure.

For the open curves, the moduli space may have codimension 1 singularities, and thereby the naive count of holomorphic curves does not give an invariant. Following \cite{Ekholm:2019yqp}, one instead imposes the skein relations (see Figure \ref{fig:Skein_relation}) and defines the skein-valued open Gromov-Witten partition function as follows:
\begin{equation}
Z_{(X,L)}=\sum_{(u,S)\in\mathcal M^{bare}}w(u,S)t^{[u(S)]}z^{-\chi(S)}a^{\text{lk}(u,L)}\langle u(\partial S)\rangle\in \text{Sk}(L),
\end{equation}
    where the skein module of $L$ is
    \begin{equation}
    \text{Sk}(L)=\mathbb{Q}[a^{\pm},z^{\pm}]\{\text{framed and oriented links in }L\}/\text{Skein relations},
    \end{equation}
and $\mathcal M^{bare}$ denotes the moduli space of \emph{bare} curves, i.e.\,only those curves that do not have any ghost bubbles.
    
In the case $L=\emptyset$, the Gromov-Witten partition function is recovered as
\begin{equation}
    Z_{GW}(X;g_s)=Z(X,\emptyset;z=e^{g_s/2}-e^{-g_s/2}).
\end{equation}

\subsection{Coloured HOMFLYPT polynomials}

We can use the geometric framework introduced above to construct coloured HOMFLYPT polynomials.
The skein of $S^3$ is isomorphic to the ground ring $\mathbb{Q}[a^{\pm},z^{\pm}]$ as shown in \cite{fyhlmo85,pt87}: any framed and oriented link $K$ can be reduced to a multiple of the empty link in a unique way. The resulting coefficient of the empty link is the HOMFLYPT polynomial of $K$ discussed in Section \ref{sec:Open_A-model}. This invariant can be generalised as follows. Such a link $K$ induces an embedding of a disjoint union of solid tori $(S^1 \times \R^2)^{\coprod |\pi_0(K)|}$ into $S^3$ which is well-defined up to homotopy. This embedding induces a map $K*\cdot:(S^1 \times \R^2)^{\otimes |\pi_0(K)|} \to \Sk(S^3)$ which is an isotopy invariant of $K$. This map can be made more explicit if we choose a basis of $\Sk(S^1 \times \R^2)$. After setting $z = q - q^{-1}$ and a base change to a suitable localisation of the ground ring, the positive part of $\Sk(S^1 \times \R^2)$ (the submodule spanned by closures of braids oriented in a fixed direction) admits a basis $\{W_\lambda\}_\lambda$ labelled by integer partitions $\lambda$, see \cite{am98}. The \emph{framed $(\lambda_1,...,\lambda_{|\pi_0(K)|})$-coloured HOMLFYPT polynomial $P_\lambda(K)$ of $K$} is then defined as
\begin{equation}
    (-1)^{\sum_i f(K_i)\lambda_i}P_{\lambda_1,...,\lambda_{|\pi_0(K)|}}(K) \coloneqq K*(W_{\lambda_1} \otimes ...\otimes W_{\lambda_{|\pi_0(K)|}}),
\end{equation}
where $f(K_i)$ denotes the framing of the $i$-th component $K_i$ of $K$, i.e.\ the linking number of $K_i$ with its push-off.

\subsubsection*{Framing change}

If $K$ is a framed knot, we will denote by $K^{f}$ the same knot but with framing $f(K) + f$. Then the coloured HOMFLYPT polynomials of $K^f$ can be computed in terms of the coloured HOMFLYPT polynomials of $K$ via the formula \cite{am98}
\begin{equation}
    P_{\lambda}(K^f) = ((-a)^{|\lambda|}q^{2c_\lambda})^fP_{\lambda}(K),
\end{equation}
where $c_\lambda$ denotes the content of the Young diagram associated to the partition $\lambda$. If we change the framing of one component of a framed link, the change of the coloured HOMFLYPT polynomials is described by an analogous formula. One particular choice of framing that will be important for our main conjecture is the \emph{blackboard framing}. Given a knot diagram in $\R^2$, the framing vector field of the blackboard framing is just the constant vector field in the missing $\R$-direction. Its framing number is given by the writhe $w(K)$ of $K$.

\subsubsection*{Reduced HOMFLYPT polynomials}

The coloured HOMFLYPT polynomials $P_{\lambda_1,...,\lambda_{|\pi_0(K)|}}(K)$ are all divisible by the coloured HOMFLYPT polynomials $P_{\lambda_i}(\bigcirc)$ of the unknot for $i = 1, ..., |\pi_0(K)|$, which are given by
\begin{equation}
    P_{\lambda}(\bigcirc) = \prod_{\square \in \lambda} \frac{a q^{c_\lambda(\square)} - a^{-1} q^{-c_\lambda(\square)}}{q^{h_\lambda(\square)}-q^{-h_\lambda(\square)}},
\end{equation}
where $c_\lambda(\square)$ and $h_\lambda(\square)$ denote the content and the hook length of a box $\square$ in the Young diagram associated to $\lambda$.
The \emph{$i$-th reduced coloured HOMFLYPT polynomials of $K$} are then defined as 
\begin{equation}
    P^{red}_{\lambda_1,...,\lambda_{|\pi_0(K)|}}(K) \coloneqq \frac{P_{\lambda_1,...,\lambda_{|\pi_0(K)|}}(K)}{P_{\lambda_i}(\bigcirc)}.
\end{equation}

\subsubsection*{Symmetric representations}

Our work in the paper focuses on colours given by symmetric representations, meaning trivial (one row) partitions $\lambda = (d)$. In this case, we will simply write $d$ for the partition to simplify the notation, and many of the products appearing in general formulas can be expressed in terms of $q$-Pochhammers. For instance, we see that
\begin{equation}
    P_d(\bigcirc) = a^{-d} q^d\frac{(a^2;q^2)_d}{(q^2;q^2)_d}.
\end{equation}

We will often combine HOMFLYPT polynomials coloured by symmetric representations into a generating series (in a formal variable $x$, indirectly depending also on $a$ and $q$), denoted by 
\begin{equation}
Z_{K}(x,a,q)=\sum_{d\geq0} P_d(K)x^d.
\end{equation}

\subsection{The worldsheet skein module}

From the view-point of holomorphic curves, the skein relations admit a natural lift, which considers the 2-dimensional curves themselves and not just their boundaries \cite{Ekholm:2024ceb}. For this, let $X$ be a 6-dimensional manifold and let $L \subseteq X$ be a 3-dimensional submanifold. We assume that both $L$ and $X$ are oriented and that there exists an almost complex structure $J$ on $X$ for which $L$ is totally real. Then the \emph{worldsheet skein module} $\Sk^{ws}(X,L)$ of $(X,L)$ is defined as the free $\Q$-module generated by isotopy classes of embeddings $u:(\Sigma,\partial \Sigma) \to (X,L)$ such that 
\begin{itemize}
    \item $\Sigma$ is a possibly disconnected Riemann surface with complex structure $j$ and (possibly empty) boundary $\partial \Sigma$,
    \item $u(\Sigma \setminus \partial \Sigma) \subseteq X \setminus L$,
    \item $u$ is $(J,j)$-holomorphic along $\partial \Sigma$,
\end{itemize}
modulo the \emph{worldsheet skein relations} which are modelled on hyperbolic and elliptic singularities of holomorphic curves with boundary:

\begin{itemize}
\item When a holomorphic curve exhibits a hyperbolic singularity, its normalisation will be an immersed curve with a single transverse self-intersection of its boundary. There are two ways of perturbing this immersed curve to obtain an embedded curve which yield two curves which are related by flipping one crossing of their boundaries in $L$ (see Figure \ref{fig:skein-relation}). The first worldsheet skein relation states that the original curve right before the singularity is equal to the difference of the two curves obtained by perturbing the normalised curve.

\item In the case of a holomorphic curve with an elliptic singularity, the normalisation is an embedded holomorphic curve whose boundary has a single transverse intersection of its interior with $L$. Again, there are two ways of perturbing this curve to obtain curves whose interior is disjoint from $L$. The second worldsheet skein relation states that, as before, a curve close to an elliptic singularity is equal to the difference of the two different perturbations of the normalisation of the singular curve.
\end{itemize}

Because hyperbolic and elliptic singularities are the only singularities of holomorphic curves that appear in generic 1-parameter families, it follows that the count of holomorphic curves in the worldsheet skein is an invariant of $(X,L)$. Their count defines the \emph{worldsheet partition function}
\begin{equation}
    Z^{ws}_{(X,L)} \coloneqq \sum_{(u,\Sigma) \in \mathcal M^{bare}} w(u) \langle u \rangle \in \Sk^{ws}(L),
\end{equation}
where $w(u)$ denotes the (orbifold) weight of $u$ and $\langle u \rangle$ denotes the class of $u$ in $\Sk^{ws}(L)$. 

Choices of additional data in the form of a 4-chain and a framing vector field for $L$ as in the definition of the skein-valued Gromov-Witten invariants of $(M,L)$ in \cite{Ekholm:2019yqp} induce a linear map $\Sk^{ws}(M,L) \to \Sk(L)$ which maps the worldsheet partition functions $Z^{ws}_{(X,L)}$ to the previously defined skein-valued partition function $Z_{(X,L)}$.
We refer to \cite{Ekholm:2024ceb} for more details.
\subsection{Knot-quiver correspondence}

The geometric construction 
of \cite{Ekholm:2019yqp}, discussed above, can be translated to the language of symmetric quivers using the knot-quiver correspondence \cite{Kucharski:2017poe,Kucharski:2017ogk}.

A quiver $Q$ is a directed graph which consists of nodes, labelled by $i=1,\ldots,n$, and arrows that connect them. We denote the~number of arrows between vertices $i$ and $j$ by $C_{ij}$, and assemble these numbers into an $n\times n$ adjacency matrix $C$. A~quiver is called symmetric if $C_{ij}=C_{ji}$.
For a~given quiver, it is important to understand the~structure of moduli spaces of its representations, i.e. spaces of linear maps $\mathbb{C}^{d_i}\to\mathbb{C}^{d_j}$, where each $\mathbb{C}^{d_i}$ is assigned to a vertex~$i$. The vector $\boldsymbol{d}=(d_1,\ldots,d_{n})\in \mathbb{N}^{n}$ is called the~dimension vector. Basic information about such spaces is encoded in their Betti numbers or their generalisations, which are captured by the motivic Donaldson-Thomas invariants $\Omega_{(d_1,\ldots,d_{n}),s}=\Omega_{\boldsymbol{d},s}$ \cite{Kontsevich:2008fj,Kontsevich:2010px,Efi12,FR1512,MR1411}. For a~symmetric quiver, these invariants are encoded in the~\emph{motivic generating series}:
\begin{equation}
    Z_Q(\boldsymbol{x},q) =
    \sum_{\boldsymbol{d}}(-q)^{\boldsymbol{d} \cdot C\cdot\boldsymbol{d}}\frac{\boldsymbol{x}^{\boldsymbol{d}}}{(q^{2};q^{2})_{\boldsymbol{d}}}
    = \sum_{d_1,\dots,d_{n}\geq 0}(-q)^{\sum_{i,j=1}^{n} C_{ij}d_i d_j}\prod_{i=1}^{n}\frac{x_i^{d_i}}{(q^{2};q^{2})_{d_i}},    \label{PQ}
\end{equation}
where a~generating parameter $x_i$ is assigned to each vertex $i$ and 
\begin{equation}
    (\alpha;q^{2})_{m} = \prod_{k=0}^{m-1}(1-\alpha q^{2k})
\end{equation}
is the~$q$-Pochhammer symbol. The~product decomposition of this series into infinite $q$-Pochhammers (a.k.a. quantum dilogarithms) determines $\Omega_{\boldsymbol{d},s}$ as follows
\begin{equation}\label{PQ-product}
    Z_Q(\boldsymbol{x},q) = \prod_{\boldsymbol{d},s}(\boldsymbol{x}^{\boldsymbol{d}}q^s;q^2)_{\infty}^{\Omega_{\boldsymbol{d},s}} = \prod_{\boldsymbol{d}\in \mathbb{N}^{n}\setminus \boldsymbol{0}} \prod_{s\in\mathbb{Z}} \prod_{k\geq 0} \Big(1 - (x_1^{d_1}\cdots x_{n}^{d_{n}}) q^{2k+s} \Big)^{\Omega_{(d_1,\ldots,d_{n}),s}}.
\end{equation}

The~\emph{knot-quiver correspondence}\,\cite{Kucharski:2017poe,Kucharski:2017ogk}
is a~conjecture that for each knot~$K$ there exist a~quiver~$Q$
and integers $m_{i}$, $l_{i}$, $k_{i}$, $i\in\{1,\dots,n\}$,
such that
\begin{equation}\label{eq:KQ-corr}
Z_{K}(x,a,q)=\left.Z_{Q}(\boldsymbol{x},q)\right|_{x_{i}=x^{m_{i}}a^{l_{i}}q^{k_{i}}}.
\end{equation}
Until now, this statement was proven for all rational \cite{Stosic:2017wno} and arborescent knots \cite{Stosic:2020xwn}. Since we will consider quivers only in the context of the motivic generating series, we will slightly abuse the notation and write $Q=(C,\boldsymbol{x})$, where $C$ is the adjacency matrix and $\boldsymbol{x}$ is the vector of variables\footnote{The vector of variables $\boldsymbol{x}$ can be just a list of independent generating parameters $x_1,x_2,\dots$, it can contain their products together with powers of $q$, or be considered after the application of $x_{i}=x^{m_{i}}a^{l_{i}}q^{k_{i}}$ and depend on $x,a,q$. In the later part of the paper, $C$ and $\boldsymbol{x}$ will often be presented in the matrix and vector form which are quite big and have their own brackets -- in those situations, we drop the bracket from $(C,\boldsymbol{x})$ and write $Q=C,\boldsymbol{x}$ for the clarity of presentation.}.

Geometrically, the partition function of a quiver with a single node agrees with the $U(1)$-partition function of a single holomorphic disk with boundary winding once around a solid torus, where the number of loops of the quiver correspond to the framing of the boundary of the disk. To understand the meaning of this, recall that the skein-valued Gromov-Witten partition function counts holomorphic curves by their boundary on the Lagrangian 3-folds. Given an embedded holomorphic curve, this partition function then has local contributions from this basic curve and all of its multiple covers, given that these contributions are isolated in the moduli space. In the case of a basic holomorphic disk, this local contribution to the partition function in the $U(1)$-skein is given by \cite{es20}
\begin{equation}
    Z_\textnormal{di}^{f} = \sum_{d>0} q^{f d(d-1)-d}\frac{x^d}{(q^2;q^2)_d},
\end{equation}
which is precisely the partition function of a one-node quiver for the right specialisation of $x$. Here, $f$ denotes a choice of framing (or equivalently, coordinates) on the solid torus given by a small tubular neighbourhood of the boundary of the disk in the Lagrangian. If the quiver has multiple nodes, then its partition function is again equal to the partition function of disks with boundary on a solid torus, where we have one disk for each node \cite{Ekholm:2018eee,Ekholm:2021gyu}. The boundaries of all of these disks are contained in a single solid torus, where they link to form the closure of some braid. The linking numbers of the different disk boundaries then correspond precisely to the number of arrows between the corresponding nodes of the quiver. In other words, the knot-quiver correspondence states that the $U(1)$-HOMFLYPT generating function of a knot can be written as the partition functions of a finite number of basic holomorphic disks.

From the physical perspective, the knot-quiver correspondence is a~duality between two 3d $\mathcal{N}=2$ theories: one associated to the~knot and denoted $T[L_K]$, and the~other associated to the~quiver and denoted $T[Q]$ \cite{Ekholm:2018eee}.

$T[L_{K}]$ arises from M-theory over the~resolved conifold~$X$ with a~single M5-brane wrapping the~conormal Lagrangian of the~knot $L_K$ \cite{Ooguri:1999bv}: 
\begin{equation}
\begin{split}
	\text{space-time}: \quad& \mathbb{R}^4 \times S^1 \times X \\
			& \cup \phantom{ \ \times S^1 \times \ \ } \cup\\
	\text{M5}: \quad & \mathbb{R}^2\times S^1 \times L_K.
\end{split}    
\end{equation}\label{eq:M-theory-setup}

The vortex partition function of $T[L_K]$ counts BPS states arising from the M2-branes wrapping embedded holomorphic curves ending on~$L_K$ \cite{Ekholm:2019yqp,Dimofte:2010tz}. Those counts for fixed powers of $x,a,q$ (which represent different quantum numbers, see \cite{Ooguri:1999bv}) depend only on topology of the knot and are known as the Labastida-Mari\~{n}o-Ooguri-Vafa invariants \cite{Ooguri:1999bv,Labastida:2000zp,Labastida:2000yw}.

The theory $T[Q]$ is easier to describe: the~gauge group is $U(1)^{n}$ and there is one charged chiral for each $U(1)$ factor.
Interactions among the~different sectors are mediated by (mixed) Chern-Simons couplings, encoded by the~quiver adjacency matrix.
The quiver variables $x_i$ encode exponentiated Fayet-Ilioupoulos couplings.
The~partition function of vortices counts the same BPS states coming from the M2-branes wrapping embedded holomorphic curves, but this time they have different labels and the interpretation of Donaldson-Thomas invariants for quivers \cite{Kontsevich:2008fj,Kontsevich:2010px}.

The duality between $T[L_K]$ and $T[Q]$ is encoded in (\ref{eq:KQ-corr}), and interpreted as an equality of their partition functions. 

\subsection{Equivalent quivers}
\label{sec:Equivalent_quivers}

The first application of the skeins of \cite{Ekholm:2019yqp} to the knot-quiver correspondence took the form of the operations on quivers which preserve the motivic generating series. Based on the geometric and physical interpretations discussed above, authors of \cite{Ekholm:2019lmb} considered skeining a~pair of disks through each other, which creates a~new disk and changes their linking numbers. In the language of quivers it can be represented as an operation called unlinking. For 
\begin{equation}
\label{eq:generic_quiver}
Q=
    \left[\begin{customsmallmatrix}
 C_{11}  &  \cdots  &  C_{1i}  &  \cdots  &  C_{1j}  &  \cdots  &  C_{1n}\\
 \vdots  &  \  &  \vdots &   &  \vdots  &   &  \vdots\\
 C_{1i} & \cdots &  C_{ii}  &  \cdots &  C_{ij}  &  \cdots &  C_{in}\\
 \vdots&  &  \vdots&  \  &  \vdots  &   &  \vdots\\
 C_{1j}&  \cdots & C_{ij} & \cdots &  C_{jj}  &  \cdots  &  C_{jn}\\
 \vdots&  & \vdots &  & \vdots &   &  \vdots\\
 C_{1n}& \cdots & C_{in} & \cdots & C_{jn} & \cdots & C_{nn}
\end{customsmallmatrix}\right]\,,
\left[\begin{customsmallmatrix}
    x_{1}\\
    \vdots\\
    x_{i}\\
    \vdots\\
    x_{j}\\
    \vdots\\
    x_{n}
\end{customsmallmatrix}\right]
\end{equation}
the \emph{unlinking} of distinct nodes $i$ and $j$ is defined as as 
\begin{equation}\label{eq:unlinking definition}
 U(ij)Q =  
    \left[\begin{customsmallmatrix}
 C_{11}  &  \cdots &  C_{1i}  &  \cdots &  C_{1j}  &  \cdots &  C_{1n}  &  C_{1i}+C_{1j}\\
\vdots  &  \  &  \vdots &   &  \vdots  &   &  \vdots &  \vdots\\
C_{1i} & \cdots &  C_{ii}  &  \cdots &  C_{ij}-1  &  \cdots &  C_{in}  &  C_{ii}+C_{ij}-1\\
 \vdots&  &  \vdots&  \  &  \vdots  &   &  \vdots &  \vdots\\
 C_{1j}&  \cdots & C_{ij}-1 & \cdots &  C_{jj}  &  \cdots  &  C_{jn}  &  C_{ij}-1+C_{jj}\\
 \vdots&  & \vdots &  & \vdots &   &  \vdots &  \vdots\\
 C_{1n}& \cdots & C_{in} & \cdots & C_{jn} & \cdots & C_{nn}  &  C_{in}+C_{jn}\\
 C_{1i}+C_{1j} & \cdots & C_{ii}+C_{ij}-1 & \cdots & C_{ij}-1+C_{jj} & \cdots &  C_{in}+C_{jn}  &  C_{ii}+C_{jj}+2C_{ij}-1 
\end{customsmallmatrix}\right]\,,
\left[\begin{customsmallmatrix}
    x_{1}\\
    \vdots\\
    x_{i}\\
    \vdots\\
    x_{j}\\
    \vdots\\
    x_{n}\\
    q^{-1}x_{i}x_{j}
\end{customsmallmatrix}\right]\,.  
\end{equation}

Another operation which preserves the motivic generating series is the addition of a \emph{redundant pair of nodes}. Interpreting $(1;q^2)_m$ as contributions from two quiver nodes, the authors of \cite{Kucharski:2017ogk} found that the motivic generating series for
\begin{equation}\label{eq:redundant-pair-general}
    \left[
	\begin{customsmallmatrix}
		 &  &  & a_{1} & a_{1}\\
		 & C &  & \vdots & \vdots\\
		 &  &  & a_{n} & a_{n}\\
		a_{1} & \ldots & a_{n} & a_{0} & a_{0}\\
		a_{1} & \ldots & a_{n} & a_{0} & a_{0}+1
	\end{customsmallmatrix}\right],
    \left[
    \begin{customsmallmatrix}
	\phantom{x}	\\
		 \boldsymbol{x} \\
		 \phantom{x}	\\
		x_{n+1} \\
		q^{-1}x_{n+1}
	\end{customsmallmatrix}\right]
\end{equation}
is the same as the one for $(C,\boldsymbol{x})$ for any $a_0,\dots,a_n,x_{n+1}$.
Geometrically, the redundant pair of nodes is represented by a pair of holomorphic disks which have the same linking to the rest of the curves, but do not link with each other; moreover, one of them has one unit of self-linking (i.e. a twist) which leads to the cancellation. From the physical point of view, we have a pair of state-anti-state which interacts with other basic BPS states in the same way and effectively annihilates each other (for more details about geometry and physics of quiver equivalences see \cite{Ekholm:2018eee,Ekholm:2019lmb,Ekholm:2021gyu}).

The third operation on quivers that preserves the motivic generating series was introduced in \cite{Ekholm:2019lmb} by considering a redundant pair of nodes such that one of them can be obtained by unlinking; an alternative approach involves analysing how reversing the orientation of the Lagrangian submanifold changes the unlinking. Both ways lead to the same operation of \emph{linking}, which for distinct nodes $i$ and $j$ is defined as
\begin{equation}\label{eq:linking definition}
L(ij)Q =  
    \left[\begin{customsmallmatrix}
 C_{11}  &  \cdots &  C_{1i}  &  \cdots &  C_{1j}  &  \cdots &  C_{1n}  &  C_{1i}+C_{1j}\\
\vdots  &  \  &  \vdots &   &  \vdots  &   &  \vdots &  \vdots\\
C_{1i} & \cdots &  C_{ii}  &  \cdots &  C_{ij}+1  &  \cdots &  C_{in}  &  C_{ii}+C_{ij}\\
 \vdots&  &  \vdots&  \  &  \vdots  &   &  \vdots &  \vdots\\
 C_{1j}&  \cdots & C_{ij}+1 & \cdots &  C_{jj}  &  \cdots  &  C_{jn}  &  C_{ij}+C_{jj}\\
 \vdots&  & \vdots &  & \vdots &   &  \vdots &  \vdots\\
 C_{1n}& \cdots & C_{in} & \cdots & C_{jn} & \cdots & C_{nn}  &  C_{in}+C_{jn}\\
 C_{1i}+C_{1j} & \cdots & C_{ii}+C_{ij} & \cdots & C_{ij}+C_{jj} & \cdots &  C_{in}+C_{jn}  &  C_{ii}+C_{jj}+2C_{ij} 
\end{customsmallmatrix}\right]\,,
\left[\begin{customsmallmatrix}
    x_{1}\\
    \vdots\\
    x_{i}\\
    \vdots\\
    x_{j}\\
    \vdots\\
    x_{n}\\
    x_{i}x_{j}
\end{customsmallmatrix}\right]\,.
\end{equation}
The physics and geometry of linking follows from the interpretations of unlinking and redundant pairs of nodes \cite{Ekholm:2019lmb}.

Subsequently, by applying the above quiver operations, the authors \cite{Jankowski:2021flt} showed that the set of minimal-size quivers corresponding to a given knot is given by the union of permutohedra polytopes. This construction was further generalised to arbitrary sequences of unlinking operations in \cite{Kucharski:2023jds} and led to a very interesting combinatorial characterisation of such equivalences; see also \cite{Kucharski:2025lcr} for the relation to (generalised) associahedra.\footnote{Also note that the unlinking operation appeared somewhat implicitly as a statement about non-uniqueness of the Nahm sum (fermionic form) of the character of some vertex operator algebra in \cite{FeiginFeiginTipunin2011}. Hence all the analysis of equivalences for symmetric quivers can be also considered for Nahm sums with non-integral symmetric matrices relevant to the Nahm conjecture. See, for example, \cite{Storzer2024}.} In parallel, the unlinking operation was used in \cite{Jankowski:2022qdp} to prove the integrality of LMOV invariants for a general rational or arborescent link whose quiver may have both positive and negative entries; the so-called quiver diagonalisation procedure, which was further categorified in \cite{DotsenkoFeiginKucharskiReineke2026}. In our work, quiver equivalences will play an instrumental role since the main conjecture will be stated for non-minimal quivers in general.

\section{HOMFLYPT generating functions in higher genera}\label{sec:HighGenusHOMFLYPT}

In this section we develop a worldsheet extension of the HOMFLYPT framework by interpreting knot invariants directly in terms of holomorphic curves and their multicover contributions. Starting from the Ooguri-Vafa formula, the ordinary HOMFLYPT polynomial is lifted to an element of the worldsheet skein by applying worldsheet skein relations to eliminate boundary components lying on $S^3$. We then propose a universal skein-valued partition function for an elementary holomorphic curve of arbitrary genus and number of boundary components, motivated by known local Gromov-Witten and disk/annulus multicover formulas, and derive its $U(1)$-skein specialisation. This naturally leads to higher genus and reduced HOMFLYPT partition functions, which generalise the generating series appearing in the knot-quiver correspondence. Finally, the multicover skein relations are related to quiver unlinking and to the quantum dilogarithm pentagon identity, providing a~geometric interpretation of wall crossing and quiver equivalences in terms of holomorphic curve contributions.

\subsection{A worldsheet generalisation of the HOMFLYPT polynomial} 

\begin{figure}[t!]
    \centering
    \includegraphics[width=0.25\linewidth]{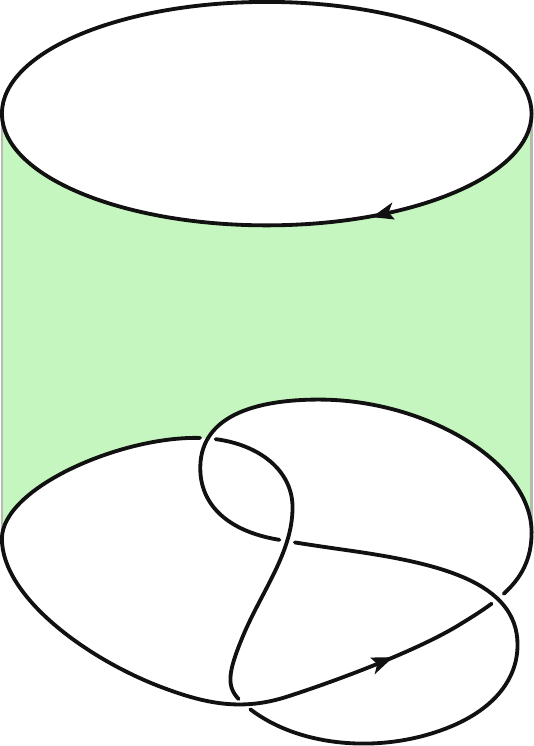}
    \caption{Holomorphic annulus, $u_{0,2}(4_1)$, with boundary component isotopic to $K=4_1$.}
    \label{fig:fig8_cylinder}
\end{figure}

In Ekholm-Shende's interpretation of the Ooguri-Vafa formula, the HOMFLYPT polynomial of a link $K$ in $S^3$ corresponds to a collection of basic annuli between $K$ and a push-off of the conormal of $K$ \cite{Ekholm:2019yqp} (see Figure \ref{fig:fig8_cylinder} as an illustration). 
Then, we apply the skein relations to the boundary components of these annuli that are embedded along $K$ until they are reduced to a multiple of the empty link.

The resulting coefficient of $W_\square$ in the skein of the solid torus is equal to the HOMFLYPT polynomial of $K$. This calculation can be generalised to the worldsheet skein: starting from the basic annuli, we apply the worldsheet skein relations until we have reduced to a linear combinations of worldsheets none of whose boundary components map to $S^3$. The resulting element in the worldsheet skein of the push-off of the conormal of $K$ (either in $T^*S^3$ or the resolved conifold $X$) is a lift of the HOMFLYPT polynomial.

\subsection{Multicover contributions of elementary holomorphic curves}
\label{sec:Multicover_contributions}

Let $u:(S, \partial S) \to (X,L)$ be a holomorphic curve. Let us make the following assumptions:
\begin{itemize}
    \item[1] Any holomorphic curve that is sufficiently $C^1$-close to a multiple cover of $u$ has to be a multiple cover of $u$.

    \item[2] For any holomorphic curve $v:(S',\partial S') \to (X,L)$ that is not a multiple cover of $u$, the boundaries $u(\partial S)$ and $v(\partial S')$ are disjoint.
\end{itemize}
This will be satisfied for instance if $u$ is super-rigid \cite{wendl23} and we have chosen a generic almost complex structure, at least below an action-cutoff. In this case, $u$ and its multiple covers will have a well-defined contribution $Z_u$ to the total Gromov-Witten partition function $Z_{(X,L)}$ of $(X,L)$, which will depend only on a holomorphic neighbourhood of $u$ in $X$. We conjecture that any Riemann surface $(S,j)$ admits an embedding into a Calabi-Yau 3-fold $X_S$ with boundary on an oriented and spin Maslov zero Lagrangian $L_S$ which is a disjoint union of $l$ solid tori satisfying 1 and 2 above whose partition function $Z_S$ is given by 
\begin{equation}\label{eq:ElementaryGWPartFct}
    Z_S = \sum_\lambda (-1)^{l |\lambda|}  q^{(2g+l-2)c_\lambda}\left( \prod_{\square \in \lambda} \left(q^{h(\square)} - q ^{- h(\square)} \right) \right)^{2g+l-2} W_\lambda \otimes ... \otimes W_\lambda \in \textnormal{Sk}\left(S^1 \times D^2\right)^{\otimes l},
\end{equation}
where $g$ and $l$ denote the genus and number of boundary components of $S$. If now a~neighbourhood of the image of $u$ is isomorphic to the local model of $(S,j)$ in $X_S$, then we can use $u$ and this identification of neighbourhoods to map a neighbourhood of $S$ into $X$ to define $u(Z_S)$, which now defines a skein-valued partition function in $X$ with boundary on $L$. By the local nature of the count of holomorphic curves, it follows that the total contribution $Z_u$ of $u$ and its multiple covers to the Gromov-Witten partition function $Z_{(X,L)}$ is precisely given by $u(Z_S)$.

\subsubsection*{Motivation}

Formula \eqref{eq:ElementaryGWPartFct} in the case that $l=0$ agrees with the formula of Bryan and Pandharipande in \cite[Corollary 7.3]{bp08} which computes the local Gromov-Witten invariant of a closed Riemann surface of genus $g$. In the case that $g=0$ and $l \in \{1,2\}$, $\eqref{eq:ElementaryGWPartFct}$ agrees with the partition function of a basic disk and a basic annulus computed by Ekholm and Shende \cite{es20,es21} for appropriate choices of orientations, spin structures and $4$-chains. In general, the skein-valued Gromov-Witten invariant of an elementary holomorphic curve is conjectured to be equivalent to relative local Gromov-Witten invariants which can be computed via the TQFT formalism of \cite{bp08}. Using the notation of \cite{bp08}, this can be made precise as follows: if we expand the partition function in \eqref{eq:ElementaryGWPartFct} (in an appropriate framing) in terms of the power sum basis $\{P_\lambda\}_\lambda$, then we expect that the coefficient of $P_{\lambda_1} \otimes ... \otimes P_{\lambda_l}$ is equal to
\begin{equation}\label{eq:LocalGWtoSkein}
    (-1)^{ld} \mathsf{GW}_{d}(g|k_1,k_2)_{\lambda_1,...,\lambda_l}|_{t_1=-t_2=-i,q=e^{\frac{iu}{2}}},
\end{equation}
where $k_1+k_2 = 2g+l-2$ and $d = |\lambda_i|$. From the TQFT formalism of \cite{bp08}, it follows that 
\begin{equation}
\begin{aligned}
    \mathsf{GW}_{d}(g|k_1,k_2)_{\lambda_1,...,\lambda_l} &= \sum_\gamma \mathsf{GW}_d(0|0,0)_{\lambda_1,...,\lambda_l}^\gamma  \mathsf{GW}(g|k_1,k_2)_\gamma,
\end{aligned}
\end{equation}
where
\begin{equation}
\begin{aligned}
    \mathsf{GW}_d(0|0,0)_{\lambda_1,...,\lambda_l}^\gamma &= (it)^{(l-1)d-\sum_i l(\lambda_i) + l(\gamma)} \sum_\rho \left(\frac{d!}{\dim\rho} \right)^{l-1} \left( \prod_{i=1}^l \frac{\chi^\rho_{\lambda_i}}{\mathfrak{z}_{\lambda_i}} \right) \chi^\rho_\gamma, \\
    \mathsf{GW}(g|k_1,k_2)_\gamma &= (it)^{-d-l(\gamma)} \sum_\rho \frac{\dim \rho}{d!}\eta_\rho^{-k_1} \overline \eta_\rho^{-k_2} \lambda_\rho^g \frac{\chi^\rho_\gamma}{\mathfrak{z}_\gamma}
\end{aligned}
\end{equation}
for 
\begin{equation}
        \lambda_\rho = (it)^{2d} \left( \frac{d!}{\dim \rho} \right)^2,\hspace{0.5cm}
        \eta_\rho = t^d q^{-c_\rho} \frac{\dim_q\rho}{\dim \rho},\hspace{0.5cm}
        \overline \eta_\rho = (-t)^d q^{c_\rho} \frac{\dim_q\rho}{\dim \rho},
\end{equation}
as long as $t_1 = -t_2 = t$. Using the identity 
\begin{equation}
    \sum_\gamma \chi^\rho_\gamma \frac{\chi^{\rho'}_\gamma}{\mathfrak{z}_\gamma} = \delta_{\rho,\rho'},
\end{equation}
it follows that 
\begin{equation}\label{eq:RelLocalGW}
    \begin{aligned}
        \mathsf{GW}_{d}(g|k_1,k_2)_{\lambda_1,...,\lambda_l} &= (it)^{(l-2)d - \sum_il(\lambda_i)} (-1)^{d(g-k_2)} t^{d(2g-k_1-k_2)} \\
       &\phantom{=} \cdot \sum_\rho \left(\frac{d!}{\dim \rho}\right)^{2g+l-2} \left( \frac{\dim_q\rho}{\dim \rho}\right)^{-k_1-k_2} q^{(k_1-k_2)c_\rho} \prod_{i=1}^l \frac{\chi^\rho_{\lambda_i}}{\mathfrak{z}_{\lambda_i}} \\
        &= (-1)^{d k_2} \sum_\rho q^{(k_1-k_2)c_\rho}\left( \prod_{\square \in \rho} \left(q^{h(\square)} - q^{- h(\square)} \right) \right)^{2g+l-2}\prod_{i=1}^l \frac{\chi^\rho_{\lambda_i}}{\mathfrak{z}_{\lambda_i}},
    \end{aligned}
\end{equation}
where we set $k_1+k_2 = 2g+l-2$ and $t = -i$ and used the fact that
\begin{equation}
    \frac{d!}{\dim_q \rho} = (-i)^d \prod_{\square \in \rho} \left(q^{h(\square)} - q^{- h(\square)} \right).
\end{equation}
Since $W_\rho = \sum_\lambda \frac{\chi^\rho_\lambda}{\mathfrak z_{\lambda}} P_\lambda$, \eqref{eq:RelLocalGW} and \eqref{eq:LocalGWtoSkein} together give \eqref{eq:ElementaryGWPartFct} for $k_1 = 2g+l-2$ and $k_2=0$. Other choices of $k_1$ and $k_2$ correspond to a different choice of framing for the skein-valued Gromov-Witten invariant.

\subsubsection*{Specialisation to the $U(1)$-skein}

In this paper, we study the behaviour of quivers associated to knots according to the knot-quiver correspondence when the knot undergoes a crossing change. Since the correspondence considers the $U(1)$-partition functions of knots, we will be mostly interested in the $U(1)$-specialisation of Gromov-Witten partition functions. For later convenience we record here the explicit expression of $Z_S$ in the $U(1)$-skein:
\begin{equation}\label{eq:U1MulticoverFormula}
    Z_S = \sum_{d\geq 0} q^{(2-2g-l)d}(q^2;q^2)_d^{2g+l-2} x_1^d \cdot...\cdot x_l^d.
\end{equation}

\subsection{Higher genus HOMFLYPT partition functions} 

In the knot-quiver correspondence for links, one considers partition functions of the form
\begin{equation}\label{eq:knotQuiverPartFact}
    Z_K = \sum_{d \geq 0} q^{(1-|\pi_0(K)|)d}(q^2;q^2)_d^{|\pi_0(K)|-1} P_{d,...,d}(K)x^d.
\end{equation}
We can now provide a geometric motivation for such partition functions. Consider a basic holomorphic curve of genus 0 with domain $\Sigma_K$ with $|\pi_0(K)| + 1$ boundary components whose boundary components are mapped to the components of $K \subseteq S^3$ and a simple curve around a solid torus. By \eqref{eq:U1MulticoverFormula}, its partition function in the $U(1)$-skein will be given by 
\begin{equation}\label{eq:linkPartFct}
    Z_{\Sigma_K} \coloneqq \sum_{d \geq 0} q^{(1-|\pi_0(K)|)d}(q^2;q^2)_d^{|\pi_0(K)|-1} P_{d,...,d}(K)x^d,
\end{equation}
which agrees with $Z_K$. If the curve also has non-trivial genus $g$, its partition function becomes
\begin{equation}\label{eq:linkPartFctWithGenus}
    Z_K^g \coloneqq Z_{\Sigma_{g,K}} \coloneqq \sum_{d \geq 0} q^{(1-|\pi_0(K)|-2g)d} (q^2;q^2)_d^{|\pi_0(K)|-1+2g} P_{d,...,d}(K)x^d.
\end{equation}
Since $Z^g_K$ is obtained from $Z_K$ by multiplying $x^d$ by $q^{-2gd}(q^2;q^2)_d^{2g}$, there exists a quiver $Q^g_K$ for $Z_K^g$ if there exists a quiver $Q_K$ for $Z_K$, and it has $4^g$ as many nodes.

\subsubsection*{The reduced HOMFLYPT partition functions}

\begin{figure}[h]
	\centering
        \begin{subfigure}{0.3\textwidth}
        \centering
            \includegraphics[width=0.3\textwidth]{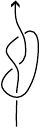}
            \caption{A representation of the trefoil in the solid torus.}
            \label{fig:trefoil in the torus}
    \end{subfigure}
    \quad\;\;\;\;\;
    \begin{subfigure}{0.3\textwidth}
    \centering
      \includegraphics[width=0.5\textwidth]{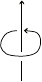}
     \caption{A representation of the Hopf link in the solid torus.}
        \label{fig:hopf link in the torus}
    \end{subfigure}
    \caption{Representation of links in the solid torus. The bottom and top of the figures are identified.}
    \label{fig:links in the solid torus}
\end{figure}

Now represent a knot $K \subseteq S^3$ as a knot in the solid torus passing once around the longitudinal direction (see Figure \ref{fig:trefoil in the torus}) and consider the partition function of a basic holomorphic disk with boundary along the knot. The resulting partition function is
\begin{equation}
    Z_K^{red} = \sum_{\lambda} \frac{(-1)^{|\lambda|}q^{-c_\lambda}}{\prod_{\square \in \lambda} (q^{h_\lambda(\square)}-q^{-h_\lambda(\square)})} P^{red}_\lambda(K) W_\lambda,
\end{equation}
which in the $U(1)$-skein (after substituting $x \to q^{-1}x$) becomes
\begin{equation}
    Z_K^{red} = \sum_{d\geq 0} \frac{P^{red}_d(K)}{(q^2;q^2)_d} x^d,
\end{equation}
which is precisely the partition function in the formulation of the knot-quiver correspondence for reduced HOMFLYPT invariants.

Similarly, the $U(1)$-partition function (after the substitution $x \to q^{-1}x$) of a genus $g$ holomorphic curve with $|\pi_0(K)|$ boundary components mapped to a link in the solid torus passing once around the longitudinal direction representing a link $K \subseteq S^3$ (see Figure \ref{fig:hopf link in the torus}) is given by
\begin{equation}
    Z^{red,g}_K = \sum_{d \geq 0}  q^{(1-|\pi_0(K)|-2g)d}(q^2;q^2)_d^{|\pi_0(K)|+2g-2} P^{red}_{d,...,d}(K) x^d.
\end{equation}

\subsubsection*{Multicover skein relations}

There is an interesting connection between the multicover contributions of holomorphic curves discussed in Section \ref{sec:Multicover_contributions}, the unlinking operation presented in Section \ref{sec:Equivalent_quivers}, and the wall-crossing relations of \cite{Kontsevich:2008fj}.

If we consider the quantum torus algebra spanned by $X_1,X_2$ which satisfy
\begin{equation}
    X_{1} X_{2} = q^2 X_{2} X_{1} 
\end{equation}
then we have the pentagon relation between their quantum dilogarithms \cite{FaddeevKashaev1994}:
\begin{equation}
\label{eq:pentagon_identity}
    (-q X_1 ; q^2)^{-1}_\infty (-q X_2 ; q^2)^{-1}_\infty = (-q X_2 ; q^2)^{-1}_\infty (-X_{1} X_{2} ; q^2)^{-1}_\infty  (-q X_1 ; q^2)^{-1}_\infty \,,
\end{equation}
which can be interpreted as the wall-crossing relation between BPS operators in two chambers of the $A_2$ Argyres-Douglas theory \cite{Kontsevich:2008fj,GaiottoMooreNeitzke2013,CecottiCordovaVafa2011}.

On the other hand, this algebra can be embedded into the quantum torus algebra spanned by two pairs of operators associated to two nodes of a quiver in such a way that \eqref{eq:pentagon_identity} can be interpreted as an equality between two motivic generating series \cite{Ekholm:2019lmb,Kucharski:2025lcr}. On the left it corresponds to a quiver with two nodes and one pair of arrows between them; on the right the corresponding quiver has three separated nodes with a loop on one of them\footnote{The generating parameter of this node is given by a $q$-shifted product of the parameters of remaining nodes.}. Comparing with equations (\ref{eq:generic_quiver}-\ref{eq:unlinking definition}) one can check that the quiver on the right can be obtained by unlinking the nodes of the quiver on the left.

From yet another perspective, the pentagon relation \eqref{eq:pentagon_identity} can be reinterpreted as a multicover $U(1)$-skein relation for the partition function of a basic holomorphic disk: one can change the linking of the boundary solid tori of two disk partition functions at the cost of adding an additional contribution of the glued disk \cite{Ekholm:2019lmb}. The relation was later extended to the full skein \cite{Hu25,Nak26,Scharitzer24}. In some cases, there also exist multicover versions of the elliptic skein relation. If the boundary of an elementary disk of one boundary of an elementary annulus is mapped to an unknot (in the latter case, we recover the HOMFLYPT partition function of an unknot), then the partition function is equal to the partition function of two spheres or two disks, respectively. For other Riemann surfaces the situation seems more complicated; for instance, Theorem 2.2 and Conjecture 2.3 in \cite{Ekholm:2024ceb} imply that the partition function of a~three-holed sphere with one boundary component mapped to an unknot can be expressed in terms of the partition functions of two annuli and eight disks. It is currently unknown how to extend the multicover skein relations to more general elementary partition functions. 

\section{Quiver decompositions from the worldsheet skein}\label{sec:Block_decompositions}

In this section, we formulate a general conjecture connecting the worldsheet skein relation with block decompositions of quivers associated with knot and link invariants. Starting from earlier observations that quivers for knots may decompose into blocks corresponding to crossing resolutions \cite{Chauhan:2025iwt}, the section proposes a~broader geometric version: 
resolving a~hyperbolic boundary-node degeneration in the worldsheet skein gives three local worldsheet configurations $S_{\pm}$ and $S_0$, whose boundary links $L_{\pm}$ and $L_0$ are related by the corresponding HOMFLYPT skein move.
In turn, we conjecture that this should induce a quiver decomposition whose diagonal blocks correspond to the above configurations with appropriate shifts in genus, framing, and monomial weights.
Unlike the earlier minimal quiver formulation, this conjecture allows non-minimal quivers and redundant pairs, which are expected to be important in the refined settings \cite{Ekholm:2021gyu}. The section also emphasises that the worldsheet skein picture carries more information than the ordinary boundary skein relation, since it keeps track of the genus and boundary structure of the contributing holomorphic curves. Finally, the proposed decomposition is interpreted physically as a decomposition of the BPS spectrum: the basic BPS states of $L_{\pm}$ are built from those of $L_{\mp}$ and $L_0$, while the off-diagonal quiver blocks encode bound state interactions between the two sectors.

\subsection{The main conjecture}

In \cite{Chauhan:2025iwt}, it was suggested that symmetric quivers corresponding to a knot $K$ can admit a block form with blocks corresponding to crossing resolutions, for some choice of crossing $c$ in the minimal diagram of $K$. Recall that the equivalent minimal-sized quivers can be described as nodes of the corresponding permutohedra graph $\Pi(K)$\cite{Jankowski:2021flt}:
\[
    \{Q_K^{(j)}\}_{j\in\Pi(K)}\,. 
\]
Then Conjecture 4.1 from \cite{Chauhan:2025iwt} states that there exists a non-empty subset $J(c)\subseteq \Pi(K)$ such that, up to a framing,
\begin{equation}\label{eq:block form old}
     \quad Q_K^{(j)} = 
     \left[\begin{array}{ c : c }
    Q_{K_\pm} & * \\
    \hdashline
    * & Q_{L_0}
    \end{array}
    \right]
\end{equation}
$\forall j\in J(c)$, where $K_\pm$ and $L_0$ are obtained from $K$ by resolving $c$. This conjecture was motivated by studying recursion relations among the $S^2$-coloured HOMFLYPT polynomials and confirmed for the figure-eight knot.

In Ekholm and Shende's interpretation of the Ooguri-Vafa formula, counts of basic holomorphic curves with boundaries on $S^3$ and push-off of conormals give the HOMFLYPT polynomial of the link \cite{Ekholm:2019yqp}. As explained in Section \ref{sec:HighGenusHOMFLYPT}, we can keep track of the genus of the curves as we apply the skein relations and we conjecture that there exists a multicover skein relation at the level of quivers.

Motivated by the geometric picture of the worldsheet skein \cite{Ekholm:2024ceb}, we formulate a more general conjecture below. In terms of the quivers in zero framing, the updated conjecture~is:

\begin{coj}[Block decomposition of quivers induced by the worldsheet skein]\label{coj:block_quiver_skein} 
    Let $L_\pm$ be a link with genus $g$ HOMFLYPT partition function $Z_{L_\pm}^g$. Assume that after applying the hyperbolic skein relation to a crossing of sign $\pm$ we get the link $L_{\mp}$ and $z$ times the link $L_0$. Then $Z_{L_\pm}^g$ admits a quiver description $Q_{L_\pm}^g$ which block decomposes as
    \begin{equation}\label{eq:block form 1}
    Q_{L_\pm}^g =  \left[\begin{array}{ c : c }
    C_{{L_{\mp}}}^{g,f_1} & * \\
    \hdashline
    * & C_{{L_{0}}}^{g',f_2}
    \end{array}
    \right], 
    \left[\begin{array}{c}
    a_1 \boldsymbol x^{f_1}_{{L_{\mp}}} \\
    \hdashline
    a_2 \boldsymbol x^{f_2}_{{L_{0}}}
    \end{array}
    \right].
    \end{equation}
    Here, $g' = g$ if the skein relation is applied to a self-crossing of a component of $L_\pm$ and $g' = g+1$ otherwise,
    \begin{equation}
        \begin{aligned}
            f_1 &= wr(L_\mp)-wr(L_\pm) = \mp 2,\\
            f_2 &= wr(L_0)-wr(L_\pm)+\sigma = \sigma \mp 1,
        \end{aligned}
    \end{equation} 
    where $\sigma \in \{\pm1\}$ if the skein relation is applied to a negative crossing and $\sigma = 0$ otherwise, and
    \begin{equation}
        \begin{aligned}
            a_1&= a^{wr(L_\mp)-wr(L_\pm)} = a^{\mp2},\\
            a_2&= a^{wr(L_0)-wr(L_\pm)} = a^{\mp1}.
        \end{aligned}
    \end{equation}
\end{coj}

\begin{rmk}
Recall that the local HOMFLYPT skein relations are usually drawn in the blackboard framing which is given by $f(K) = wr(K)$. Correspondingly, the block decomposition using the blackboard framing is
\begin{equation}\label{eq:block form 2}
    Q_{bb,L_\pm}^g =  \left[\begin{array}{ c : c }
    C_{bb,L_{\mp}}^{g} & * \\
    \hdashline
    * & C_{{bb,L_{0}}}^{g',\sigma}
    \end{array}
    \right], 
    \left[\begin{array}{c}
    \boldsymbol x_{{bb,L_{\mp}}} \\
    \hdashline
    \boldsymbol x^\sigma_{{bb,L_{0}}}
    \end{array}
    \right]
\end{equation}
For our examples in Section \ref{sec:Examples}, however, the zero quiver-framing form \eqref{eq:block form 1} will be more convenient.
\end{rmk}

\noindent The key features of the above decompositions in comparison with the older conjecture \eqref{eq:block form old}, are:
\begin{itemize}
    \item First, minimality of the quiver block is not assumed, i.e.\,it may contain some number of redundant pairs which can be removed via a suitable sequence of symmetry transpositions in the sense of \cite{Jankowski:2021flt}. 
    \item Second, the worldsheet skein picture contains more information than just the boundary skein. Keeping track of both genus and the number of boundary components will be essential for our case studies.
\end{itemize}
\begin{rmk}\label{rmk:off_diagonal}
Conjecture \ref{coj:block_quiver_skein} does not clarify the structure of the off-diagonal blocks in a block decomposition. In some cases, however, these can be derived directly by using the redundant pair condition; see the discussion in Section \ref{sec:Examples}.
\end{rmk}

\begin{rmk}\label{rmk:reduced vs unreduced}
    While the conjecture is stated for the unreduced invariants, we expect it to also to hold for the reduced invariants. As was noted in \cite{Kucharski:2017ogk}, the relation between quivers corresponding to reduced and unreduced HOMFLYPT invariants is especially simple: multiplying by an unknot factor corresponds to doubling the quiver size, where the first block is the same as the initial quiver, and the second block is that quiver shifted by one unit of framing. 
    Therefore, if our conjecture is tested for the reduced invariants, it will also hold for the unreduced invariants. In Section \ref{sec:Examples} for simplicity we will be dealing with the reduced case.
\end{rmk}

We believe that Conjecture \ref{coj:block_quiver_skein} is an important step towards geometric understanding of the link generalisation of the knot-quiver correspondence. In what follows, we will reformulate twist recursions from \cite{Chauhan:2025iwt} using this new approach and provide some examples supporting our main conjecture.

\begin{rmk}
\label{rmk:arbitrary_colours}
    The main theorem of \cite{Ekholm:2024ceb} in the case of the Hopf link suggests that perhaps there is a generalisation of the knot-quiver correspondence to arbitrary colours, where the role of the quiver nodes is now played by the partition functions of holomorphic curves. The transformation between the all-colour quiver-like decomposition to the quiver for the partition function with the same symmetric colour in the case of Hopf link is detailed in Appendix \ref{app:hopf}.
\end{rmk}

\subsection{Physical interpretation}

Conjecture \ref{coj:block_quiver_skein} has interesting physical interpretation in terms of BPS states in the 3d $\mathcal{N}=2$ theories associated to the knot and the quiver \cite{Ekholm:2018eee}. (The knot-quiver correspondence \cite{Kucharski:2017poe,Kucharski:2017ogk} implies that the BPS spectrum for both theories -- encoded in Labastida-Mari\~{n}o-Ooguri-Vafa invariants for knots \cite{Ooguri:1999bv,Labastida:2000zp,Labastida:2000yw} and Donaldson-Thomas invariants for quivers \cite{Kontsevich:2008fj,Kontsevich:2010px} -- is the same, so for simplicity we will refer to links, having in mind that the same statements are true for the corresponding quivers.) The block decomposition of the quiver corresponding to $L_{\pm}$ into subquivers corresponding to $L_{\mp}$ and $L_{0}$ implies that the BPS spectrum for $L_{\pm}$ is composed of the spectra for $L_{\mp}$ and $L_{0}$ plus the bound states encoded in the non-diagonal terms denoted by $*$ in \eqref{eq:block form 1}. This means that the structure of BPS states and their interactions is quite rigid and intertwined between different cases: the basic states (which are encoded in the quiver nodes) for $L_{\pm}$ are given by the basic states for $L_{\mp}$ and $L_{0}$. If two basic states for $L_{\pm}$ are also the basic states for $L_{\mp}$ (or $L_0$), their interactions are the same. The only free parameters are the arrows encoding the interactions between the pair composed of one basic state for $L_{\mp}$ and one for $L_0$ -- they are hidden in $*$ in \eqref{eq:block form 1}.


\section{Knot operations and quiver partition functions}
\label{sec:knot_operations}

There are many different operations that can be performed on a collection of links, some of which behave naturally with respect to the associated quivers. In this section, we describe the action on the quivers of several different operations on links and their partition functions. Whenever applicable, we will also relate the corresponding operations to the worldsheet skein relation.

\subsection{Knot sums}

For a link $K$, denote by $P_{d}^{red}(K)$ its reduced, symmetrically $(d,...,d)$-coloured HOMFPLYPT polynomial. Assume that we have a quiver description for genus $g_i$-curves on two links $K_i$ for $i \in \{1,2\}$ with $l_i$ components as follows:

\begin{equation}
    \sum_{d \geq 0} q^{-d(2g_i+l_i-1)} (q^2;q^2)_d^{l_i+2g_i-2} P_{d}^{red}(K_i) x^d = \sum_{\boldsymbol{d}_i} (-q)^{\boldsymbol{d}_i \cdot C_i \cdot \boldsymbol{d}_i} \frac{\boldsymbol{x}_i^{\boldsymbol{d}_i}}{(q^2;q^2)_{\boldsymbol d_i}},
\end{equation}
where $\boldsymbol{d}_i = (d_i^1,d_i^2,...,d_i^{n_i})^T$, $\boldsymbol{x}_i = (x_i^1,x_i^2,...,x_i^{n_i})^T$, and $C_i$ is an $n_i \times n_i$-matrix.

The reduced HOMFLYPT polynomials of the connected sum $K_1 \# K_2$ (with the induced framing) are given by
\begin{equation}
    P_{d}^{red}(K_1 \# K_2) = P_{d}^{red}(K_1) P_{d}^{red}(K_2).
\end{equation}
From this, one computes that the reduced, genus $g=g_1+g_2$, symmetrically coloured HOMFLYPT polynomials of $K_1\# K_2$ admit a quiver description with $n_1 n_2$ nodes with linking matrix of block form
\begin{equation}
    \renewcommand\arraystretch{1.25}
\left[
\begin{array}{c|c|c|c}
 C_1 + (C_2)_{1,1} & C_1 + (C_2)_{1,2} + U &  C_1 + (C_2)_{1,3} + U & \dots\\
 \hline
 C_1 + (C_2)_{2,1} + U^T & C_1 + (C_2)_{2,2} &  C_1 + (C_2)_{2,3} + U & \dots\\
 \hline
 C_1 + (C_2)_{3,1} + U^T & C_1 + (C_2)_{3,2} + U^T &  C_1 + (C_2)_{3,3} & \dots\\
 \hline
 \vdots & \vdots &  \vdots & \ddots\\
\end{array}
\right],
\end{equation}
where $U$ denotes the strictly upper triangular matrix of 1's,
there are $n_2 \times n_2$ blocks, and the block in position $(k_1,k_2)$ is given by $C_1 + (C_2)_{k_1,k_2}$ if $k_1 = k_2$, by $C_1 + (C_2)_{k_1,k_2} + U$ if $k_1 < k_2$, and by $C_1 + (C_2)_{k_1,k_2} + U^T$ if $k_1 > k_2$.

The linear terms are given by 
\begin{equation}
    \boldsymbol x = (x_1^1 x_2^1, x_1^2 x_2^1,..., x_1^{n_1} x_2^1, x_1^1 x_2^2, x_1^2 x_2^2,..., x_1^{n_1} x_2^{n_2}).
\end{equation}

Assume now that Conjecture \ref{coj:block_quiver_skein} holds for a triple $K_1^\circ, K_1^+, K_1^-$. Then it also holds for the triple $K_1^\circ \# K_2, K_1^+ \# K_2, K_1^- \# K_2$ for any link $K_2$.

\subsection{Adding genus}
The effect of increasing the genus on the partition function is to multiply $x^d$ by a factor of $q^{-2d}(q^2;q^2)_d^2$. As explained in \cite[Section 4]{Kucharski:2017ogk}, it follows that if a partition function $Z$ has associated quiver $Q = (C, \boldsymbol{{x}})$, then $Z^{g=1}$ is equal to the quiver partition function of 
\begin{equation}
    Q^{g=1}  = \left[
\begin{array}{c|c|c|c}
C & C + U & C+U & C+U+1\\ 
 \hline 
 C + U^T &  C+1 & C+U^T & C+U+1\\
 \hline 
 C+U^T & C+U &  C+1 & C+U+1 \\
 \hline
 C+U^T+1 & C+U^T+1 & C+U^T+1 & C+2 \\
\end{array}
\right], \left[
\begin{array}{c}
 q^{-2} \boldsymbol{x}\\ 
 \hline 
 q^{-1} \boldsymbol{x}\\
 \hline 
 q^{-1} \boldsymbol{x} \\
 \hline
 \boldsymbol{x} \\
\end{array}
\right],
\end{equation}

where $U$ denotes the strictly upper triangular matrix with entries 1.

This shows that if Conjecture \ref{coj:block_quiver_skein} holds for links $L_\pm$ and $L_0$ for some genus $g$, then it also holds for all genera $g' \geq g$.

\subsection{Adding full twists}\label{sec:Twist_recursions}

The stable growth conjecture of \cite{Chauhan:2025iwt} attempts to construct quivers for knots which differ by a number of full twists. Below we will obtain its reformulation in terms of Conjecture \ref{coj:block_quiver_skein}. Furthermore, we will find a new linking identity for augmented quivers which will be useful for our case studies.

\subsubsection*{Augmented quivers}

Augmented quivers were defined in \cite{Chauhan:2025iwt} and are instrumental in implementing the twist recursion using operations of linking or unlinking \cite{Ekholm:2019lmb,Kucharski:2023jds}.
We slightly modify the previous definition by including one more node of $x$-degree 0; the reason for that will become clear shortly. When the orientation of strands is opposite and the $a$-degrees of the uncoloured HOMFLYPT homology generators stablilise in the positive half-plane $\deg(a)>0$ as the number of twists goes to infinity\footnote{This ensures that the uncoloured homology grows vertically via (0,2)-splitting, see also \cite{Jankowski:2021flt,Chauhan:2025iwt}. In the opposite situation $\deg(a)<0$ we need to apply mirror operation $C\mapsto I-C$.}, we define
\begin{equation}\label{eq:aug quiver normalised}
    Q_{K_i}^+ := 
    \left[
\begin{array}{cc|cc}
 0 & 0 & 0 & 0 \\
 0 & 1 & 0 & 1 \\
 \hline
 0 & 0 & Q_{K_{i-1}} & * \\
 0 & 1 & * & R \\
\end{array}
\right],
\left[
\begin{array}{c}
  a^2 \\
  a^2 q^{-1} \\
  \hline
 \boldsymbol x_{{K_{i-1}}} \\
 \boldsymbol x_{{R}} \\
\end{array}
\right]\,
\end{equation}
whose nodes we label as $\left(0',0,1,2,\dots,n\right)$.
Here $Q_{K_i}=Q_{K_i}^+\setminus \{0,0'\}$ is a quiver corresponding to a knot $K_i$ in the sequence $\{K_0,K_1,\dots\}$, where $K_{i+1}$ differs from $K_i$ by one extra full twist in a fixed twist region. In order to define the augmentation~\eqref{eq:aug quiver normalised}, one first has to find the block decomposition
\begin{equation}
    Q_{K_i} = 
    \left[
\begin{array}{cc}
 Q_{K_{i-1}} & * \\
 * & R \\
\end{array}
\right],
\left[
\begin{array}{c}
 \boldsymbol x_{{K_{i-1}}} \\
 \boldsymbol x_{{R}} \\
\end{array}
\right]\,.
\end{equation}
Once the proper block form is found (usually for $i=1$), quivers for other knots in the family are obtained via unlinking all arrows connecting the 1-loop augmentation node to other nodes in $R$:
\begin{equation}
    Q_{K_{i+1}}^+ = \prod_{\iota \in R}U(0,\iota)Q_{K_i}^+.
\end{equation}
By iterating this procedure, one obtains $Q_{K_{i+1}}^+$, $Q_{K_{i+2}}^+$, and so on from $Q^+_{K_{i}}$. Note that the nodes $\left(0',0\right)$ become a redundant pair after an infinite number of unlinkings.

When the orientation of the strands in a twist region is the same (for example, for torus knots $T_{2,2p+1}$),\footnote{This convention also assumes that the homological zig-zag is oriented upwards, otherwise one has to consider the knot mirror operation on the augmentation. For simplicity and compactness of expressions, in this section we sometimes abuse our convention and use $Q$ to denote both a pair $(C,\boldsymbol{x})$ and the adjacency matrix $C$.} we define
\begin{equation}\label{eq:aug quiver normalised same}
    Q_{K_i}^+ := 
    \left[
\begin{array}{cc|cc}
 1 & 0 & 2 & 2 \\
0 & 0 & 2 & 1 \\
 \hline
 2 & 2 & Q_{K_{i-1}} & * \\
 2 & 1 & * & R \\
\end{array}
\right],
\left[
\begin{array}{c}
  q \\
  q^2 \\
  \hline
 \boldsymbol x_{{K_{i-1}}} \\
 \boldsymbol x_{{R}} \\
\end{array}
\right]
\end{equation}
with the same labelling of nodes as above. Analogously to the above case, once the proper block form is found, quivers for other knots in the family are obtained via \emph{linking}:
\begin{equation}
    Q_{K_{i+1}}^+ = \prod_{\iota \in R}L(0,\iota)Q_{K_i}^+.
\end{equation}

\begin{rmk}
    Note that augmented quivers can be interpreted directly as those corresponding to an infinitely twisted link or a projector inserted in the twist region, depending on the twist orientation.
    By considering properly the normalised augmentation quiver \eqref{eq:aug quiver normalised}, one can write its partition function as
    \begin{equation}\label{eq:aug quiver pf}
            Z_{Q_{K_i}^+}(x,a,q) = \sum_{r \geq 0} \frac{P_{r;K_\infty}(a,q)}{(q^2;q^2)_r}x^r.
    \end{equation}
    Similarly, for the same orientation, denoting $K_\infty^\parallel$ the corresponding projector, we can write
    \begin{equation}\label{eq:aug quiver pf same}
            Z_{Q_{K_i}^+}(x,a,q) = \sum_{r \geq 0} \frac{P_{r;K_\infty^\parallel}(a,q)}{(q^2;q^2)_r}x^r.
    \end{equation}
    Note that the nodes $\left(0',0\right)$ in the infinite (un)linking limit for both \eqref{eq:aug quiver normalised} and \eqref{eq:aug quiver normalised same} become a~redundant pair.
\end{rmk}

\subsubsection*{The linking identity}\label{sec:linking identity}

In \cite{Chauhan:2025iwt} it was observed that the augmented quivers for those knot families where the twisting strands have opposite orientations, correspond to a 0-resolution of that twist region, i.e. $L_0$. Here we prove this in the case of figure-eight knot and conjecture a more general identity which implies this correspondence.
To this end, we will make use of augmented quivers written in the form \eqref{eq:aug quiver normalised} or \eqref{eq:aug quiver normalised same}.

Consider the augmented quiver for $K=4_1$, written in the form \eqref{eq:aug quiver normalised}:
\begin{equation}\renewcommand\arraystretch{1.25
}\label{eq:4_1 augmented quiver}
Q^+_{4_1}= 
\left[
\begin{array}{cc|ccccc}
 0 & 0 & 0 & 0 & 0 & 0 & 0\\
 0 & 1 & 0 & 1 & 1 & 1 & 1 \\
 \hline
 0 & 0 & 0 & -1 & -1 & 0 & 0 \\
 0 & 1 & -1 & -2 & -2 & -1 & 0 \\
 0 & 1 & -1 & -2 & -1 & -1 & 0 \\
 0 & 1 & 0 & -1 & -1 & 1 & 1 \\
 0 & 1 & 0 & 0 & 0 & 1 & 2 \\
\end{array}
\right],
\left[
\begin{array}{c}
  a^2 \\
  a^2 q^{-1} \\
  \hline
 x \\
 a^{-2} q^{2} x \\
 q^{-1} x \\
 q x \\
 a^2 q^{-2} x\\
\end{array}
\right]\,.
\end{equation}
According to the stable growth conjecture \cite{Chauhan:2025iwt}, due to the presence of the augmentation node, this quiver encodes the (unreduced) Hopf link invariants. Previously that was only checked up to a few orders in $x$. Below we provide a proof based on quiver equivalences. First, link one pair of arrows:
\begin{equation}\renewcommand\arraystretch{1.25
}
L(01)Q^+_{4_1}= 
\left[
\begin{array}{cc|cccc|cc}
 0 & 0 & 0 & 0 & 0 & 0 & 0 & 0 \\
 0 & 1 & 1 & 1 & 1 & 1 & 1 & 1 \\
 \hline
 0 & 1 & 0 & -1 & -1 & 0 & 0 & 0 \\
 0 & 1 & -1 & -2 & -2 & -1 & 0 & 0 \\
 0 & 1 & -1 & -2 & -1 & -1 & 0 & 0 \\
 0 & 1 & 0 & -1 & -1 & 1 & 1 & 1 \\
 \hline
 0 & 1 & 0 & 0 & 0 & 1 & 2 & 1 \\
 0 & 1 & 0 & 0 & 0 & 1 & 1 & 1 \\
\end{array}
\right],
\left[
\begin{array}{c}
  a^2 \\
  a^2 q^{-1} \\
  \hline
 x \\
 a^{-2} q^{2} x \\
 q^{-1} x \\
 q x \\
 \hline
 a^2 q^{-2} x\\
 a^2 q^{-1} x
\end{array}
\right]\,.
\end{equation}

This produces a new node which forms a redundant pair with one of the generators in $Q_{4_1}$. It is easy to see that the remaining part corresponds to the reduced invariants of $L_{2,2}$, see also Appendix \ref{app:hopf}:
\begin{equation}\renewcommand\arraystretch{1.25
}
Q^{f.r.}_{L_{2,2}}=
\left[
\begin{array}{cc|cccc}
 0 & 0 & 0 & 0 & 0 & 0\\
 0 & 1 & 1 & 1 & 1 & 1\\
 \hline
 0 & 1 & 0 & -1 & -1 & 0\\
 0 & 1 & -1 & -2 & -2 & -1\\
 0 & 1 & -1 & -2 & -1 & -1\\
 0 & 1 & 0 & -1 & -1 & 1\\
\end{array}
\right],
\left[
\begin{array}{c}
  a^2 \\
  a^2 q^{-1} \\
  \hline
 x \\
 a^{-2} q^{2} x \\
 q^{-1} x \\
 q x
\end{array}
\right]\,.
\end{equation}
Indeed, the first two nodes contribute the normalisation factor $(a^2;q^2)_r^{-1} = (a^2;q^2)^{-1}_{\infty}(a^2q^{2r};q^2)_\infty$. We can therefore write
\begin{equation}\label{eq:aug quiver pf 2}
            Z_{Q_{K_i}^+}(x,a,q) = \sum_{r \geq 0} (a^2;q^2)_r^{-1}\frac{P^{\rm fin}_{r;L_0}(a,q)}{(q^2;q^2)_r}x^r = \sum_{r \geq 0} \frac{P_{r,L_0}(a,q)}{P_{r,0_1}(a,q)} x^r
\end{equation}
with $K_i=4_1$ and $L_0=L_{2,2}$.

This example suggests the following generalisation:
\begin{equation}\label{eq:linking identity ap}
\prod_{\iota \in Q_{K_{i-1}}} L(0,\iota)
    \left[
\begin{array}{cc|cc}
 0 & 0 & 0 & 0 \\
 0 & 1 & 0 & 1 \\
 \hline
 0 & 0 & Q_{K_{i-1}} & * \\
 0 & 1 & * & R \\
\end{array}
\right],
\left[
\begin{array}{c}
  a^2 \\
  a^2 q^{-1} \\
  \hline
 \boldsymbol x_{{K_{i-1}}} \\
 \boldsymbol x_{{R}} \\
\end{array}
\right]\,
\overset{?}{\simeq}
\left[
\begin{array}{cc|c}
 0 & 0 & 0 \\
 0 & 1 & 1 \\
 \hline
 0 & 1 & Q_{L_0} \\
\end{array}
\right],
\left[
\begin{array}{c}
  a^2 \\
  a^2 q^{-1} \\
  \hline
 \boldsymbol x_{{L_0}} \\
\end{array}
\right]\,,
\end{equation}
where $Q_{L_0}$ is minimal, i.e.\ its nodes are in 1:1 correspondence with generators of finite HOMFLYPT link homology, and $\simeq$ denotes quiver equivalence via cancellation of redundant pairs.

A similar conjectural identity can be written for the same orientation of strands:
\begin{equation}
\prod_{\iota \in Q_{K_{i-1}}} U(0,\iota)
    \left[
\begin{array}{cc|cc}
 1 & 0 & 2 & 2 \\
 0 & 0 & 2 & 1 \\
 \hline
 2 & 2 & Q_{K_{i-1}} & * \\
 2 & 1 & * & R \\
\end{array}
\right],
\left[
\begin{array}{c}
  q \\
  q^2 \\
  \hline
 \boldsymbol x_{{K_{i-1}}} \\
 \boldsymbol x_{{R}} \\
\end{array}
\right]\,
\overset{?}{\simeq}
\left[
\begin{array}{cc|c}
 1 & 0 & 2 \\
 0 & 0 & 1 \\
 \hline
 2 & 1 & Q_{K_\infty^\parallel} \\
\end{array}
\right],
\left[
\begin{array}{c}
  q \\
  q^2 \\
  \hline
 \boldsymbol x_{{K_\infty^\parallel}} \\
\end{array}
\right]\,.
\end{equation}

Here the overall normalisation factor on the right side is different from \eqref{eq:aug quiver pf 2}:
\begin{equation}\label{eq:aug quiver pf 2 same}
            Z_{Q_{K_i}^+}(x,a,q) = \sum_{r \geq 0} (q^{2r+2};q^2)_r^{-1}\frac{\bar{P}_{r;K_\infty^\parallel}(a,q)}{(q^2;q^2)_r}x^r = \sum_{r \geq 0} \frac{\bar{P}_{r;K_\infty^\parallel}(a,q)}{(q^2;q^2)_{2r}}x^r.
\end{equation}

\begin{rmk}
    As we will see in Section \ref{sec:Examples}, the identity \eqref{eq:linking identity ap} will be extremely useful for verifying Conjecture \ref{coj:block_quiver_skein} in specific cases. On the other hand, identity \eqref{eq:aug quiver pf 2 same} can be used to find minimal quivers corresponding to limits of parallel twists, further improving the results of \cite{Chauhan:2025iwt}. The existence of such finite quivers is not manifest at all, and we believe that these identities will help to understand their meaning.
\end{rmk}

\subsubsection*{Twist recursions via Conjecture \ref{coj:block_quiver_skein}}

We proceed with our main result in this section, which connects the twist recursions of~\cite{Chauhan:2025iwt} with block decomposition of quivers from the worldsheet skein in Section \ref{sec:Block_decompositions}. We will consider the two orientations separately, and then interpret the linking identity using such decompositions.
More concretely, if we ignore the off-diagonal terms, we can derive the twist recursion by using the block decompositions of quivers in Conjecture \ref{coj:block_quiver_skein}.

Let $Q_i$ denote a quiver description of a link $K_i$ in a twist family, where the twisting is applied to strands of opposite orientations belonging to the same component of $K_i$, and we apply positive twists (meaning that the crossings in the twist region will be negative). Let $L_0$ denote the link obtained by splicing $K_0$. Following the skein equation (in blackboard framing)
\begin{equation}
    K_i = K_{i-1} - z a^{-2i+1} L_0 
\end{equation}
for $i \geq 2$,
we can write $Q_i$ as
\begin{equation}\label{eq:TwistBlockDec1}
    Q_{i} = \left[
\begin{array}{c|c}
 a^{2} Q_{i-1} & * \\ 
 \hline
 * & a Q^{f=-1}_{L_0} \\
\end{array}
\right],
\end{equation}
where all quivers are in framing zero.
Here, we make use of the following notation:
Given a~quiver description $Q = (C, \boldsymbol x)$, we define $a^{k_1} q^{k_2} Q^{f=k_3} \coloneqq (C^{f=k_1+k_3}, a^{k_1} q^{-k_1+k_2-k_3}\boldsymbol x)$.

We can solve this block decomposition for $Q_{L_0}$ to obtain
\begin{equation}\label{eq:TwistBlockDec2}
    Q_{L_0} = \left[
\begin{array}{c|c}
 a^{-1} Q_{i}^{f=1} & * \\ 
 \hline
 * & a Q^{f=2}_{i-1} \\
\end{array}
\right]
=
\left[
\begin{array}{c|c}
 a^{-1} Q_{i-1}^{f=1} & * \\ 
 \hline
 * & a Q^{f=2}_{i-2} \\
\end{array}
\right].
\end{equation}
This block from for $Q_{L_0}$ can then be substituted in \eqref{eq:TwistBlockDec1} to obtain (after a reordering of the blocks)
\begin{equation}\label{eq:TwistBlockDec3}
    Q_{i} =
\left[
\begin{array}{c|c|c}
  Q_{i-1} & * & *\\ 
 \hline
 * & a^{2}Q_{i-1} & * \\
 \hline
 * & * & a^{2} Q_{i-2}^{f=1}
\end{array}
\right].
\end{equation}

Then, as above, we define the twice augmented quiver $Q_{i}^+$ as
\begin{align}
    Q_{i}^+ &=  
\left[
\begin{array}{cc|c|c|c}
 0 & 0 & 0 & 0 & 0 \\
 0 & 1 & 0 & 1 & 1 \\
 \hline
 0 & 0 & Q_{i-1} & * & *\\ 
 \hline
 0 & 1 & * & a^{2} Q_{i-1} & * \\
 \hline
 0 & 1 & * & * & a^{2} Q_{i-2}^{f=1}
\end{array}
\right]\,,
\end{align}
where $x_{0'} = a^2$ and $x_0 = a^2 q^{-1}$.

After unlinking the last two blocks from the auxiliary node, we obtain
\begin{align}
\left[
\begin{array}{cc|c|c|c|c|c}
 0 & 0 & 0 & 0 & 0 & 0 & 0 \\
 0 & 1 & 0 & 0 & 0 & 1 & 1 \\
 \hline
 0 & 0 &  Q_{i-1} & * & * & * & *\\ 
 \hline
 0 & 0 & * & a^{2}Q_{i-1} & * & * & *\\
 \hline
 0 & 0 & * & * & a^{2} Q_{i-2}^{f=1} & * & *\\
 \hline
 0 & 1 & * & * & * & a^4 Q_{i-1} & * \\
 \hline 
 0 & 1 & * & * & * & * & a^4 Q_{i-2}^{f=1}
\end{array}
\right].
\end{align}
Three of the blocks combine to give $Q_i$, and after adding a cancelling pair of blocks corresponding to $L_0$, the quiver description becomes
\begin{align}
&\left[
\begin{array}{cc|c|c|c|c|c}
 0 & 0 & 0 & 0 & 0 & 0 & 0 \\
 0 & 1 & 0 & 1 & 1 & 1 & 1 \\
 \hline
 0 & 0 &  Q_{i} & * & * & * & *\\ 
 \hline
 0 & 1 & * & a^4 Q_{i-1} & * & * & *\\
 \hline
 0 & 1 & * & * & a^3 Q_{L_0}^{f=-1} & * & *\\
 \hline
 0 & 1 & * & * & * & a^3 Q_{L_0} & * \\
 \hline 
 0 & 1 & * & * & * & * & a^{4} Q_{i-2}^{f=1}
\end{array}
\right]\\
&= \left[
\begin{array}{cc|c|c|c}
 0 & 0 & 0 & 0 & 0 \\
 0 & 1 & 0 & 1 & 1 \\
 \hline
 0 & 0 & Q_{i} & * & *\\ 
 \hline
 0 & 1 & * & a^2 Q_{i} & * \\
 \hline
 0 & 1 & * & * & a^2 Q_{i-1}^{f=1}
\end{array}
\right] = Q_{i+1}^+,
\end{align}
where in going from the first to the second line, we used \eqref{eq:TwistBlockDec1} to combine the last four blocks to $a^2 Q_{i}$ and $Q_{i-1}^{f=1}$.

\begin{rmk}
    In the same way one can analyse the case that the two strands in the twist region belong to two different components of $K_i$. In this case, $Q_{L_0}$ in the above calculation needs to be replaced by $Q_{L_0}^{g=1}$.
\end{rmk}

Now we assume that in the twist region for the family of knots $K_i$, the strands are oriented in the same direction. In this case, the skein relation has the form
\begin{equation}
    K_i = K_{i-1} - z L_i, 
\end{equation}
which we lift to the quiver decomposition
\begin{equation}\label{eq:TwistBlockDec1SameOr}
    Q_{i} = \left[
\begin{array}{c|c}
 a^{2} Q_{i-1} & * \\ 
 \hline
 * & a Q^{f=-1}_{L_i} \\
\end{array}
\right].
\end{equation}
Similarly, we lift the skein relation
\begin{equation}
     L_i = L_{i-1} - z K_{i-1}
\end{equation}
to the quiver decomposition
\begin{equation}\label{eq:TwistBlockDec2SameOr}
    Q_{L_i} = \left[
\begin{array}{c|c}
 a^{2} Q_{L_{i-1}} & * \\ 
 \hline
 * & a Q^{g=1,f=-1}_{{i-1}} \\
\end{array}
\right].
\end{equation}
By substituting \eqref{eq:TwistBlockDec2SameOr} in \eqref{eq:TwistBlockDec1SameOr}, we obtain
\begin{equation}
\begin{aligned}
    Q_{i} &= \left[
\begin{array}{c|c|c}
 a^{2} Q_{i-1} & * & *\\ 
 \hline 
 * & a^2 Q_{{i-1}}^{g=1,f=-2} & *\\
 \hline
 * & * & a^3 Q^{f=-1}_{L_{i-1}} \\
\end{array}
\right] \\
&= \left[
\begin{array}{c|c|c|c}
 a^{2} q^{-2} Q_{i-1}^{f=-2} & * & * & *\\ 
 \hline 
 * & a^2 Q_{{i-1}}^{f=-1} & * & *\\
 \hline 
 * & * & a^2 q^2 Q_{{i-1}} & * \\
 \hline
 * & * & * & a^3 Q^{f=-1}_{L_{i-1}} \\
\end{array}
\right],
\end{aligned}
\end{equation}
where we used 
\begin{equation}
    Q^{g=1}  = \left[
\begin{array}{c|c|c|c}
q^{-2} Q & * & * & *\\ 
 \hline 
 * &  Q^{f=1} & * & *\\
 \hline 
 * & * &  Q^{f=1} & * \\
 \hline
 * & * & * & q^2 Q^{f=2} \\
\end{array}
\right]
\end{equation}
and removed a cancelling pair of blocks. Using \eqref{eq:TwistBlockDec1SameOr}, we can simplify further to 
\begin{equation}\label{eq:TwistBlockDec3SameOr}
    Q_{i} = \left[
\begin{array}{c|c|c}
 a^{2} q^{-2} Q_{i-1}^{f=-2} & * & *\\ 
 \hline 
 * & a^4 Q_{{i-2}}^{f = - 1} & *\\
 \hline 
 * & * & a^2 q^2 Q_{i-1}
\end{array}
\right].
\end{equation}
The twice augmented quiver is 
\begin{equation}\label{eq:TwistBlockDec4SameOr}
    Q^+_{i} = \left[
\begin{array}{c c|c|c|c}
 1 & 0 & 2 & 2 & 2 \\
 0 & 0 & 2 & 1 & 1 \\
 \hline 
 2 & 2 & a^{2} q^{-2} Q_{i-1}^{f=-2} & * & *\\ 
 \hline 
 2 & 1 & * & a^4 Q_{{i-2}}^{f = - 1} & *\\
 \hline 
 2 & 1 & * & * & a^2 q^2 Q_{i-1}
\end{array}
\right],
\end{equation}
where $x_1 = q$ and $x_2 = q^2$. If we now link the second auxiliary node with the last two blocks, we obtain
\begin{equation}
    Q^+_{i} = \left[
\begin{array}{c c|c|c|c}
 1 & 0 & 2 & 2 & 2 \\
 0 & 0 & 2 & 1 & 1 \\
 \hline 
 2 & 2 & Q_i & * & *\\ 
 \hline 
 2 & 1 & * & a^4 q^4 Q_{{i-2}}^{f = 1} & *\\
 \hline 
 2 & 1 & * & * & a^2 q^6 Q_{i-1}^{f=2}
\end{array}
\right]
\end{equation}

It follows from \eqref{eq:TwistBlockDec3SameOr} that 
\begin{equation}
\left[
\begin{array}{c|c}
 a^2 Q_{i-1}^{f=-1} & * \\ 
 \hline
 * & q^2 Q_{i} \\
\end{array}
\right]
= 
\left[
\begin{array}{c|c}
 a^4 q^2 Q_{i-2}^{f=-1} & * \\ 
 \hline
 * & a^2 q^4 Q_{i-1} \\
\end{array}
\right],
\end{equation}
which implies that 
\begin{equation}
    Q^+_{i} = \left[
\begin{array}{c c|c|c|c}
 1 & 0 & 2 & 2 & 2 \\
 0 & 0 & 2 & 1 & 1 \\
 \hline 
 2 & 2 & Q_i & * & *\\ 
 \hline 
 2 & 1 & * & a^2 q^2 Q_{{i-1}}^{f = 1} & *\\
 \hline 
 2 & 1 & * & * & q^4 Q_{i}^{f=2}
\end{array}
\right] = \left(a^{-2} q^2 Q_{i+1}^{f=2}\right)^+
\end{equation}

 \begin{rmk}
     In the case that the skein relation is applied to two strands belonging to two different components of a link, a similar argument can be used to explain the twist recursion from the point of view of Conjecture \ref{coj:block_quiver_skein}. 
 \end{rmk}

We can also understand the linking identity from the point of view of block decompositions. We first discuss the case of opposite orientations of the strands in the twist region. Starting from \eqref{eq:TwistBlockDec3},
\begin{align}
    Q_{i}^+ &=  
\left[
\begin{array}{cc|c|c|c}
 0 & 0 & 0 & 0 & 0 \\
 0 & 1 & 0 & 1 & 1 \\
 \hline
 0 & 0 & Q_{i-1} & * & *\\ 
 \hline
 0 & 1 & * & a^{2} Q_{i-1} & * \\
 \hline
 0 & 1 & * & * & a^{2} Q_{i-2}^{f=1}
\end{array}
\right],
\end{align}
we link the second block and the auxiliary node to obtain
\begin{align}
\left[
\begin{array}{cc|c|c|c|c}
 0 & 0 & 0 & 0 & 0 & 0\\
 0 & 1 & 1 & 1 & 1 & 1\\
 \hline
 0 & 1 & Q_{i-1} & * & * & *\\ 
 \hline
 0 & 1 & * & a^{2} Q_{i-1} & * & *\\
 \hline
 0 & 1 & * & * & a^{2} Q_{i-2}^{f=1} & *\\
 \hline
 0 & 1 & * & * & * & a^2 Q_{i-1}^{f=-1}
\end{array}
\right].
\end{align}
Then the third and the fifth block cancel, and we are left with
\begin{align}\label{eq:L_0 normalized quiver}
\left[
\begin{array}{cc|c|c}
 0 & 0 & 0 & 0 \\
 0 & 1 & 1  & 1 \\
 \hline
 0 & 1 & Q_{i-1} & *\\
 \hline
 0 & 1 & *  & a^{2} Q_{i-2}^{f=1}
\end{array}
\right] = \left[
\begin{array}{cc|c}
 0 & 0 & 0 \\
 0 & 1 & 1 \\
 \hline
 0 & 1 & Q_{L_0} \\
\end{array}
\right],
\end{align}
where we used \eqref{eq:TwistBlockDec2}.\\

Now consider the case of parallel orientation of the strands. By applying the unlinking identity to \eqref{eq:TwistBlockDec4SameOr}, it follows that
\begin{equation}
    Q^+_{i} 
= \left[
\begin{array}{c c|c|c}
 1 & 0 & 2 & 2 \\
 0 & 0 & 1 & 1 \\
 \hline 
 2 & 1 & Q_i & * \\ 
 \hline 
 2 & 1 & * & a^2 q^2 Q_{{i-1}}^{f = 1} 
\end{array}
\right].
\end{equation}
In addition, Formula \eqref{eq:TwistBlockDec3SameOr} implies that 
\begin{equation}
    Q_i^+ = \left(a^{2} q^{-2} Q_{i-1}^{f=-2} \right)^+ = ... = \left( a^{2(i-1)} q^{-2(i-1)} Q_1^{f=-2(i-1)}\right)^+,
\end{equation}
from which we can conclude that  
\begin{equation}
    Q^+_{i} = \left[
\begin{array}{c c|c|c}
 1 & 0 & 2 & 2 \\
 0 & 0 & 1 & 1 \\
 \hline 
 2 & 1 & a^{2(i-1)} q^{-2(i-1)} Q_1^{f=-2(i-1)} & * \\ 
 \hline 
 2 & 1 & * & a^{2i} q^{-2(i-2)} Q_{0}^{f = -2i-1} 
\end{array}
\right].
\end{equation}

\begin{rmk}
    In the case of opposite orientation of the strands in the twist region, the normalised partition function of $L_0$ whose quiver is given by \eqref{eq:L_0 normalized quiver} admits a generalisation to the full skein, where the partition function of $L_0$ is acted on by the partition function of two disks with contractible boundary in the solid torus, one of which links once and the others zero times, see Appendix \ref{app:full skein linking identity}. However, this does not seem to be the case for the normalised partition function if the strands have parallel orientations in the twist region. For the interested reader, we included in Appendix \ref{app:full skein linking identity} also a formula in the full skein of two disks with contractible boundary acting on a partition function, which they link once and twice, respectively.
\end{rmk}


\section{Examples}\label{sec:Examples}

In this section, we confirm Conjecture \ref{coj:block_quiver_skein} in various special cases. Here, we work with the reduced invariants (see also Remark \ref{rmk:reduced vs unreduced}). 
Our main case studies are links whose symmetric quivers corresponding to $S^r$-coloured HOMFLYPT invariants are known for an arbitrary number of twists\footnote{All quiver examples presented here are consistent with the large-coloured HOMFLYPT data reported in ~\cite{RSS2026}.}. This includes 2-bridge torus knots and links as well as twist knots \cite{Kucharski:2017ogk}. For these simple links, quivers have a particularly nice block form which greatly simplifies verification of Conjecture \ref{coj:block_quiver_skein}. Apart from these, we will also consider the two more complicated examples of the knots $6_2$ and $8_{13}$, whose quivers have more intricate structure \cite{Kucharski:2017ogk,Chauhan:2025iwt}. The partition functions of the links considered here can be found in \cite{Gukov:2015gmm,Kucharski:2017ogk} and the quivers can be obtained via the methods of \cite{Kucharski:2017ogk} together with suitable transpositions \cite{Jankowski:2021flt}.

\subsection{\texorpdfstring{$T_{2,2p+1} = T_{2,2p-1} - z T_{2,2p}$}{T{2,2p+1} = T{2,2p-1} - z T{2,2p}}}

In this section, we consider the skein relation for the negative torus knot $T_{2,2p+1}$ for $p \geq 1$.

The quiver for $T_{2,2p+1}$ is given by \cite{Kucharski:2017ogk}:

\begin{equation}\label{eq:torus knot minimal quiver}
Q_{T_{2,2p+1}}=    \left[
\begin{array}{ccccccc}
 F_0 & F_1 & F_2 & F_3 & \dots & F_{p-1} & F_{p} \\
 F_1^T & D_1 & U_2 & U_3 & \dots & U_{p-1} & U_p \\
 F_2^T & U_2^T & D_2 & U_3 & \dots & U_{p-1} & U_p \\
 F_3^T & U_3^T & U_3^T & D_3 & \dots & U_{p-1} & U_p \\
 \vdots & \vdots & \vdots & \vdots & \ddots & \vdots & \vdots \\
 F_{p-1}^T & U_{p-1}^T & U_{p-1}^T & U_{p-1}^T & \dots & D_{p-1} & U_p \\
 F_p^T & U_p^T & U_p^T & U_p^T & \dots & U_p^T & D_p \\
\end{array}
\right],
\quad
\left[
\begin{array}{c}
    \boldsymbol{x}_0 \\
    \boldsymbol{x}_1 \\
    \boldsymbol{x}_2 \\
    \boldsymbol{x}_3 \\
    \vdots \\
    \boldsymbol{x}_{p-1} \\
    \boldsymbol{x}_p 
\end{array}
\right],
\end{equation}
where
\begin{equation}
\setlength{\arraycolsep}{3pt}
\begin{aligned}
F_0 &= [\,0\,],
&
F_k &=
\left[
\begin{customsmallmatrix}
2k-1 & 2k-1
\end{customsmallmatrix}
\right],
\\[1.2ex]
D_k &=
\left[
\begin{customsmallmatrix}
2k & 2k\\
2k & 2k+1
\end{customsmallmatrix}
\right],
&
U_k &=
\left[
\begin{customsmallmatrix}
2k-1 & 2k-1\\
2k   & 2k
\end{customsmallmatrix}
\right],
\\[1.2ex]
\boldsymbol{x}_0 &= \left[q^{-2p}a^{2p}x \right],
&
\boldsymbol{x}_k &=
\left[
\begin{customsmallmatrix}
q^{2k-2p}a^{2p}x\\
q^{2k-2p-{\color{blue}3}}a^{2p+2}x
\end{customsmallmatrix}
\right].
\end{aligned}
\end{equation}
for $k = 1, ... , p$.

The quiver for the link $T_{2,2p}$ is given by 
\begin{equation}
Q_{T_{2,2p}}=    \left[ 
    \begin{array}{ccccccc}
 F_0 & F_1 & F_2 & F_3 & \dots & F_{p-1} & F_{p}^e \\
 F_1^T & D_1 & U_2 & U_3 & \dots & U_{p-1} & U_p^e \\
 F_2^T & U_2^T & D_2 & U_3 & \dots & U_{p-1} & U_p^e \\
 F_3^T & U_3^T & U_3^T & D_3 & \dots & U_{p-1} & U_p^e \\
 \vdots & \vdots & \vdots & \vdots & \ddots & \vdots & \vdots \\
 F_{p-1}^T & U_{p-1}^T & U_{p-1}^T & U_{p-1}^T & \dots & D_{p-1} & U_p^e \\
 F_p^T & U_p^{eT} & U_p^{eT} & U_p^{eT} & \dots & U_p^{eT} & D_p^e \\
\end{array}
    \right], 
    \quad
    \left[
\begin{array}{c}
    \boldsymbol{x}_0 \\
    \boldsymbol{x}_1 \\
    \boldsymbol{x}_2 \\
    \boldsymbol{x}_3 \\
    \vdots \\
    \boldsymbol{x}_{p-1} \\
    \boldsymbol{x}_p 
\end{array}
\right],
\end{equation}
where
\begin{equation}
\setlength{\arraycolsep}{3pt}
\begin{aligned}
F_0 &=
\left[
\begin{customsmallmatrix}
0 & 0\\
0 & 1
\end{customsmallmatrix}
\right],
&
F_k &=
\left[
\begin{customsmallmatrix}
2k-1 & 2k-1 & 2k & 2k\\
2k-1 & 2k-1 & 2k & 2k
\end{customsmallmatrix}
\right],
\\[1.2ex]
D_k &=
\left[
\begin{customsmallmatrix}
2k   & 2k   & 2k   & 2k+1\\
2k   & 2k+1 & 2k   & 2k+1\\
2k   & 2k   & 2k+1 & 2k+1\\
2k+1 & 2k+1 & 2k+1 & 2k+2
\end{customsmallmatrix}
\right],
&
U_k &=
\left[
\begin{customsmallmatrix}
2k-1 & 2k-1 & 2k   & 2k\\
2k   & 2k   & 2k+1 & 2k+1\\
2k-1 & 2k-1 & 2k   & 2k\\
2k   & 2k   & 2k+1 & 2k+1
\end{customsmallmatrix}
\right],
\\[1.2ex]
F_p^e &=
\left[
\begin{customsmallmatrix}
2p-1 & 2p-1\\
2p-1 & 2p-1
\end{customsmallmatrix}
\right],
&
D_p^e &=
\left[
\begin{customsmallmatrix}
2p & 2p\\
2p & 2p+1
\end{customsmallmatrix}
\right],
\\[1.2ex]
U_p^e &=
\left[
\begin{customsmallmatrix}
2p-1 & 2p-1\\
2p   & 2p\\
2p-1 & 2p-1\\
2p   & 2p
\end{customsmallmatrix}
\right],
&
\boldsymbol{x}_0 &=
\left[
\begin{customsmallmatrix}
q^{-2p}a^{2p-1}x\\
q^{-2p+1}a^{2p-1}x
\end{customsmallmatrix}
\right],
\\[1.2ex]
\boldsymbol{x}_k &=
\left[
\begin{customsmallmatrix}
q^{2k-2p}a^{2p-1}x\\
q^{2k-2p-3}a^{2p+1}x\\
q^{2k-2p+1}a^{2p-1}x\\
q^{2k-2p-2}a^{2p+1}x
\end{customsmallmatrix}
\right],
&
\boldsymbol{x}_p &=
\left[
\begin{customsmallmatrix}
a^{2p-1}x\\
q^{-3}a^{2p+1}x
\end{customsmallmatrix}
\right].
\end{aligned}
\end{equation}
for $k = 1, ... , p-1$.

Let us show that the quiver matrix for $T_{2,2p}$ can be written in block form, where the two diagonal blocks are given by the quiver matrices for $a T_{2,2p-1}$ and $a^{-1}T_{2,2p+1}$. First, observe that the submatrix for $T_{2,2p}$ corresponding to the quiver nodes $\{2,(4k+1,4k+2)_{k=1}^{p-1}\}$ is precisely the quiver matrix of $a T_{2,2p-1}$. The submatrix of the remaining nodes is then given by the quiver matrix of $a^{-1} T_{2,2p+1}$ with all entries shifted by 1.

So far we have checked the block decomposition only for one component of a fixed skein relation. Accordingly, using the above block form for $Q_{T_{2,2p}}$ one can obtain the one for other components of the skein relation, by including the sufficient number of redundant pairs (see Section \ref{sec:Equivalent_quivers}). For example, to get the block decomposition for $T_{2,2p+1}$, one has to add one generator per each node in the $T_{2,2p-1}$-block of $Q_{T_{2,2p}}$, which forms the redundant pair with that node. As a result, one obtains a non-minimal quiver for the torus knot $T_{2,2p+1}$ which however has the desired block form. By cancelling redundant pairs, one usually gets a minimal quiver \eqref{eq:torus knot minimal quiver} for the corresponding knot. This already stresses one important point that the quiver block decompositions generally require considering non-minimal quivers. We will explore this interesting phenomenon in other examples as well.

\subsection{\texorpdfstring{$zT_{2,2p+2} = zT_{2,2p} - T_{2,2p+1}^{g=1}$}{zT2,2p+2 = zT2,2p - T2,2p+1 g=1}}

\subsubsection*{Case $p=0$}

In the case $p=0$, the link $T_{2,2}$ is the Hopf link, $T_{2,0}$ is the unlink, and $T_{2,1}$ is a framed unknot. The quiver for the Hopf link is given by 
\begin{equation}
\renewcommand\arraystretch{1.25}
\left[
\begin{array}{cccc}
 0 & 0 & 1 & 1 \\
 0 & 1 & 1 & 1 \\
 1 & 1 & 2 & 2 \\
 1 & 1 & 2 & 3 
\end{array}
\right],
\quad
\left[
\begin{array}{c}
    a q^{-2} x \\
    a  q^{-1} x \\
    a x \\
    a^3 q^{-3}x 
\end{array}
\right]\,,
\end{equation}
the quiver for the unlink is given by 
\begin{equation}\label{quiver:RedUnlink}
\renewcommand\arraystretch{1.25}
Q_{00} = \left[
\begin{array}{cc}
 0 & 0 \\
 0 & 1
\end{array}
\right],
\;
\left[
\begin{array}{c}
    a^{-1} x \\
    a q^{-1} x
\end{array}
\right]\,,
\end{equation}
and the quiver for the genus 1 unknot is given by 
\begin{equation}\label{quiver:genus1Unknot}
\renewcommand\arraystretch{1.25}
Q_0^{g=1} = \left[
\begin{array}{cccc}
 0 & 0 & 0 & 1 \\
 0 & 1 & 0 & 1 \\
 0 & 0 & 1 & 1 \\
 1 & 1 & 1 & 2  
\end{array}
\right],
\;\;
\left[
\begin{array}{c}
    q^{-2} x \\
    q^{-1} x \\
    q^{-1} x \\
    x 
\end{array}
\right]\,.
\end{equation}

Using equivalences of quivers, the quiver of the Hopf link can be written as 
\begin{equation}\label{eq:hopf quiver decomposition}
\begin{split}
\renewcommand\arraystretch{1.25}
\left[
\begin{array}{cccc}
 0 & 0 & 1 & 1 \\
 0 & 1 & 1 & 1 \\
 1 & 1 & 2 & 2 \\
 1 & 1 & 2 & 3
\end{array}
\right],
\;\;
\left[
\begin{array}{c}
    a q^{-2} x \\
    a q^{-1}  x \\
    a x \\
    a^3 q^{-3}x 
\end{array}
\right]\, 
&\sim \left[
\begin{array}{cccc|cc}
 0 & 0 & 1 & 1 & 0 & 0 \\
 0 & 1 & 1 & 1 & 0 & 0 \\
 1 & 1 & 2 & 2 & 1 & 1 \\
 1 & 1 & 2 & 3 & 2 & 2 \\
 \hline
 0 & 0 & 1 & 2 & 1 & 1 \\
 0 & 0 & 1 & 2 & 1 & 2
\end{array}
\right],
\;\;
\left[
\begin{array}{c}
    a q^{-2} x \\
    a q^{-1}  x \\
    a x \\
    a^3 q^{-3}x \\
    \hline
    a q^{-1} x \\
    a q^{-2} x
\end{array}
\right]\,\\
&\sim \left[
\begin{array}{cc|cccc}
 2 & 2 & 0 & 0 & 1 & 1 \\
 2 & 3 & 1 & 1 & 2 & 2 \\
 \hline
 0 & 1 & 0 & 0 & 0 & 1 \\
 0 & 1 & 0 & 1 & 0 & 1 \\
 1 & 2 & 0 & 0 & 1 & 1 \\
 1 & 2 & 1 & 1 & 1 & 2 
\end{array}
\right],
\;\;
\left[
\begin{array}{c}
    a q^{-2} x \\
    a^3 q^{-3} x \\
    \hline
    q^{-2} a x \\
    a q^{-1} x \\
    a q^{-1} x \\
    a x  
\end{array}
\right]\,
\end{split}
\end{equation}
where in the first step we added a redundant pair of nodes, and in the last step we applied the permutation $(6,4,1,2,5,3)$. The resulting quiver description decomposes into two blocks given by $a^2 Q_{00}$ and $a Q_0^{g=1,f=-1}$. Similarly to the first case study, by adding a sufficient number of redundant pairs, one can then get block decompositions for the other components of the skein relation.

\subsubsection*{Case $p\geq 1$}

In this paragraph, we consider the skein relation for the negative torus link $T_{2,2p+2}$ for $p \geq 1$.

The quiver for $T_{2,2p+1}^{g=1}$ is given by 
\begin{equation}
\begingroup
\setlength{\arraycolsep}{4pt}
\renewcommand{\arraystretch}{0.95}
\left[
\begin{array}{c|ccccccc}
F_0       & F_1       & F_2       & F_3       & \cdots & F_{p-2}       & F_{p-1}       & F_p \\ \hline
F_1^T     & D_1       & U_2       & U_3       & \cdots & U_{p-2}       & U_{p-1}       & U_p \\
F_2^T     & U_2^T     & D_2       & U_3       & \cdots & U_{p-2}       & U_{p-1}       & U_p \\
F_3^T     & U_3^T     & U_3^T     & D_3       & \cdots & U_{p-2}       & U_{p-1}       & U_p \\
\vdots    & \vdots    & \vdots    & \vdots    & \ddots & \vdots        & \vdots        & \vdots \\
F_{p-2}^T & U_{p-2}^T & U_{p-2}^T & U_{p-2}^T & \cdots & D_{p-2}       & U_{p-1}       & U_p \\
F_{p-1}^T & U_{p-1}^T & U_{p-1}^T & U_{p-1}^T & \cdots & U_{p-1}^T     & D_{p-1}       & U_p \\
F_p^T     & U_p^T     & U_p^T     & U_p^T     & \cdots & U_{p-1}^T     & U_p^T         & D_p^e
\end{array}
\right],
\quad
\left[
\begin{array}{c}
\boldsymbol{x}_0\\
\boldsymbol{x}_1\\
\boldsymbol{x}_2\\
\boldsymbol{x}_3\\
\vdots\\
\boldsymbol{x}_{p-2}\\
\boldsymbol{x}_{p-1}\\
\boldsymbol{x}_p
\end{array}
\right].
\endgroup
\end{equation}

where
\begin{equation}
\setlength{\arraycolsep}{3pt}
\begin{aligned}
F_0 &=
\left[
\begin{customsmallmatrix}
0 & 0 & 0 & 1\\
0 & 1 & 0 & 1\\
0 & 0 & 1 & 1\\
1 & 1 & 1 & 2
\end{customsmallmatrix}
\right],
&
F_k &=
\left[
\begin{array}{@{}cccc|cccc@{}}
2k-1 & 2k & 2k & 2k+1 & 2k-1 & 2k & 2k & 2k+1\\
2k-1 & 2k & 2k & 2k+1 & 2k-1 & 2k & 2k & 2k+1\\
2k-1 & 2k & 2k & 2k+1 & 2k-1 & 2k & 2k & 2k+1\\
2k-1 & 2k & 2k & 2k+1 & 2k-1 & 2k & 2k & 2k+1
\end{array}
\right].
\end{aligned}
\end{equation}

\begin{equation}
\begingroup
\setlength{\arraycolsep}{2.5pt}
\renewcommand{\arraystretch}{0.9}
D_k=
\left[
\begin{array}{cccc|cccc}
2k   & 2k   & 2k   & 2k+1 & 2k   & 2k+1 & 2k+1 & 2k+2\\
2k   & 2k+1 & 2k   & 2k+1 & 2k   & 2k+1 & 2k+1 & 2k+2\\
2k   & 2k   & 2k+1 & 2k+1 & 2k   & 2k+1 & 2k+1 & 2k+2\\
2k+1 & 2k+1 & 2k+1 & 2k+2 & 2k   & 2k+1 & 2k+1 & 2k+2\\ \hline
2k   & 2k   & 2k   & 2k   & 2k+1 & 2k+1 & 2k+1 & 2k+2\\
2k+1 & 2k+1 & 2k+1 & 2k+1 & 2k+1 & 2k+2 & 2k+1 & 2k+2\\
2k+1 & 2k+1 & 2k+1 & 2k+1 & 2k+1 & 2k+1 & 2k+2 & 2k+2\\
2k+2 & 2k+2 & 2k+2 & 2k+2 & 2k+2 & 2k+2 & 2k+2 & 2k+3
\end{array}
\right].
\endgroup
\end{equation}

\begin{equation}
\begingroup
\setlength{\arraycolsep}{2.5pt}
\renewcommand{\arraystretch}{0.9}
U_k=
\left[
\begin{array}{cccc|cccc}
2k-1 & 2k   & 2k   & 2k+1 & 2k-1 & 2k   & 2k   & 2k+1\\
2k-1 & 2k   & 2k   & 2k+1 & 2k-1 & 2k   & 2k   & 2k+1\\
2k-1 & 2k   & 2k   & 2k+1 & 2k-1 & 2k   & 2k   & 2k+1\\
2k-1 & 2k   & 2k   & 2k+1 & 2k-1 & 2k   & 2k   & 2k+1\\ \hline
2k   & 2k+1 & 2k+1 & 2k+2 & 2k   & 2k+1 & 2k+1 & 2k+2\\
2k   & 2k+1 & 2k+1 & 2k+2 & 2k   & 2k+1 & 2k+1 & 2k+2\\
2k   & 2k+1 & 2k+1 & 2k+2 & 2k   & 2k+1 & 2k+1 & 2k+2\\
2k   & 2k+1 & 2k+1 & 2k+2 & 2k   & 2k+1 & 2k+1 & 2k+2
\end{array}
\right].
\endgroup
\end{equation}

\begin{equation}
\begingroup
\setlength{\arraycolsep}{2.5pt}
\renewcommand{\arraystretch}{0.9}
D_p^e=
\left[
\begin{array}{cccc|cccc}
2p   & 2p   & 2p   & 2p+1 & 2p   & 2p+1 & 2p   & 2p+1\\
2p   & 2p+1 & 2p   & 2p+1 & 2p   & 2p+1 & 2p   & 2p+1\\
2p   & 2p   & 2p+1 & 2p+1 & 2p+1 & 2p+1 & 2p+1 & 2p+1\\
2p+1 & 2p+1 & 2p+1 & 2p+2 & 2p+2 & 2p+2 & 2p+2 & 2p+2\\ \hline
2p   & 2p   & 2p+1 & 2p+2 & 2p+1 & 2p+1 & 2p+1 & 2p+2\\
2p+1 & 2p+1 & 2p+1 & 2p+2 & 2p+1 & 2p+2 & 2p+1 & 2p+2\\
2p   & 2p   & 2p+1 & 2p+2 & 2p+1 & 2p+1 & 2p+2 & 2p+2\\
2p+1 & 2p+1 & 2p+1 & 2p+2 & 2p+2 & 2p+2 & 2p+2 & 2p+3
\end{array}
\right].
\endgroup
\end{equation}

and
\begin{equation}
\setlength{\arraycolsep}{3pt}
\begin{aligned}
\boldsymbol{x}_0 &=
\left[
\begin{customsmallmatrix}
q^{-2p-2}a^{-2p-2}x\\
q^{-2p-1}a^{-2p-2}x\\
q^{-2p-1}a^{-2p-2}x\\
q^{-2p}a^{-2p-2}x
\end{customsmallmatrix}
\right],
&
\boldsymbol{x}_k &=
\left[
\begin{customsmallmatrix}
q^{2k-2p-2}a^{-2p-2}x\\
q^{2k-2p-1}a^{-2p-2}x\\
q^{2k-2p-1}a^{-2p-2}x\\
q^{2k-2p}a^{-2p-2}x\\
q^{2k-2p-5}a^{-2p}x\\
q^{2k-2p-4}a^{-2p}x\\
q^{2k-2p-4}a^{-2p}x\\
q^{2k-2p-3}a^{-2p}x
\end{customsmallmatrix}
\right].
\end{aligned}
\end{equation}
for $k = 1, ... , p$.

This quiver contains the quiver for $a T_{2,2p+2}$ with $+1$ framing as a subquiver corresponding to the nodes $\{\{4k+1,4k+2\}_{k=0}^{2p},8p,8p+4\}$, and the remaining nodes form a quiver for $a^2 T_{2,2p}$.

\subsection{\texorpdfstring{$(4+2p)_1 = (2+2p)_1 - z a^{-2p-1} T_{2,2}$}{(4+2p)1 = (2+2p)1 - z a{-2p-1} T{2,2}}}

By applying the skein relation to a negative crossing in the twist region of the knots $4_1$, $6_1$,..., we obtain another knot from this family and a right-handed Hopf link in framing $-2p-1$. We show that this also works at the level of associated quivers. Here, we interpret ``$2_1$" as a framed unknot.

The quiver for $(2p+2)_1$ is given by
\begin{equation}
\left[
\begin{array}{c|ccccc}
F_0    & F       & F       & F       & \cdots & F \\ \hline
F^T    & D_1     & U_1     & U_1     & \cdots & U_1 \\
F^T    & U_1^T   & D_2     & U_2     & \cdots & U_2 \\
F^T    & U_1^T   & U_2^T   & D_3     & \cdots & U_3 \\
\vdots & \vdots  & \vdots  & \vdots  & \ddots & \vdots \\
F^T    & U_1^T   & U_2^T   & U_3^T   & \cdots & D_p
\end{array}
\right],
\quad
\left[
\begin{array}{c}
\boldsymbol{x}_0\\
\boldsymbol{x}_1\\
\boldsymbol{x}_2\\
\boldsymbol{x}_3\\
\vdots\\
\boldsymbol{x}_p
\end{array}
\right],
\end{equation}

where
\begin{equation}
\setlength{\arraycolsep}{3pt}
\begin{aligned}
F_0 &=
\left[ 
\begin{customsmallmatrix}
0
\end{customsmallmatrix}
\right],
&
F &=
\left[
\begin{customsmallmatrix}
-1 & -1 & 0 & 0
\end{customsmallmatrix}
\right],
\\[1.2ex]
D_k &=
\left[
\begin{customsmallmatrix}
2k-4 & 2k-4 & 2k-3 & 2k-2\\
2k-4 & 2k-3 & 2k-3 & 2k-2\\
2k-3 & 2k-3 & 2k-1 & 2k-1\\
2k-2 & 2k-2 & 2k-1 & 2k
\end{customsmallmatrix}
\right],
&
U_k &=
\left[
\begin{customsmallmatrix}
2k-4 & 2k-4 & 2k-3 & 2k-2\\
2k-3 & 2k-3 & 2k-3 & 2k-2\\
2k-2 & 2k-2 & 2k-1 & 2k-1\\
2k-1 & 2k-1 & 2k   & 2k
\end{customsmallmatrix}
\right],
\\[1.2ex]
\boldsymbol{x}_0 &= \left[x\right],
&
\boldsymbol{x}_k &=
\left[
\begin{customsmallmatrix}
q^{-2k+4}a^{2k-4}x\\
q^{-2k+1}a^{2k-2}x\\
q^{-2k+3}a^{2k-2}x\\
q^{-2k}a^{2k}x
\end{customsmallmatrix}
\right].
\end{aligned}
\end{equation}
for $k = 1, ... , p$.

The quiver description of the positive Hopf link reads
\begin{equation}
\renewcommand\arraystretch{1.25}
\left[
\begin{array}{cccc}
 -2 & -2 & -1 & -1 \\
 -2 & -1 & -1 & -1 \\
 -1 & -1 & 0 & 0 \\
 -1 & -1 & 0 & 1 
\end{array}
\right],
\;\;
\left[
\begin{array}{c}
    a^{-3} q^2  x \\
    a^{-1} q^{-1}x \\
    a^{-1} q^{-1} x \\
    a^{-1} q x 
\end{array}
\right]\,.
\end{equation}

The quiver matrix for $(4+2p)_1$ contains the quiver matrix for $aT_{2,2}$ in framing $-1$ as the submatrix corresponding to the quiver nodes $\{1,2,3,4\}$, and the remaining nodes give the quiver matrix for $a^2(2+2p)_1$.

\subsection{\texorpdfstring{$(2+2p)_1 = a^{-2p} 0_1 + z a \widetilde T_{2,2p}$}{(2+2p)1 = a{-2p} 01 + z a T{2,2p}}}

By applying the skein relation to one of the two crossings that are not in the twist region of the knots $4_1$, $6_1$,..., we obtain a framed unknot and a framed torus knot $T_{2,2p}$ with opposite orientations on its components, which we denote by $\widetilde T_{2,2p}$. We show that this also works at the level of associated quivers.

We use the following alternative quiver for $(2p+2)_1$:
\begin{equation}
\left[
\begin{array}{c|ccccc}
F_0    & F       & F       & F       & \cdots & F \\ \hline
F^T    & D_1     & U_1     & U_1     & \cdots & U_1 \\
F^T    & U_1^T   & D_2     & U_2     & \cdots & U_2 \\
F^T    & U_1^T   & U_2^T   & D_3     & \cdots & U_3 \\
\vdots & \vdots  & \vdots  & \vdots  & \ddots & \vdots \\
F^T    & U_1^T   & U_2^T   & U_3^T   & \cdots & D_p
\end{array}
\right],
\quad
\left[
\begin{array}{c}
\boldsymbol{x}_0\\
\boldsymbol{x}_1\\
\boldsymbol{x}_2\\
\boldsymbol{x}_3\\
\vdots\\
\boldsymbol{x}_p
\end{array}
\right].
\end{equation}

where
\begin{equation}
\setlength{\arraycolsep}{3pt}
\begin{aligned}
F_0 &=
\left[ 
\begin{customsmallmatrix}
0
\end{customsmallmatrix}
\right],
&
F &=
\left[
\begin{customsmallmatrix}
-1 & -1 & 0 & 0
\end{customsmallmatrix}
\right],
\\[1.2ex]
D_k &=
\left[
\begin{customsmallmatrix}
2k-4 & 2k-4 & 2k-3 & 2k-3\\
2k-4 & 2k-3 & 2k-2 & 2k-2\\
2k-3 & 2k-2 & 2k-1 & 2k-1\\
2k-3 & 2k-2 & 2k-1 & 2k
\end{customsmallmatrix}
\right],
&
U_k &=
\left[
\begin{customsmallmatrix}
2k-4 & 2k-4 & 2k-3 & 2k-3\\
2k-3 & 2k-3 & 2k-2 & 2k-2\\
2k-2 & 2k-2 & 2k-1 & 2k-1\\
2k-1 & 2k-1 & 2k   & 2k
\end{customsmallmatrix}
\right],
\\[1.2ex]
\boldsymbol{x}_0 &= \left[x\right],
&
\boldsymbol{x}_k &=
\left[
\begin{customsmallmatrix}
q^{-2k+4}a^{2k-4}x\\
q^{-2k+1}a^{2k-2}x\\
q^{-2k+3}a^{2k-2}x\\
q^{-2k}a^{2k}x
\end{customsmallmatrix}
\right].
\end{aligned}
\end{equation}

The quiver for $\widetilde T_{2,2p}$ is given by
\begin{equation}\label{eq:QuivT22pTilde}
\left[
\begin{array}{c|ccccc}
F_0       & F_1     & F_2     & F_3     & \cdots & F_{p-1} \\ \hline
F_1^T     & D_1     & U_1     & U_1     & \cdots & U_1 \\
F_2^T     & U_1^T   & D_2     & U_2     & \cdots & U_2 \\
F_3^T     & U_1^T   & U_2^T   & D_3     & \cdots & U_3 \\
\vdots    & \vdots  & \vdots  & \vdots  & \ddots & \vdots \\
F_{p-1}^T & U_1^T   & U_2^T   & U_3^T   & \cdots & D_{p-1}
\end{array}
\right],
\;\;
\left[
\begin{array}{c}
\boldsymbol{x}_0\\
\boldsymbol{x}_1\\
\boldsymbol{x}_2\\
\boldsymbol{x}_3\\
\vdots\\
\boldsymbol{x}_{p-1}
\end{array}
\right].
\end{equation}

where, for \(r_k:=2p-2k\),
\begin{equation}
\setlength{\arraycolsep}{3pt}
\begin{aligned}
F_0 &=
\left[
\begin{customsmallmatrix}
2p+1 & 2p   & 2p-1 & 2p-1\\
2p   & 2p   & 2p-1 & 2p-1\\
2p-1 & 2p-1 & 2p-1 & 2p-2\\
2p-1 & 2p-1 & 2p-2 & 2p-2
\end{customsmallmatrix}
\right],
&
F_k &=
\left[
\begin{customsmallmatrix}
r_k   & r_k   & r_k-1 & r_k-1\\
r_k   & r_k   & r_k-1 & r_k-1\\
r_k   & r_k   & r_k-1 & r_k-1\\
r_k-1 & r_k-1 & r_k-2 & r_k-2
\end{customsmallmatrix}
\right],
\\[1.2ex]
D_k &=
\left[
\begin{customsmallmatrix}
r_k   & r_k-1 & r_k-1 & r_k-1\\
r_k-1 & r_k-1 & r_k-2 & r_k-2\\
r_k-1 & r_k-2 & r_k-1 & r_k-2\\
r_k-1 & r_k-2 & r_k-2 & r_k-2
\end{customsmallmatrix}
\right],
&
U_k &=
\left[
\begin{customsmallmatrix}
r_k   & r_k   & r_k-1 & r_k-1\\
r_k-1 & r_k-1 & r_k-2 & r_k-2\\
r_k   & r_k   & r_k-1 & r_k-1\\
r_k-1 & r_k-1 & r_k-2 & r_k-2
\end{customsmallmatrix}
\right],
\\[1.2ex]
\boldsymbol{x}_0 &=
\left[
\begin{customsmallmatrix}
q^{-1} a^{2p+1}x\\
q^{-2p+2}a^{2p-1}x\\
q^{-2p+1}a^{2p-1}x\\
q^{-2p}a^{2p-1}x
\end{customsmallmatrix}
\right],
&
\boldsymbol{x}_k &=
\left[
\begin{customsmallmatrix}
q^{2-r_k}a^{r_k-1}x\\
q^{1-r_k}a^{r_k-1}x\\
q^{1-r_k}a^{r_k-1}x\\
q^{-r_k}a^{r_k-1}x
\end{customsmallmatrix}
\right].
\end{aligned}
\end{equation}
for $k = 1, ... , p-1$.

The second node of the quiver for $(2+2p)_1$ gives $a^{-2}$ times the quiver of the unknot and the remaining nodes give the quiver for $a^{-1} \widetilde{T}_{2,2p}$.

\subsection{\texorpdfstring{$z\widetilde T_{2,2p+2} = z\widetilde T_{2,2p} - z^2 a^{-2p-1}0_1$}{z T{2,2p+2} = z T{2,2p} - z2 a{-2p-1}01}}

If we apply the skein relation to one of the crossings of $\widetilde T_{2,2p+2}$ for $p \geq 1$, we obtain $\widetilde T_{2,2p}$ and a framed unknot.

The last four nodes of the quiver of $\widetilde T_{2,2p+2}$ described in \eqref{eq:QuivT22pTilde} give rise to $a Q_0^{g=1,f=-1}$, and the remaining nodes give the quiver for $a^2 \widetilde T_{2,2p}$.

\subsection{\texorpdfstring{$(2p+1)_2 = a^{-2p+1} 0_1 - z \widetilde T_{2,2p}$}{(2p+1)2 = a{-2p+1} 01 - z T{2,2p}}}

By applying the skein relation to one of the two crossings that are not in the twist region of the knots $3_1$, $5_2$, $7_2$, ..., we obtain a framed unknot and a torus knot $T_{2,2p}$ with opposite orientations on its components. We show that this also works at the level of associated quivers.

We use the following quiver for $(2p+1)_2$ (or $3_1$ if $p = 1$):
\begin{equation}
\left[
\begin{array}{cccccc}
F_0   & F       & F       & F       & \cdots & F \\
F^T   & D_1     & U_1     & U_1     & \cdots & U_1 \\
F^T   & U_1^T   & D_2     & U_2     & \cdots & U_2 \\
F^T   & U_1^T   & U_2^T   & D_3     & \cdots & U_3 \\
\vdots& \vdots  & \vdots  & \vdots  & \ddots & \vdots \\
F^T   & U_1^T   & U_2^T   & U_3^T   & \cdots & D_{p-1}
\end{array}
\right],
\quad
\left[
\begin{array}{c}
\boldsymbol{x}_0\\
\boldsymbol{x}_1\\
\boldsymbol{x}_2\\
\boldsymbol{x}_3\\
\vdots\\
\boldsymbol{x}_{p-1}
\end{array}
\right].
\end{equation}

\begin{equation}
\setlength{\arraycolsep}{3pt}
\begin{aligned}
F_0 &=
\left[
\begin{customsmallmatrix}
0&1&1\\
1&2&2\\
1&2&3
\end{customsmallmatrix}
\right],
&
F &=
\left[
\begin{customsmallmatrix}
0&0&1&1\\
1&1&2&2\\
2&2&3&3
\end{customsmallmatrix}
\right],
\\[1ex]
D_k &=
\left[
\begin{customsmallmatrix}
2k-1&2k-1&2k&2k\\
2k-1&2k&2k+1&2k+1\\
2k&2k+1&2k+2&2k+2\\
2k&2k+1&2k+2&2k+3
\end{customsmallmatrix}
\right],
&
U_k &=
\left[
\begin{customsmallmatrix}
2k-1&2k-1&2k&2k\\
2k&2k&2k+1&2k+1\\
2k+1&2k+1&2k+2&2k+2\\
2k+2&2k+2&2k+3&2k+3
\end{customsmallmatrix}
\right],
\\[1ex]
\boldsymbol{x}_0 &=
\left[
\begin{customsmallmatrix}
q^{-2}a^2x\\
a^2x\\
q^{-3}a^4x
\end{customsmallmatrix}
\right],
&
\boldsymbol{x}_k &=
\left[
\begin{customsmallmatrix}
q^{-2k+1}a^{2k}x\\
q^{-2k-2}a^{2k+2}x\\
q^{-2k}a^{2k+2}x\\
q^{-2k-3}a^{2k+4}x
\end{customsmallmatrix}
\right].
\end{aligned}
\end{equation}
for $k = 1, ... , p-1$.

The quiver description of $\widetilde T_{2,2p}$ is given by \eqref{eq:QuivT22pTilde}.

The second to last node of the quiver for $\widetilde T_{2,2p}$ gives $a Q_0$, and the remaining nodes give the quiver for $a^{-1} (2p+1)_2$ (or $a^{-1}3_1$ if $p=1$) in framing $+1$.

\subsection{\texorpdfstring{$(2p+3)_2 = (2p+1)_2 - z a^{-2p} T_{2,2}$}{(2p+3)2 = (2p+1)2 - z a{-2p} T{2,2}}}

By applying the skein relation to a crossing in the twist region of the knots $3_1$, $5_2$, $7_2$, ..., we obtain another knot in this family and a left-handed Hopf link $T_{2,2}$. We show that this also works at the level of associated quivers.

We use the following alternative quiver for $(2p+1)_2$:
\begin{equation}
\left[
\begin{array}{c|ccccc}
F_0    & F       & F       & F       & \cdots & F \\ \hline
F^T    & D_1     & U_1     & U_1     & \cdots & U_1 \\
F^T    & U_1^T   & D_2     & U_2     & \cdots & U_2 \\
F^T    & U_1^T   & U_2^T   & D_3     & \cdots & U_3 \\
\vdots & \vdots  & \vdots  & \vdots  & \ddots & \vdots \\
F^T    & U_1^T   & U_2^T   & U_3^T   & \cdots & D_{p-1}
\end{array}
\right],
\;\;
\left[
\begin{array}{c}
\boldsymbol{x}_0\\
\boldsymbol{x}_1\\
\boldsymbol{x}_2\\
\boldsymbol{x}_3\\
\vdots\\
\boldsymbol{x}_{p-1}
\end{array}
\right].
\end{equation}

where
\begin{equation}
\setlength{\arraycolsep}{3pt}
\begin{aligned}
F_0 &=
\left[
\begin{customsmallmatrix}
0&1&1\\
1&2&2\\
1&2&3
\end{customsmallmatrix}
\right],
&
F &=
\left[
\begin{customsmallmatrix}
0&0&1&1\\
1&2&2&2\\
1&2&3&3
\end{customsmallmatrix}
\right],
\\[1.2ex]
D_k &=
\left[
\begin{customsmallmatrix}
2k-1&2k-1&2k&2k\\
2k-1&2k&2k+1&2k+1\\
2k&2k+1&2k+2&2k+2\\
2k&2k+1&2k+2&2k+3
\end{customsmallmatrix}
\right],
&
U_k &=
\left[
\begin{customsmallmatrix}
2k-1&2k-1&2k&2k\\
2k&2k&2k+1&2k+1\\
2k+1&2k+2&2k+2&2k+2\\
2k+1&2k+2&2k+3&2k+3
\end{customsmallmatrix}
\right],
\\[1.2ex]
\boldsymbol{x}_0 &=
\left[
\begin{customsmallmatrix}
q^{-2}a^2x\\
a^2x\\
q^{-3}a^4x
\end{customsmallmatrix}
\right],
&
\boldsymbol{x}_k &=
\left[
\begin{customsmallmatrix}
q^{-2k+1}a^{2k}x\\
q^{-2k-2}a^{2k+2}x\\
q^{-2k}a^{2k+2}x\\
q^{-2k-3}a^{2k+4}x
\end{customsmallmatrix}
\right].
\end{aligned}
\end{equation}
for $k = 1, ... , p-1$.

The first four nodes of the quiver for $(3+2p)_2$ give the quiver for $a T_{2,2}$ in framing $-1$, and the remaining nodes give the quiver for $a^2(2p+1)_2$ (or $a^23_1$ if $p=1$).

\subsection{\texorpdfstring{$6_{2} = 3_{1} + zT_{2,4}$}{62 = 31 + zT24}}

The quiver for the $6_2$ knot is given by:
\begin{equation}\label{eq:6_2 quiver}
\renewcommand\arraystretch{1.25}
    \left[
\begin{array}{ccccccccccc}
 -2 & -2 & -1 & -1 & -1 & -1 & 0 & -1 & 1 & 1 & 1 \\
 -2 & -1 & -1 & 0 & 0 & 0 & 1 & 0 & 1 & 2 & 2 \\
 -1 & -1 & 0 & 1 & 0 & 0 & 1 & 0 & 1 & 2 & 2 \\
 -1 & 0 & 1 & 0 & 0 & 0 & 1 & 0 & 2 & 1 & 1 \\
 -1 & 0 & 0 & 0 & 1 & 1 & 1 & 1 & 2 & 2 & 2 \\
 -1 & 0 & 0 & 0 & 1 & 1 & 1 & 1 & 2 & 2 & 2 \\
 0 & 1 & 1 & 1 & 1 & 1 & 2 & 1 & 2 & 2 & 2 \\
 -1 & 0 & 0 & 0 & 1 & 1 & 1 & 2 & 2 & 3 & 3 \\
 1 & 1 & 1 & 2 & 2 & 2 & 2 & 2 & 3 & 3 & 3 \\
 1 & 2 & 2 & 1 & 2 & 2 & 2 & 3 & 3 & 3 & 3 \\
 1 & 2 & 2 & 1 & 2 & 2 & 2 & 3 & 3 & 3 & 4 \\
\end{array}
\right]\,,
\quad
\left[
\begin{array}{c}
 x \\
 a^2 q^{-3} x \\
 a^2 q^{-2} x \\
 q^2 x \\
 a^2 q^{-1} x \\
 a^2 q^{-1} x \\
 a^2 x \\
 a^4 q^{-4} x \\
 a^4 q^{-3} x \\
 a^2 q x \\
 a^4 q^{-2} x \\
\end{array}
\right]
\end{equation}

Applying the particular worldsheet skein relation, we expect to get holomorphic surfaces whose boundaries are the trefoil and $T_{2,4}$. We can see directly that the quiver entries 1, 4, and 6 correspond to the trefoil in framing $-2$, while the remaining block corresponds to $T_{2,4}$ in framing $-1$. The trefoil quiver reads:

\begin{equation}\renewcommand\arraystretch{1.25}
    Q_{3_1}^{f=-2} = 
    \left[
\begin{array}{ccc}
 -2 & -1 & -1 \\
 -1 & 0 & 0 \\
 -1 & 0 & 1 \\
\end{array}
\right]\,,
\;\;
\left[
\begin{array}{c}
 x \\
 q^2 x \\
 a^2 q^{-1} x \\
\end{array}
\right]
\end{equation}
The quiver for the link $T_{2,4}$ is given below. It is obtained by deleting nodes 1, 4, and 6, as well as permuting nodes 4 and 5 to make the block form manifest.

\begin{equation}\renewcommand\arraystretch{1.25}
Q_{T_{2,4}}^{f=-1} \; = \;
    \left[
\begin{array}{cccc|cccc}
 -1 & -1 & 0 & 0 & 1 & 1 & 2 & 2 \\
 -1 & 0 & 0 & 0 & 1 & 1 & 2 & 2 \\
 0 & 0 & 1 & 1 & 1 & 2 & 2 & 2 \\
 0 & 0 & 1 & 2 & 1 & 2 & 3 & 3 \\
 \hline
 1 & 1 & 1 & 1 & 2 & 2 & 2 & 2 \\
 1 & 1 & 2 & 2 & 2 & 3 & 3 & 3 \\
 2 & 2 & 2 & 3 & 2 & 3 & 3 & 3 \\
 2 & 2 & 2 & 3 & 2 & 3 & 3 & 4 \\
\end{array}
\right]\,,
\;\;
\left[
\begin{array}{c}
 a^2 q^{-3} x \\
 a^2 q^{-2} x \\
 a^2 q^{-1} x \\
 a^4 q^{-4} x \\
 \hline
 a^2 x \\
 a^4 q^{-3} x \\
 a^2 q x \\
 a^4 q^{-2} x \\
\end{array}
\right]
\end{equation}

Comparing with Conjecture \ref{coj:block_quiver_skein}, we identify

$$
f_1=-2,\qquad f_2=-1,\qquad a_1=a^{-2},\qquad a_2=a^{-1}.
$$

\subsection{\texorpdfstring{$8_{13}=6_3-z\text{Wh}$}{8{13}=63-z\text{Wh}}}\label{sec:8_13 skein}
The knot $8_{13}$ corresponds to sequence $\{6_3,8_{13},10_{10},12_{a_{744}},\dots\}$, where each knot is obtained by twisting a particular region, whose 0-resolution produces the~Whitehead link -- the simplest hyperbolic link. In \cite{Chauhan:2025iwt}, these properties allowed to confirm the unlinking--twist recursion for this family of knots. Here we will confirm Conjecture \ref{coj:block_quiver_skein} for a suitable quiver $\tilde{Q}_{8_{13}}$, which blocks decomposes into $Q_{6_3}$ and $Q_{\rm Wh}$:

\begin{equation}\label{eq:8_13 block form coj}
\tilde{Q}_{8_{13}} \; = \; \left[\begin{array}{ c : c }
    C_{{6_3}}^{0,f_1=2} & * \\
    \hdashline
    * & \tilde{C}_{{\rm Wh}}^{0,f_2=0}
    \end{array}
    \right], 
    \;\;
    \left[\begin{array}{c}
    a_1 \boldsymbol x_{{6_3}} \\
    \hdashline
    a_2 \boldsymbol x_{{\rm Wh}}
    \end{array}
    \right]\,.    
\end{equation}
To this end, consider as a starting point $Q_{8_{13}}^+$ found in \cite{Chauhan:2025iwt}. Adapting our convention for augmented quivers with two degree-0 generators, we write
\begin{equation*}
Q_{8_{13}}^+ =
\renewcommand{\arraystretch}{1.2}
\Resize{12.5cm}{
\setlength{\arraycolsep}{1pt}
\left[
\begin{array}{cc|ccccccccccccccccccccccccccccc}
 0 & 0 & 0 & 0 & 0 & 0 & 0 & 0 & 0 & 0 & 0 & 0 & 0 & 0 & 0 & 0 & 0 & 0 & 0 & 0 & 0 & 0 & 0 & 0 & 0 & 0 & 0 & 0 & 0 & 0 & 0 \\
 0 & 1 & 0 & 0 & 0 & 0 & 0 & 0 & 0 & 0 & 0 & 0 & 0 & 0 & 0 & 1 & 1 & 1 & 1 & 1 & 1 & 1 & 1 & 1 & 1 & 1 & 1 & 1 & 1 & 1 & 1 \\
 \hline
 0 & 0 & 0 & 0 & 0 & -1 & -1 & 1 & 0 & 0 & -1 & 0 & 0 & -1 & -1 & 0 & 0 & -1 & -1 & 1 & 1 & 0 & 0 & 0 & 0 & 0 & -1 & -1 & 0 & -1 & -1 \\
 0 & 0 & 0 & 1 & 0 & -1 & -2 & 1 & 0 & -1 & -1 & 1 & 1 & 0 & -1 & 1 & 1 & 0 & -1 & 2 & 1 & 1 & -1 & 1 & 2 & 1 & 0 & 0 & 0 & 0 & -2 \\
 0 & 0 & 0 & 0 & 0 & -1 & -2 & 0 & 0 & -1 & -2 & 1 & 1 & 0 & -1 & 0 & 1 & -1 & -1 & 0 & 1 & -1 & -1 & 1 & 2 & 0 & 0 & 0 & 0 & -1 & -2 \\
 0 & 0 & -1 & -1 & -1 & -2 & -3 & 0 & -1 & -2 & -2 & -1 & 0 & -2 & -2 & -1 & 0 & -2 & -2 & 0 & 1 & -1 & -2 & -1 & 0 & -1 & -1 & -2 & 0 & -2 & -3 \\
 0 & 0 & -1 & -2 & -2 & -3 & -3 & -2 & -1 & -3 & -3 & -1 & 0 & -2 & -2 & -2 & -1 & -3 & -2 & -2 & -1 & -3 & -2 & -1 & 0 & -2 & -1 & -2 & -1 & -3 & -3 \\
 0 & 0 & 1 & 1 & 0 & 0 & -2 & 2 & 1 & 0 & -1 & 2 & 2 & 1 & -1 & 1 & 2 & 0 & 0 & 2 & 2 & 1 & 0 & 2 & 3 & 1 & 1 & 1 & 0 & 0 & -2 \\
 0 & 0 & 0 & 0 & 0 & -1 & -1 & 1 & 1 & 0 & -1 & 1 & 1 & 0 & -1 & 0 & 0 & -1 & -1 & 1 & 1 & 0 & 0 & 1 & 1 & 0 & 0 & 0 & 0 & -1 & -1 \\
 0 & 0 & 0 & -1 & -1 & -2 & -3 & 0 & 0 & -1 & -2 & 0 & 1 & -1 & -2 & -1 & 0 & -2 & -1 & 0 & 1 & -1 & -1 & 0 & 1 & -1 & 0 & -1 & 0 & -2 & -3 \\
 0 & 0 & -1 & -1 & -2 & -2 & -3 & -1 & -1 & -2 & -2 & 0 & 0 & -1 & -2 & -1 & -1 & -2 & -2 & -1 & -1 & -2 & -2 & 0 & 0 & -2 & -1 & -1 & -2 & -3 & -3 \\
 0 & 0 & 0 & 1 & 1 & -1 & -1 & 2 & 1 & 0 & 0 & 3 & 2 & 1 & 0 & 2 & 1 & 1 & -1 & 3 & 2 & 2 & 0 & 3 & 3 & 2 & 1 & 2 & 1 & 1 & -1 \\
 0 & 0 & 0 & 1 & 1 & 0 & 0 & 2 & 1 & 1 & 0 & 2 & 2 & 1 & 0 & 1 & 1 & 0 & 0 & 2 & 2 & 1 & 1 & 2 & 2 & 1 & 1 & 1 & 1 & 0 & 0 \\
 0 & 0 & -1 & 0 & 0 & -2 & -2 & 1 & 0 & -1 & -1 & 1 & 1 & 0 & -1 & 0 & 1 & -1 & -2 & 1 & 2 & 0 & -1 & 1 & 2 & 0 & 0 & 0 & 1 & -1 & -2 \\
 0 & 0 & -1 & -1 & -1 & -2 & -2 & -1 & -1 & -2 & -2 & 0 & 0 & -1 & -1 & -1 & -1 & -2 & -2 & -1 & -1 & -2 & -2 & 0 & 0 & -1 & -1 & -1 & -1 & -2 & -2 \\
 0 & 1 & 0 & 1 & 0 & -1 & -2 & 1 & 0 & -1 & -1 & 2 & 1 & 0 & -1 & 3 & 2 & 1 & 0 & 3 & 2 & 2 & 0 & 3 & 3 & 2 & 1 & 1 & 1 & 1 & -1 \\
 0 & 1 & 0 & 1 & 1 & 0 & -1 & 2 & 0 & 0 & -1 & 1 & 1 & 1 & -1 & 2 & 2 & 1 & 0 & 3 & 2 & 2 & 0 & 2 & 2 & 2 & 1 & 1 & 1 & 1 & -1 \\
 0 & 1 & -1 & 0 & -1 & -2 & -3 & 0 & -1 & -2 & -2 & 1 & 0 & -1 & -2 & 1 & 1 & 0 & -1 & 1 & 1 & 0 & -1 & 2 & 2 & 0 & 0 & 0 & 0 & -1 & -2 \\
 0 & 1 & -1 & -1 & -1 & -2 & -2 & 0 & -1 & -1 & -2 & -1 & 0 & -2 & -2 & 0 & 0 & -1 & -1 & 1 & 1 & 0 & -1 & 0 & 0 & 0 & -1 & -1 & 0 & -1 & -2 \\
 0 & 1 & 1 & 2 & 0 & 0 & -2 & 2 & 1 & 0 & -1 & 3 & 2 & 1 & -1 & 3 & 3 & 1 & 1 & 4 & 3 & 2 & 1 & 4 & 4 & 2 & 2 & 2 & 1 & 1 & -1 \\
 0 & 1 & 1 & 1 & 1 & 1 & -1 & 2 & 1 & 1 & -1 & 2 & 2 & 2 & -1 & 2 & 2 & 1 & 1 & 3 & 3 & 2 & 1 & 3 & 3 & 2 & 2 & 2 & 1 & 1 & -1 \\
 0 & 1 & 0 & 1 & -1 & -1 & -3 & 1 & 0 & -1 & -2 & 2 & 1 & 0 & -2 & 2 & 2 & 0 & 0 & 2 & 2 & 1 & 0 & 3 & 3 & 0 & 1 & 1 & 0 & -1 & -2 \\
 0 & 1 & 0 & -1 & -1 & -2 & -2 & 0 & 0 & -1 & -2 & 0 & 1 & -1 & -2 & 0 & 0 & -1 & -1 & 1 & 1 & 0 & 0 & 1 & 1 & 0 & 0 & 0 & 0 & -1 & -2 \\
 0 & 1 & 0 & 1 & 1 & -1 & -1 & 2 & 1 & 0 & 0 & 3 & 2 & 1 & 0 & 3 & 2 & 2 & 0 & 4 & 3 & 3 & 1 & 5 & 4 & 3 & 2 & 3 & 2 & 2 & 0 \\
 0 & 1 & 0 & 2 & 2 & 0 & 0 & 3 & 1 & 1 & 0 & 3 & 2 & 2 & 0 & 3 & 2 & 2 & 0 & 4 & 3 & 3 & 1 & 4 & 4 & 3 & 2 & 3 & 2 & 2 & 0 \\
 0 & 1 & 0 & 1 & 0 & -1 & -2 & 1 & 0 & -1 & -2 & 2 & 1 & 0 & -1 & 2 & 2 & 0 & 0 & 2 & 2 & 0 & 0 & 3 & 3 & 2 & 1 & 1 & 1 & 0 & -1 \\
 0 & 1 & -1 & 0 & 0 & -1 & -1 & 1 & 0 & 0 & -1 & 1 & 1 & 0 & -1 & 1 & 1 & 0 & -1 & 2 & 2 & 1 & 0 & 2 & 2 & 1 & 1 & 1 & 1 & 0 & -1 \\
 0 & 1 & -1 & 0 & 0 & -2 & -2 & 1 & 0 & -1 & -1 & 2 & 1 & 0 & -1 & 1 & 1 & 0 & -1 & 2 & 2 & 1 & 0 & 3 & 3 & 1 & 1 & 2 & 1 & 0 & -1 \\
 0 & 1 & 0 & 0 & 0 & 0 & -1 & 0 & 0 & 0 & -2 & 1 & 1 & 1 & -1 & 1 & 1 & 0 & 0 & 1 & 1 & 0 & 0 & 2 & 2 & 1 & 1 & 1 & 1 & 0 & -1 \\
 0 & 1 & -1 & 0 & -1 & -2 & -3 & 0 & -1 & -2 & -3 & 1 & 0 & -1 & -2 & 1 & 1 & -1 & -1 & 1 & 1 & -1 & -1 & 2 & 2 & 0 & 0 & 0 & 0 & -1 & -2 \\
 0 & 1 & -1 & -2 & -2 & -3 & -3 & -2 & -1 & -3 & -3 & -1 & 0 & -2 & -2 & -1 & -1 & -2 & -2 & -1 & -1 & -2 & -2 & 0 & 0 & -1 & -1 & -1 & -1 & -2 & -2
\end{array}
\right]
,
\left[
\begin{array}{c}
a^2 \\
a^2q^{-1} \\
\hline
x \\
a^2q^{-3}x \\
x \\
q^{-2}x \\
qa^{-2}x \\
a^2q^{-2}x \\
qx \\
q^{-1}x \\
q^2a^{-2}x \\
a^2q^{-1}x \\
q^2x \\
x \\
q^3a^{-2}x \\
a^4q^{-5}x \\
a^2q^{-2}x \\
a^2q^{-4}x \\
q^{-1}x \\
a^4q^{-4}x \\
a^2q^{-1}x \\
a^2q^{-3}x \\
x \\
a^4q^{-3}x \\
a^2x \\
a^2q^{-2}x \\
qx \\
a^2q^{-2}x \\
qx \\
q^{-1}x \\
q^2a^{-2}x
\end{array}
\right]
}\,.
\end{equation*}

\noindent Applying the linking identity of Section \ref{sec:linking identity}, i.e.\ linking all 0's sequentially in the third row/column of the above matrix, we get\footnote{We will omit the explicit form of this large quiver for brevity here. The quiver data is available upon request.}
\begin{equation}
    L(2,15)\dots L(2,4)L(2,3)\,Q_{8_{13}}^+ \simeq \overline{Q}_{\rm Wh}\,,
\end{equation}
where $\overline{Q}_{\rm Wh}$ is the quiver corresponding to the Whitehead link whose generators have degree~1; thanks to linking, degree-0 generators get absorbed into the normalisation factor \eqref{eq:aug quiver pf}. 
The~resulting quiver is quite large -- much larger than the expected minimal size, 16, which is the size of the finite-dimensional link homology \cite{Gukov:2015gmm}. Remarkably enough, this 42 node quiver can be simplified into a smaller quiver via a~sequence of transpositions (which corresponds to a path in the permutohedra graph from \cite{Jankowski:2021flt}) which generates the sufficient number of redundant pairs. More concretely, we apply the following sequence of operations (read left to right, line-by-line):\footnote{This particular sequence was found by \texttt{GPT-5.6 Sol} and verified by the authors using \texttt{Mathematica}.}
%
\newcommand{\Tau}[4]{%
\ensuremath{\tau_{(#1,#2);(#3,#4)}}%
}
\begingroup
\fontsize{11}{7.2}\selectfont
\setlength{\tabcolsep}{2pt}
\renewcommand{\arraystretch}{1.25}
\setlength{\LTleft}{0pt plus 1fill}
\setlength{\LTright}{0pt plus 1fill}

\begin{longtable}{@{}llllll@{}}
\Tau{1}{31}{2}{30}, & \Tau{9}{39}{10}{38}, & \Tau{8}{39}{10}{37}, & \Tau{7}{39}{10}{36}, & \Tau{6}{39}{10}{35}, & \Tau{5}{39}{10}{34}, \\
\Tau{4}{39}{10}{33}, & \Tau{3}{39}{10}{32}, & \Tau{1}{39}{10}{30}, & \Tau{2}{22}{10}{14}, & \(D_{(39,22)}\), & \(D_{(31,14)}\), \\
\Tau{5}{35}{6}{34}, & \Tau{4}{35}{6}{33}, & \Tau{3}{35}{6}{32}, & \Tau{1}{35}{6}{30}, & \(D_{(35,18)}\), & \Tau{1}{32}{3}{30}, \\
\(D_{(32,24)}\), & \Tau{1}{33}{4}{30}, & \Tau{2}{26}{10}{16}, & \Tau{15}{26}{16}{23}, & \Tau{2}{23}{10}{15}, & \Tau{6}{26}{10}{20}, \\
\Tau{2}{20}{6}{16}, & \Tau{3}{26}{10}{28}, & \Tau{2}{28}{3}{16}, & \Tau{3}{20}{6}{28}, & \Tau{2}{23}{10}{15}, & \Tau{19}{26}{20}{23}, \\
\Tau{15}{20}{16}{19}, & \Tau{23}{28}{26}{27}, & \Tau{15}{28}{16}{27}, & \Tau{11}{41}{12}{40}, & \Tau{9}{41}{12}{38}, & \Tau{8}{41}{12}{37}, \\
\Tau{7}{41}{12}{36}, & \Tau{5}{41}{12}{34}, & \Tau{1}{41}{12}{30}, & \Tau{4}{41}{12}{33}, & \(D_{(41,26)}\), & \(D_{(33,16)}\), \\
\Tau{19}{28}{20}{27}, & \Tau{7}{37}{8}{36}, & \Tau{5}{37}{8}{34}, & \Tau{1}{37}{8}{30}, & \(D_{(37,20)}\), & \Tau{1}{34}{5}{30}, \\
\(D_{(34,28)}\), & \Tau{7}{38}{9}{36}, & \Tau{7}{38}{9}{36}, & \Tau{2}{19}{6}{15}, & \Tau{2}{8}{4}{6}, & \Tau{4}{23}{12}{15}, \\
\Tau{5}{19}{8}{27}, & \Tau{5}{23}{12}{27}, & \Tau{7}{38}{9}{36}, & \Tau{8}{23}{12}{19}, & \Tau{2}{21}{6}{17}, & \Tau{1}{38}{9}{30}, \\
\Tau{8}{23}{12}{19}, & \Tau{1}{38}{9}{30}, & \Tau{2}{21}{6}{17}, & \Tau{7}{40}{11}{36}, & \Tau{9}{40}{11}{38}, & \Tau{6}{23}{10}{19}, \\
\Tau{9}{40}{11}{38}, & \Tau{4}{27}{5}{15}, & \Tau{15}{25}{17}{23}, & \Tau{6}{12}{8}{10}, & \Tau{4}{25}{12}{17}, & \Tau{2}{21}{6}{17}, \\
\Tau{9}{40}{11}{38}, & \Tau{6}{12}{8}{10}, & \Tau{4}{25}{12}{17}, & \Tau{4}{21}{8}{17}, & \Tau{7}{38}{9}{36}, & \Tau{2}{5}{3}{4}, \\
\Tau{2}{5}{3}{4}, & \Tau{6}{23}{10}{19}, & \Tau{2}{5}{3}{4}, & \Tau{8}{25}{12}{21}, & \Tau{19}{25}{21}{23}, & \Tau{9}{42}{13}{38}, \\
\Tau{7}{40}{11}{36}, & \Tau{19}{25}{21}{23}, & \Tau{2}{5}{3}{4}, & \Tau{19}{25}{21}{23}, & \Tau{9}{42}{13}{38}, & \Tau{6}{25}{10}{21}, \\
\Tau{1}{36}{7}{30}, & \Tau{9}{42}{13}{38}, & \Tau{8}{23}{12}{19}, & \Tau{9}{42}{13}{38}, & \Tau{9}{40}{11}{38}, & \Tau{6}{25}{10}{21}, \\
\Tau{2}{5}{3}{4}, & \Tau{2}{5}{3}{4}, & \Tau{1}{6}{2}{7}, & \Tau{8}{25}{12}{21}, & \Tau{9}{40}{11}{38}, & \Tau{23}{29}{25}{27}, \\
\Tau{1}{8}{4}{7}, & \Tau{9}{42}{13}{38}, & \Tau{1}{21}{7}{17}, & \Tau{9}{42}{13}{38}, & \Tau{2}{5}{3}{4}, & \Tau{3}{8}{5}{6}, \\
\Tau{1}{19}{7}{15}, & \Tau{2}{36}{6}{30}, & \Tau{1}{19}{7}{15}, & \Tau{4}{36}{8}{30}, & \Tau{17}{36}{21}{30}, & \Tau{1}{36}{7}{30}, \\
\Tau{7}{40}{11}{36}, & \Tau{3}{8}{5}{6}, & \Tau{3}{8}{5}{6}, & \Tau{3}{8}{5}{6}, & \Tau{23}{29}{25}{27}, & \Tau{6}{12}{8}{10}, \\
\Tau{9}{42}{13}{38}, & \Tau{6}{25}{10}{21}, & \Tau{2}{5}{3}{4}, & \Tau{7}{42}{13}{36}, & \Tau{4}{21}{8}{17}, & \Tau{7}{42}{13}{36}, \\
\Tau{4}{21}{8}{17}, & \Tau{1}{23}{11}{15}, & \Tau{7}{42}{13}{36}, & \Tau{2}{5}{3}{4}, & \Tau{15}{40}{23}{30}, & \Tau{2}{5}{3}{4}, \\
\Tau{11}{42}{13}{40}, & \Tau{7}{40}{11}{36}, & \Tau{17}{42}{29}{30}, & \(D_{(30,15)}\), & \Tau{2}{5}{3}{4}, & \Tau{8}{25}{12}{21}, \\
\Tau{7}{23}{11}{19}, & \Tau{19}{40}{23}{36}, & \Tau{6}{12}{8}{10}, & \Tau{2}{5}{3}{4}, & \Tau{1}{29}{13}{17}, & \Tau{6}{25}{10}{21}, \\
\Tau{1}{29}{13}{17}, & \Tau{7}{12}{8}{11}, & \Tau{7}{25}{11}{21}, & \Tau{1}{29}{13}{17}, & \Tau{11}{29}{13}{25}, & \Tau{19}{40}{23}{36}, \\
\Tau{8}{23}{12}{19}, & \Tau{2}{5}{3}{4}, & \Tau{8}{23}{12}{19}, & \Tau{11}{29}{13}{25}, & \Tau{7}{25}{11}{21}, & \Tau{8}{23}{12}{19}, \\
\Tau{8}{23}{12}{19}, & \Tau{7}{25}{11}{21}, & \Tau{11}{29}{13}{25}, & \Tau{11}{29}{13}{25}, & \Tau{7}{25}{11}{21}, & \Tau{19}{40}{23}{36}, \\
\Tau{1}{29}{13}{17}, & \Tau{19}{40}{23}{36}, & \Tau{1}{29}{13}{17}, & \Tau{8}{23}{12}{19}, & \Tau{7}{23}{11}{19}, & \Tau{2}{5}{3}{4}, \\
\Tau{1}{29}{13}{17}, & \Tau{1}{29}{13}{17}, & \Tau{2}{5}{3}{4}, & \Tau{7}{23}{11}{19}, & \Tau{7}{25}{11}{21}, & \Tau{2}{5}{3}{4}, \\
\Tau{8}{23}{12}{19}, & \Tau{1}{29}{13}{17}, & \Tau{7}{25}{11}{21}, & \Tau{1}{29}{13}{17}, & \Tau{7}{12}{8}{11}, & \Tau{19}{40}{23}{36}, \\
\Tau{9}{40}{11}{38}, & \Tau{8}{25}{12}{21}, & \Tau{2}{5}{3}{4}, & \Tau{9}{40}{11}{38}, & \Tau{2}{5}{3}{4}, & \Tau{7}{29}{13}{21}, \\
\Tau{9}{40}{11}{38}, & \Tau{2}{8}{4}{6}, & \Tau{7}{29}{13}{21}, & \Tau{3}{8}{5}{6}, & \Tau{1}{29}{13}{17}, & \Tau{2}{5}{3}{4}, \\
\Tau{2}{8}{4}{6}, & \Tau{1}{29}{13}{17}, & \Tau{7}{38}{9}{36}, & \Tau{3}{8}{5}{6}, & \Tau{9}{42}{13}{38}, & \Tau{3}{8}{5}{6}, \\
\Tau{7}{25}{11}{21}, & \Tau{19}{40}{23}{36}, & \Tau{19}{40}{23}{36}, & \Tau{3}{12}{5}{10}, & \Tau{3}{12}{5}{10}, & \Tau{2}{8}{4}{6}, \\
\Tau{21}{40}{25}{36}, & \Tau{7}{42}{13}{36}, & \Tau{2}{8}{4}{6}, & \Tau{19}{25}{21}{23}, & \Tau{2}{8}{4}{6}, & \Tau{19}{25}{21}{23}, \\
\Tau{3}{12}{5}{10}, & \Tau{19}{25}{21}{23}, & \Tau{11}{42}{13}{40}, & \Tau{19}{29}{21}{27}, & \Tau{7}{38}{9}{36}, & \(D_{(36,19)}\), \\
\Tau{11}{42}{13}{40}, & \Tau{9}{42}{13}{38}, & \Tau{1}{29}{13}{17}, & \Tau{2}{8}{4}{6}, & \Tau{6}{13}{9}{10}, & \Tau{8}{13}{9}{12}, \\
\Tau{8}{13}{9}{12}, & \Tau{6}{13}{9}{10}, & \Tau{6}{12}{8}{10}, & \Tau{6}{12}{8}{10}, & \Tau{2}{8}{4}{6}, & \Tau{2}{12}{4}{10}, \\
\Tau{1}{29}{13}{17}, & \Tau{2}{5}{3}{4}, & \Tau{23}{29}{25}{27}, & \Tau{9}{42}{13}{38}, & \Tau{11}{42}{13}{40}, & \Tau{9}{40}{11}{38}, \\
\(D_{(40,23)}\), & \Tau{2}{5}{3}{4}, & \Tau{2}{5}{3}{4}, & \Tau{7}{25}{11}{21}, & \Tau{3}{12}{5}{10}, & \Tau{7}{25}{11}{21}, \\
\Tau{7}{25}{11}{21}, & \Tau{3}{12}{5}{10}, & \Tau{8}{25}{12}{21}, & \Tau{9}{42}{13}{38}, & \(D_{(42,27)}\), & \Tau{8}{25}{12}{21}, \\
\Tau{3}{12}{5}{10}, & \Tau{3}{8}{5}{6}, & \(D_{(38,3)}\) & {} & {} & {} \\
\end{longtable}
\endgroup
Here each $\tau_{(a,b);(c,d)}$ is a symmetry transposition in the sense of \cite{Jankowski:2021flt} -- it leaves the motivic generating series intact while modifying the quiver by trading one unit of linking between $C_{ab}$ and $C_{cd}$. Cancelling the 13 redundant pairs produced along the way (operations $D_{(i,j)}$) results in the expected 16-node quiver for the Whitehead link:\footnote{The generating series of the $S^r$-coloured HOMFLYPT invariants of the Whitehead link can be deduced from the properties of homology differentials \cite{Gukov:2015gmm}.}
\begin{equation}\label{eq:wh link quiver}
\overline{Q}_{\rm Wh}\;\; \simeq \;\;
\renewcommand{\arraystretch}{1.2}
\setlength{\arraycolsep}{1pt}
\left[
\begin{customsmallmatrix}
 0 & 0 & -1 & -1 & 1 & 0 & 0 & -1 & 0 & 0 & -1 & -1 & -1 & 0 & -1 & -1 \\
 0 & 1 & -1 & -2 & 1 & 0 & -1 & -1 & 1 & 1 & 0 & -1 & -1 & -1 & 0 & -2 \\
 -1 & -1 & -2 & -3 & 0 & -1 & -2 & -2 & -1 & 0 & -2 & -2 & -2 & -2 & -1 & -3 \\
 -1 & -2 & -3 & -3 & -2 & -1 & -3 & -3 & -1 & 0 & -2 & -2 & -2 & -2 & -1 & -3 \\
 1 & 1 & 0 & -2 & 2 & 1 & 0 & -1 & 2 & 2 & 1 & -1 & 0 & 0 & 1 & -2 \\
 0 & 0 & -1 & -1 & 1 & 1 & 0 & -1 & 1 & 1 & 0 & -1 & -1 & 0 & 0 & -1 \\
 0 & -1 & -2 & -3 & 0 & 0 & -1 & -2 & 0 & 1 & -1 & -2 & -1 & -1 & 0 & -3 \\
 -1 & -1 & -2 & -3 & -1 & -1 & -2 & -2 & 0 & 0 & -1 & -2 & -2 & -2 & -1 & -3 \\
 0 & 1 & -1 & -1 & 2 & 1 & 0 & 0 & 3 & 2 & 1 & 0 & -1 & 0 & 1 & -1 \\
 0 & 1 & 0 & 0 & 2 & 1 & 1 & 0 & 2 & 2 & 1 & 0 & 0 & 1 & 1 & 0 \\
 -1 & 0 & -2 & -2 & 1 & 0 & -1 & -1 & 1 & 1 & 0 & -1 & -2 & -1 & 0 & -2 \\
 -1 & -1 & -2 & -2 & -1 & -1 & -2 & -2 & 0 & 0 & -1 & -1 & -2 & -2 & -1 & -2 \\
 -1 & -1 & -2 & -2 & 0 & -1 & -1 & -2 & -1 & 0 & -2 & -2 & -1 & -1 & -1 & -2 \\
 0 & -1 & -2 & -2 & 0 & 0 & -1 & -2 & 0 & 1 & -1 & -2 & -1 & 0 & 0 & -2 \\
 -1 & 0 & -1 & -1 & 1 & 0 & 0 & -1 & 1 & 1 & 0 & -1 & -1 & 0 & 1 & -1 \\
 -1 & -2 & -3 & -3 & -2 & -1 & -3 & -3 & -1 & 0 & -2 & -2 & -2 & -2 & -1 & -2 \\
\end{customsmallmatrix}
\right]
,
\quad
\left[
\begin{customsmallmatrix}
x \\
a^2q^{-3}x \\
q^{-2}x \\
qa^{-2}x \\
a^2q^{-2}x \\
qx \\
q^{-1}x \\
q^2a^{-2}x \\
a^2q^{-1}x \\
q^2x \\
x \\
q^3a^{-2}x \\
q^{-1}x \\
x \\
qx \\
q^2a^{-2}x
\end{customsmallmatrix}
\right]
\,.
\end{equation}
Since the starting quiver was $Q_{8_{13}}$ and the newly generated nodes after linking are cancelled with some other nodes in $Q_{8_{13}}$, we can filter out what is left -- the node subset
\begin{equation}\label{eq:wh block in 8_13}
\{1, 2, 4, 5, 6, 7, 8, 9, 10, 11, 12, 13, 17, 21, 25, 29\}\subset (Q_{8_{13}})_0\,.
\end{equation}
The above manipulations exactly determine the Whitehead block \eqref{eq:wh link quiver} in $Q_{8_{13}}$. However, our work is not yet done. The problem is that the complement block
\begin{equation}\label{eq:6-3 block in 8_13}
    \{3, 14, 15, 16, 18, 19, 20, 22, 23, 24, 26, 27, 28\}\subset (Q_{8_{13}})_0\,
\end{equation}
does not correspond to $6_3$ (already the second colour does not match). We need to remedy this by finding a suitable path in the permutohedra graph of $Q_{8_{13}}$. Indeed, after applying symmetry transpositions to $Q_{8_{13}}$ which do not involve nodes \eqref{eq:wh block in 8_13},
\begingroup \fontsize{11}{7.2}\selectfont 
\setlength{\tabcolsep}{2pt} \renewcommand{\arraystretch}{1.25} \setlength{\LTleft}{0pt plus 1fill} \setlength{\LTright}{0pt plus 1fill} 
\begin{longtable}{@{}llllll@{}}
\Tau{3}{18}{6}{24}, &
\Tau{2}{24}{3}{14}, &
\Tau{3}{22}{10}{24}, &
\Tau{14}{20}{16}{18}, &
\Tau{18}{26}{20}{22}, &
\Tau{3}{20}{6}{28}, \\

\Tau{14}{26}{16}{22}, &
\Tau{2}{28}{3}{16}, &
\Tau{3}{26}{10}{28} &
{} & {} & {}
\end{longtable}
\endgroup
\noindent we can finally see that the resulting quiver admits the block form \eqref{eq:8_13 block form coj} corresponding to skein relation, as expected:
\begin{equation*}
    \tilde{Q}_{8_{13}} = 
\renewcommand{\arraystretch}{1.2}
\Resize{12.5cm}{
\setlength{\arraycolsep}{1pt}
\left[
\begin{array}{ccccccccccccccccccccccccccccc}
0 & 0 & 0 & -1 & -1 & 1 & 0 & 0 & -1 & 0 & 0 & -1 & -1 & 0 & 0 & -1 & -1 & 1 & 1 & 0 & 0 & 0 & 0 & 0 & -1 & -1 & 0 & -1 & -1 \\
 0 & 1 & 0 & -1 & -2 & 1 & 0 & -1 & -1 & 1 & 1 & 0 & -1 & 1 & 1 & 0 & -1 & 2 & 1 & 1 & -1 & 1 & 2 & 0 & 0 & 0 & 0 & -1 & -2 \\
 0 & 0 & 0 & -1 & -2 & 0 & 0 & -1 & -2 & 1 & 1 & 0 & -1 & 1 & 1 & 0 & -1 & 1 & 1 & 0 & -1 & 2 & 2 & 0 & 0 & 1 & 0 & -1 & -2 \\
 -1 & -1 & -1 & -2 & -3 & 0 & -1 & -2 & -2 & -1 & 0 & -2 & -2 & -1 & 0 & -2 & -2 & 0 & 1 & -1 & -2 & -1 & 0 & -1 & -1 & -2 & 0 & -2 & -3 \\
 -1 & -2 & -2 & -3 & -3 & -2 & -1 & -3 & -3 & -1 & 0 & -2 & -2 & -2 & -1 & -3 & -2 & -2 & -1 & -3 & -2 & -1 & 0 & -2 & -1 & -2 & -1 & -3 & -3 \\
 1 & 1 & 0 & 0 & -2 & 2 & 1 & 0 & -1 & 2 & 2 & 1 & -1 & 1 & 2 & 0 & 0 & 2 & 2 & 1 & 0 & 2 & 3 & 0 & 1 & 1 & 0 & -1 & -2 \\
 0 & 0 & 0 & -1 & -1 & 1 & 1 & 0 & -1 & 1 & 1 & 0 & -1 & 0 & 0 & -1 & -1 & 1 & 1 & 0 & 0 & 1 & 1 & 0 & 0 & 0 & 0 & -1 & -1 \\
 0 & -1 & -1 & -2 & -3 & 0 & 0 & -1 & -2 & 0 & 1 & -1 & -2 & -1 & 0 & -2 & -1 & 0 & 1 & -1 & -1 & 0 & 1 & -1 & 0 & -1 & 0 & -2 & -3 \\
 -1 & -1 & -2 & -2 & -3 & -1 & -1 & -2 & -2 & 0 & 0 & -1 & -2 & -1 & -1 & -2 & -2 & -1 & -1 & -2 & -2 & 0 & 0 & -2 & -1 & -1 & -2 & -3 & -3 \\
 0 & 1 & 1 & -1 & -1 & 2 & 1 & 0 & 0 & 3 & 2 & 1 & 0 & 2 & 1 & 1 & -1 & 3 & 2 & 2 & 0 & 3 & 3 & 1 & 1 & 2 & 1 & 0 & -1 \\
 0 & 1 & 1 & 0 & 0 & 2 & 1 & 1 & 0 & 2 & 2 & 1 & 0 & 1 & 1 & 0 & 0 & 2 & 2 & 1 & 1 & 2 & 2 & 1 & 1 & 1 & 1 & 0 & 0 \\
 -1 & 0 & 0 & -2 & -2 & 1 & 0 & -1 & -1 & 1 & 1 & 0 & -1 & 0 & 1 & -1 & -2 & 1 & 2 & 0 & -1 & 1 & 2 & 0 & 0 & 0 & 1 & -1 & -2 \\
 -1 & -1 & -1 & -2 & -2 & -1 & -1 & -2 & -2 & 0 & 0 & -1 & -1 & -1 & -1 & -2 & -2 & -1 & -1 & -2 & -2 & 0 & 0 & -1 & -1 & -1 & -1 & -2 & -2 \\
 0 & 1 & 1 & -1 & -2 & 1 & 0 & -1 & -1 & 2 & 1 & 0 & -1 & 3 & 2 & 1 & 0 & 3 & 2 & 1 & 0 & 3 & 3 & 2 & 1 & 2 & 1 & 1 & -1 \\
 0 & 1 & 1 & 0 & -1 & 2 & 0 & 0 & -1 & 1 & 1 & 1 & -1 & 2 & 2 & 1 & 0 & 3 & 2 & 2 & 0 & 2 & 2 & 2 & 1 & 1 & 1 & 1 & -1 \\
 -1 & 0 & 0 & -2 & -3 & 0 & -1 & -2 & -2 & 1 & 0 & -1 & -2 & 1 & 1 & 0 & -1 & 2 & 1 & 0 & -1 & 1 & 2 & 0 & 0 & 0 & 0 & -1 & -2 \\
 -1 & -1 & -1 & -2 & -2 & 0 & -1 & -1 & -2 & -1 & 0 & -2 & -2 & 0 & 0 & -1 & -1 & 1 & 1 & 0 & -1 & 0 & 0 & 0 & -1 & -1 & 0 & -1 & -2 \\
 1 & 2 & 1 & 0 & -2 & 2 & 1 & 0 & -1 & 3 & 2 & 1 & -1 & 3 & 3 & 2 & 1 & 4 & 3 & 2 & 1 & 4 & 4 & 2 & 2 & 3 & 1 & 1 & -1 \\
 1 & 1 & 1 & 1 & -1 & 2 & 1 & 1 & -1 & 2 & 2 & 2 & -1 & 2 & 2 & 1 & 1 & 3 & 3 & 2 & 1 & 3 & 3 & 2 & 2 & 2 & 1 & 1 & -1 \\
 0 & 1 & 0 & -1 & -3 & 1 & 0 & -1 & -2 & 2 & 1 & 0 & -2 & 1 & 2 & 0 & 0 & 2 & 2 & 1 & 0 & 2 & 3 & 0 & 1 & 1 & 0 & -1 & -2 \\
 0 & -1 & -1 & -2 & -2 & 0 & 0 & -1 & -2 & 0 & 1 & -1 & -2 & 0 & 0 & -1 & -1 & 1 & 1 & 0 & 0 & 1 & 1 & 0 & 0 & 0 & 0 & -1 & -2 \\
 0 & 1 & 2 & -1 & -1 & 2 & 1 & 0 & 0 & 3 & 2 & 1 & 0 & 3 & 2 & 1 & 0 & 4 & 3 & 2 & 1 & 5 & 4 & 3 & 2 & 3 & 2 & 2 & 0 \\
 0 & 2 & 2 & 0 & 0 & 3 & 1 & 1 & 0 & 3 & 2 & 2 & 0 & 3 & 2 & 2 & 0 & 4 & 3 & 3 & 1 & 4 & 4 & 3 & 2 & 3 & 2 & 2 & 0 \\
 0 & 0 & 0 & -1 & -2 & 0 & 0 & -1 & -2 & 1 & 1 & 0 & -1 & 2 & 2 & 0 & 0 & 2 & 2 & 0 & 0 & 3 & 3 & 2 & 1 & 1 & 1 & 0 & -1 \\
 -1 & 0 & 0 & -1 & -1 & 1 & 0 & 0 & -1 & 1 & 1 & 0 & -1 & 1 & 1 & 0 & -1 & 2 & 2 & 1 & 0 & 2 & 2 & 1 & 1 & 1 & 1 & 0 & -1 \\
 -1 & 0 & 1 & -2 & -2 & 1 & 0 & -1 & -1 & 2 & 1 & 0 & -1 & 2 & 1 & 0 & -1 & 3 & 2 & 1 & 0 & 3 & 3 & 1 & 1 & 2 & 1 & 0 & -1 \\
 0 & 0 & 0 & 0 & -1 & 0 & 0 & 0 & -2 & 1 & 1 & 1 & -1 & 1 & 1 & 0 & 0 & 1 & 1 & 0 & 0 & 2 & 2 & 1 & 1 & 1 & 1 & 0 & -1 \\
 -1 & -1 & -1 & -2 & -3 & -1 & -1 & -2 & -3 & 0 & 0 & -1 & -2 & 1 & 1 & -1 & -1 & 1 & 1 & -1 & -1 & 2 & 2 & 0 & 0 & 0 & 0 & -1 & -2 \\
 -1 & -2 & -2 & -3 & -3 & -2 & -1 & -3 & -3 & -1 & 0 & -2 & -2 & -1 & -1 & -2 & -2 & -1 & -1 & -2 & -2 & 0 & 0 & -1 & -1 & -1 & -1 & -2 & -2 \\
\end{array}
\right]
,
\left[
\begin{array}{c}
x \\
a^2q^{-3}x \\
x \\
q^{-2}x \\
qa^{-2}x \\
a^2q^{-2}x \\
qx \\
q^{-1}x \\
q^2a^{-2}x \\
a^2q^{-1}x \\
q^2x \\
x \\
q^3a^{-2}x \\
a^4q^{-5}x \\
a^2q^{-2}x \\
a^2q^{-4}x \\
q^{-1}x \\
a^4q^{-4}x \\
a^2q^{-1}x \\
a^2q^{-3}x \\
x \\
a^4q^{-3}x \\
a^2x \\
a^2q^{-2}x \\
qx \\
a^2q^{-2}x \\
qx \\
q^{-1}x \\
q^2a^{-2}x
\end{array}
\right]
}\,.
\end{equation*}
Note that we preferred to keep the ordering of nodes intact (it is the same as in the augmented quiver in \cite{Chauhan:2025iwt}); it also shows that the skein-block decomposition is highly nontrivial as these blocks do not coincide with the blocks (denoted by dashed lines in $Q_{8_{13}}^+$) required for the twist recursion. Instead, the two blocks in $\widetilde{Q}_{18}$ are determined by the partitioning \eqref{eq:6-3 block in 8_13} and \eqref{eq:wh block in 8_13}. The framing of the blocks agrees with Conjecture \ref{coj:block_quiver_skein} and can be computed as follows:
\begin{equation}
    \begin{aligned}
            f_1 &= wr(L_\mp)-wr(L_\pm) = 0-(-2)=2,\\
            f_2 &= wr(L_0)-wr(L_\pm)+\sigma = -1-(-2)-1=0.
    \end{aligned}
\end{equation}

\begin{rmk}
    This example is interesting for yet another reason: it also shows how sensitive the block form can be to quiver equivalences. This observation suggests that there may be some form of ``stability clusters'' in the set of all equivalent quivers, labelled by one or more crossings of the minimal diagram of $K$, where destabilisation means a quiver equivalence which destroys the block form. We leave this direction for future study.
\end{rmk}

\section{Future directions}
\label{sec:Future}

In this section we discuss research directions that come from the results presented in this work. 

\begin{itemize}
    \item The most interesting long-term goal would be a formulation of the skein theory for quivers, which would allow for constructing them in a systematic way using the quivers for the unknot. 
    \item An important step in this direction would be the proof of Conjecture \ref{coj:block_quiver_skein}. It would require good control of higher-colour behaviour of links that enter the skein relation, which is quite difficult. Note that some form of categorified skein relations for anti-symmetric representations was considered in \cite{hogancamp2026skein,HogancampRoseWedrich2024}. It would be extremely interesting to investigate the relation between their construction and ours.
    \item An intermediate step would be
    understanding the off-diagonal blocks (see Remark \ref{rmk:off_diagonal}), which also depend on higher-colour HOMFLYPT.    
    \item The intriguing relation between quivers and knot homologies \cite{Ekholm:2021gyu} brings a question about the $t$-deformation of the results presented in this work in the context of generalised quiver partition functions.
    \item Generalised quiver descriptions of links are expected to incorporate holomorphic curves with multiple boundary components, such as annuli, in addition to basic disks. For the Hopf link, for example, the HOMFLYPT polynomial coloured by different symmetric representations on the two components admits a quiver description that includes an annulus, an anti-annulus, and disk contributions \cite{Ekholm:2024ceb}. It would be very interesting to identify further examples of this type, understand their structure, and formulate a general geometric description of the knot-quiver correspondence for links. 
    \item Another promising direction is to investigate the worldsheet skein decomposition of generalised quivers associated with knot complements. In \cite{Chauhan:2026pdv}, the authors express the $\mathfrak{gl}_{1}$ partition function of a $(2,2p+1)$-torus knot complement in terms of the twisted Hopf link partition function via a Lagrangian cobordism, which induces a specialisation of homology variables and modifies the linking data. This construction yields a~generalised quiver description of the complement partition function. We expect that applying worldsheet skein relations to the twisted Hopf link configuration will induce a block decomposition of the corresponding complement quiver.
\end{itemize}

\section*{AI statement}
Large language models (GPT-5.6 Sol) were used for finding references, small notational improvements, and, most importantly, generation of simplifying sequences of quiver operations $\tau_{(ij),(kl)},D_{(m,n)}$ in Section \ref{sec:8_13 skein} and Appendix \ref{sec:linking identity appendix}. All outputs were cross-checked for correctness by the authors. The manuscript was written solely by the authors.

\section*{Acknowledgements}
The work of S.C. is supported by the project “Quiver Structures in Knot Invariants from String Theory and Enumerative Geometry”, funded by Vergstiftelsen and the Swedish Research Council (VR 2022-06593), through the Centre of Excellence in Geometry and Physics at Uppsala University. S.C. is grateful to the string theory group at IMSc for hosting him as a visiting researcher. The work of P.K. is supported by the Polish National Science Centre through SONATA grant (2022/47/D/ST2/02058). P.K. would like to thank the Center for Quantum and Topological Systems at NYU Abu Dhabi and the School of Theoretical Physics at the Dublin Institute for Advanced Studies for hospitality during visits in which part of this work was carried out. The work of L.N. is supported by the Fondation Sciences Mathématiques de Paris. The work of V.K.S is supported by {\it Tamkeen} under the {\it NYU Abu Dhabi Research Institute grant} {\tt CG008}. The work of M.S. is done within the activities of the Center for Mathematical Studies, University of Lisbon (CEMS.UL), and was supported by FCT --  Fundaç\~ao para a Ci\^encia e a Tecnologia, I.P., through national funds, under the project UID/04561/2025 (DOI no. 10.54499/UID/04561/2025). The work of M.S. is supported by FCT through Exploratory Grant no. 2024.16612.PEX (DOI no. 10.54499/2024.16612.PEX). 

\section*{Data Availability Statement}
All datasets generated and analyzed during the current study are available in the arXiv repository, https://arxiv.org/abs/2611.07455v1.
\section*{Declarations}
\noindent\textbf{Conflict of interest} On behalf of all authors, the corresponding author states that there is no conflict of interest.\\

\noindent\textbf{Ethical Statement} Not applicable.\\

\noindent\textbf{Informed Consent} Not applicable.

\appendix

\section{Direct comparison of two quivers for the Hopf link}
\label{app:hopf}

In this appendix we confirm that the 4-node quiver corresponding to the normalised Hopf link invariants coloured by the same representation on both components (this quiver was first considered in \cite{Kucharski:2017ogk}), is equivalent in the sense of \cite{Jankowski:2021flt} to the Hopf link quiver based on the annuli description \cite{Ekholm:2024ceb}.

The familiar quiver from the knot-quiver correspondence is given by,
\begin{equation}\label{eq:hopf quiver a}
\renewcommand\arraystretch{1.25}
    Q_{\tilde{T}_{2,2}}=\left[
\begin{array}{cccc}
 0 & 0 & -1 & -1 \\
 0 & 1 & -1 & -1 \\
 -1 & -1 & -1 & -2 \\
 -1 & -1 & -2 & -2 \\
\end{array}
\right],
\;\;
\left[
\begin{array}{c}
    q^4 x \\
    a^2q x \\
    q^3x \\
    q^2 x
\end{array}
\right].
\end{equation}

From the geometry of the Hopf link conormal, consider a system with two pairs of basic discs (one per each component of the link), plus a pair annulus-(anti)annulus \cite{Ekholm:2024ceb}. This corresponds to the generating function of HOMFLYPT invariants which carry colours $r_1,r_2$ on each component. The corresponding quiver adjacency matrix is,

\begin{equation}\label{eq:hopf quiver fully unreduced}
\renewcommand\arraystretch{1.25}
C_{\rm{unred}} = \left[
\begin{array}{cc|cc|cc}
 0 & 0 & 0 & 0 & -1 & 0 \\
 0 & 1 & 0 & 0 & -1 & 0 \\
 \hline
 0 & 0 & 0 & 0 & -1 & 0 \\
 0 & 0 & 0 & 1 & -1 & 0 \\
 \hline
 -1 & -1 & -1 & -1 & -2 & -2 \\
 0 & 0 & 0 & 0 & -2 & 0 \\
\end{array}
\right]\,.
\end{equation}
(Here we have the identifications $d_1+d_2+d_5+d_6=r_1$, $d_3+d_4+d_5+d_6=r_2$).
Now consider the bicoloured Hopf link invariants divided by $(a^2;q^2)_{r_1}$:
\begin{equation}
\begin{aligned}\label{eq:above}
    \sum_{r_1,r_2\geq 0} \frac{P_{r_1,r_2}(a,q)}{(a^2;q^2)_{r_1}}x_1^{r_1}x_2^{r_2}=&\ \sum_{d_1,d_2,d_3,d_4\geq 0}\frac{q^{\boldsymbol{d}^T(C_{\rm red})\boldsymbol{d}+d_1+d_2+2d_4}}{(q^2;q^2)_{d_1}(q^2;q^2)_{d_2}(q^2;q^2)_{d_3}}(-a^2)^{d_3}x_1^{d_1+d_4}x_2^{d_2+d_3+d_4} \\
    = &\ 1-\frac{q}{1-q^2}x_1-\frac{q(1-a^2)}{1-q^2}x_2+\frac{(1-q^2-a^2a^2+q^4)}{(1-q^2)^2}x_1x_2+\dots
\end{aligned}
\end{equation}
Here the reduced quiver adjacency matrix is given by,
\begin{equation}
\renewcommand\arraystretch{1.25}
C_{\rm red}=\left[
\begin{array}{c|cc|c}
 0 & 0 & 0 & -1 \\
 \hline
 0 & 0 & 0 & -1 \\
 0 & 0 & 1 & 0 \\
 \hline
 -1 & -1 & 0 & -2 \\
\end{array}
\right]
\end{equation}
with the identifications $d_1+d_4=r_1$, $d_2+d_3+d_4=r_2$.

In order to have a generating series for $r_1=r_2=r$, we need to substitute $x_1=x_2=x$ and $d_1=d_2+d_3$ in \eqref{eq:above}. After renaming $d_2\mapsto d_1$, $d_3\mapsto d_2$ and replacing $x^2$ by $x$, we get
\begin{gather}
    \sum_{r\geq 0} \frac{P_{r,r}(a,q)}{(a^2;q^2)_{r}}x^r = \sum_{d_1,d_2,d_3\geq 0}\frac{q^{\boldsymbol{d}^T(C'_{\rm red})\boldsymbol{d}+2d_1+d_2+2d_3}}{(q^2;q^2)_{d_1+d_2}(q^2;q^2)_{d_1}(q^2;q^2)_{d_2}}(-a^2)^{d_2}x^{d_1+d_2+d_3} \\
    = 1+\frac{(-(1+a^2)q^2+q^4+1)}{(1-q^2)^2}x+O(x^2) \nonumber
\end{gather}
with
\begin{equation}\label{eq:intermediate quiver}
\renewcommand\arraystretch{1.25}
C'_{\rm red} = \left[
\begin{array}{cc|c}
 0 & 0 & -2 \\
 0 & 1 & -1 \\
 \hline
 -2 & -1 & -2 \\
\end{array}
\right]\,.
\end{equation}
Notice that the annulus with self-linking number $-2$ has to be treated separately. Using the $q$-binomial theorem, we can write the annulus contribution in the usual quiver form (here $a,b$ may depend on other summation indices in the larger sum) as a combination of basic discs
\begin{equation}
    \sum_{d_i\geq 0}q^{a d_i^2+b d_i}x^{d_i} = 
    \sum_{\alpha,\beta\geq 0}q^{a(\alpha+\beta)^2+b(\alpha+\beta)}x^{\alpha+\beta}\frac{(-1)^{\beta}q^{\beta^2+\beta}}{(q^2;q^2)_{\alpha}(q^2;q^2)_{\beta}}.
\end{equation}
Applying this to \eqref{eq:intermediate quiver}, we get
\begin{equation}
\begin{aligned}
    & \sum_{d_1,d_2,d_3,d_4\geq 0}\frac{(-1)^{d_2+d_3}q^{\boldsymbol{d}^T(\check{C}_b)\boldsymbol{d}+2d_1+d_2+3d_3+2d_4}}{(q^2;q^2)_{d_1+d_2}(q^2;q^2)_{d_1}(q^2;q^2)_{d_2}(q^2;q^2)_{d_3}(q^2;q^2)_{d_4}}a^{2d_2}x^{d_1+d_2+d_3+d_4} \\
    & \qquad \qquad = 1+\frac{\left(-\left(a^2+1\right) q^2+q^4+1\right)}{\left(q^2-1\right)^2}x+O(x^2)\,.
\end{aligned}
\end{equation}
with $\check{C}_b$ given by
\begin{equation}
\renewcommand\arraystretch{1.25}
\check{C}_b=\left[
\begin{array}{cc|cc}
 0 & 0 & -2 & -2 \\
 0 & 1 & -1 & -1 \\
 \hline
 -2 & -1 & -1 & -2 \\
 -2 & -1 & -2 & -2 \\
\end{array}
\right].
\end{equation}
It remains to confirm that
\begin{equation}\label{eq:the new identity}
\begin{aligned}
    & \sum_{\boldsymbol{d}\geq 0}\frac{(-q)^{\sum (C_a)_{ij} d_id_j}}{(q^2;q^2)_{d_1}\dots (q^2;q^2)_{d_4}}\boldsymbol{x}_a^{\boldsymbol{d}} =
    \sum_{\boldsymbol{d}\geq 0}(q^{2(d_1+d_2)+2};q^2)_{d_3+d_4}\frac{(-q)^{\sum (\check{C}_b)_{ij} d_id_j}}{(q^2;q^2)_{d_1}\dots (q^2;q^2)_{d_4}}\boldsymbol{x}_b^{\boldsymbol{d}} \\
    & \qquad \qquad = 1 + \frac{x \left(-\left(a^2+1\right) q^2+q^4+1\right)}{1-q^2} + O(x^2)\,.
\end{aligned}
\end{equation}
Let us briefly sketch the derivation of $\check{C}_b$ and the equality \eqref{eq:the new identity}. 
We need to take into account normalisation factor $(q^2;q^2)_{d_1+d_2+d_3+d_4}$ which is yet missing. Including this factor and cancelling with $(q^2;q^2)_{d_1+d_2}^{-1}$, we get right hand side of the prequiver form \eqref{eq:the new identity}.

Note that $\check{C}_b$ contains two blocks -- the first one comes from the second pair of basic discs supported on the second component of the Hopf link, and the other block comes from the annulus contribution. After splitting $q$-Pochhammer $(q^{2(d_1+d_2)+2};q^2)_{d_3+d_4}$ in \eqref{eq:the new identity}, it takes form
\begin{equation}
\renewcommand\arraystretch{1.25}
Q_b=\left[
\begin{array}{cc|cccc}
 0 & 0 & -2 & -2 & -1 & -1 \\
 0 & 1 & -1 & -1 & 0 & 0 \\
 \hline
 -2 & -1 & -1 & -2 & -1 & -1 \\
 -2 & -1 & -2 & -2 & -2 & -2 \\
 -1 & 0 & -1 & -2 & 0 & -1 \\
 -1 & 0 & -1 & -2 & -1 & -1 \\
\end{array}
\right],
\;\;
\left[
\begin{array}{c}
    q^2 x \\
    a^2q x \\
    \hline
    q^3x \\
    q^2 x \\
    q^4 x \\
    q^3 x
\end{array}
\right]\,.
\end{equation}
Consider the permutohedra graph which consists of two quivers related by a transposition of a pair of arrows:\footnote{Here we have first applied the permutation $(5, 2, 3, 4, 1, 6)$.}
\begin{equation}
\renewcommand\arraystretch{1.25}
\left[
\begin{array}{cccccc}
 0 & 0 & -1 & \textcolor{red}{-2} & -1 & -1 \\
 0 & 1 & -1 & -1 & 0 & 0 \\
 -1 & -1 & -1 & -2 & -2 & \textcolor{red}{-1} \\
 \textcolor{red}{-2} & -1 & -2 & -2 & -2 & -2 \\
 -1 & 0 & -2 & -2 & 0 & -1 \\
 -1 & 0 & \textcolor{red}{-1} & -2 & -1 & -1 \\
\end{array}
\right]
\quad
\longleftrightarrow
\quad
\left[
\begin{array}{cccc:cc}
 0 & 0 & -1 & \textcolor{red}{-1} & -1 & -1 \\
 0 & 1 & -1 & -1 & 0 & 0 \\
 -1 & -1 & -1 & -2 & -2 & \textcolor{red}{-2} \\
 \textcolor{red}{-1} & -1 & -2 & -2 & -2 & -2 \\
 \hdashline
 -1 & 0 & -2 & -2 & 0 & -1 \\
 -1 & 0 & \textcolor{red}{-2} & -2 & -1 & -1 \\
\end{array}
\right].
\end{equation}
Note that in the rightmost quiver, the last two nodes form a redundant pair, and the resulting 4-node quiver agrees with \eqref{eq:hopf quiver a}, as claimed.

\section{The Linking identity for \texorpdfstring{$\tilde{T}_{2,4}$}{tT24}}\label{sec:linking identity appendix}

Here we apply the linking identity of Section \ref{sec:linking identity} to one particular quiver for the negative torus link $\tilde{T}_{2,4}$, found in Section \ref{sec:Examples}. We begin with the quiver
\begin{equation}\renewcommand{\arraystretch}{1.1}
Q_{\tilde{T}_{2,4}} =
\left[
\begin{array}{cccccccc}
 5 & 4 & 3 & 3 & 2 & 2 & 1 & 1 \\
 4 & 4 & 3 & 3 & 2 & 2 & 1 & 1 \\
 3 & 3 & \textcolor{red}{3} & 2 & 2 & 2 & 1 & 1 \\
 3 & 3 & 2 & 2 & 1 & 1 & 0 & 0 \\
 2 & 2 & 2 & 1 & \textcolor{red}{2} & 1 & 1 & 1 \\
 2 & 2 & 2 & 1 & 1 & \textcolor{red}{1} & 0 & 0 \\
 1 & 1 & 1 & 0 & 1 & 0 & 1 & 0 \\
 1 & 1 & 1 & 0 & 1 & 0 & 0 & \textcolor{red}{0} \\
\end{array}
\right],
\;\;
\left[
\begin{array}{c}
 a^6 x \\
 a^4 q^3 x \\
 a^4 q^2 x \\
 a^4 q x \\
 a^2 q^5 x \\
 a^2 q^4 x \\
 a^2 q^4 x \\
 a^2 q^3 x \\
\end{array}
\right]\,.
\end{equation}
Notice that $Q_{\tilde{T}_{2,4}}$ admits exactly two subquivers corresponding to the Hopf link (in different framing): one is given (manifestly by construction) by the first $4\times 4$ block, i.e.\ the nodes $\{1,2,3,4\}$. The other one is less manifest -- nodes $\{3,5,6,8\}$. The latter one will be relevant for us, since it has smaller framing and therefore corresponds to the initial step of the twist recursion \cite{Chauhan:2025iwt}. We write the corresponding augmented quiver:
\begin{equation}\renewcommand{\arraystretch}{1.15}
Q_{\tilde{T}_{2,4}}^+ \;\; = \;\;
\left[
\begin{array}{cc|cccccccc}
 0 & 0 & 0 & 0 & 0 & 0 & 0 & 0 & 0 & 0 \\
 0 & 1 & 1 & 1 & 0 & 1 & 0 & 1 & 0 & 0 \\
 \hline
 0 & 1 & 5 & 4 & 3 & 3 & 2 & 2 & 1 & 1 \\
 0 & 1 & 4 & 4 & 3 & 3 & 2 & 2 & 1 & 1 \\
 0 & 0 & 3 & 3 & 3 & 2 & 2 & 2 & 1 & 1 \\
 0 & 1 & 3 & 3 & 2 & 2 & 1 & 1 & 0 & 0 \\
 0 & 0 & 2 & 2 & 2 & 1 & 2 & 1 & 1 & 1 \\
 0 & 1 & 2 & 2 & 2 & 1 & 1 & 1 & 0 & 0 \\
 0 & 0 & 1 & 1 & 1 & 0 & 1 & 0 & 1 & 0 \\
 0 & 0 & 1 & 1 & 1 & 0 & 1 & 0 & 0 & 0 \\
\end{array}
\right],
\quad
\left[
\begin{array}{c}
 a^2 \\
 a^2q^{-1} \\
 \hline
 a^6 x \\
 a^4 q^3 x \\
 a^4 q^2 x \\
 a^4 q x \\
 a^2 q^5 x \\
 a^2 q^4 x \\
 a^2 q^4 x \\
 a^2 q^3 x \\
\end{array}
\right]\,.
\end{equation}
Below we show that the linking identity produces the quiver for the 0-resolution, i.e.\ the genus-1 unknot. After linking the appropriate entries in the above quiver, we get
\begin{equation}
\renewcommand{\arraystretch}{1.15}
\begin{split}
    & L(2,10)L(2,9)L(2,7)L(2,5)Q_{\tilde{T}_{2,4}}^+ \\ & =
    \left[
\begin{array}{cc|cccccccccccc}
 0 & 0 & 0 & 0 & 0 & 0 & 0 & 0 & 0 & 0 & 0 & 0 & 0 & 0 \\
 0 & 1 & 1 & 1 & 1 & 1 & 1 & 1 & 1 & 1 & 1 & 1 & 1 & 1 \\
 \hline
 0 & 1 & 5 & 4 & 3 & 3 & 2 & 2 & 1 & 1 & 4 & 3 & 2 & 2 \\
 0 & 1 & 4 & 4 & 3 & 3 & 2 & 2 & 1 & 1 & 4 & 3 & 2 & 2 \\
 0 & 1 & 3 & 3 & 3 & 2 & 2 & 2 & 1 & 1 & 3 & 3 & 2 & 2 \\
 0 & 1 & 3 & 3 & 2 & 2 & 1 & 1 & 0 & 0 & 3 & 2 & 1 & 1 \\
 0 & 1 & 2 & 2 & 2 & 1 & 2 & 1 & 1 & 1 & 2 & 2 & 2 & 2 \\
 0 & 1 & 2 & 2 & 2 & 1 & 1 & 1 & 0 & 0 & 3 & 2 & 1 & 1 \\
 0 & 1 & 1 & 1 & 1 & 0 & 1 & 0 & 1 & 0 & 1 & 1 & 1 & 1 \\
 0 & 1 & 1 & 1 & 1 & 0 & 1 & 0 & 0 & 0 & 1 & 1 & 0 & 0 \\
 0 & 1 & 4 & 4 & 3 & 3 & 2 & 3 & 1 & 1 & 4 & 3 & 2 & 2 \\
 0 & 1 & 3 & 3 & 3 & 2 & 2 & 2 & 1 & 1 & 3 & 3 & 2 & 2 \\
 0 & 1 & 2 & 2 & 2 & 1 & 2 & 1 & 1 & 0 & 2 & 2 & 2 & 1 \\
 0 & 1 & 2 & 2 & 2 & 1 & 2 & 1 & 1 & 0 & 2 & 2 & 1 & 1 \\
\end{array}
\right]
\quad
\left[
\begin{array}{c}
 a^2 \\
 a^2q^{-1} \\
 \hline
 a^6 x \\
 a^4 q^3 x \\
 a^4 q^2 x \\
 a^4 q x \\
 a^2 q^5 x \\
 a^2 q^4 x \\
 a^2 q^4 x \\
 a^2 q^3 x \\
 a^6 q x \\
 a^4 q^4 x \\
 a^4 q^3 x \\
 a^4 q^2 x
\end{array}
\right]
\,.
\end{split}
\end{equation}
As we have done before, we can absorb the first two nodes of dergee-0 into a prefactor for the series coefficients. The remaining 12-node quiver can be transformed via the following sequence of operations:
\begin{equation}
\begin{split}
& \Tau{1}{6}{2}{4}
\;\longrightarrow\;
D_{(1,9)}
\;\longrightarrow\;
D_{(2,10)}
\;\longrightarrow\;
\Tau{3}{6}{4}{5}\\
& \qquad \;\longrightarrow\;
\Tau{3}{8}{4}{7}
\;\longrightarrow\;
D_{(4,12)}
\;\longrightarrow\;
D_{(3,11)} \,. 
\end{split}
\end{equation}
Here $D_{(i,j)}$ denotes the removal of the redundant pair $(i,j)$. The resulting quiver has 4 nodes: 
\begin{equation}\renewcommand{\arraystretch}{1.1}
Q_{\tilde{T}_{2,4}}^+ \; \simeq  \;
\left[
\begin{array}{cccc}
 2 & 1 & 1 & 1 \\
 1 & 1 & 0 & 0 \\
 1 & 0 & 1 & 0 \\
 1 & 0 & 0 & 0 \\
\end{array}
\right] ,
\;\;
\left[
\begin{array}{c}
 a^2 q^5 x \\
 a^2 q^4 x \\
 a^2 q^4 x \\
 a^2 q^3 x \\
\end{array}
\right]\,.
\end{equation}
After the admissible transposition $\Tau{1}{4}{2}{3}$, the quiver is identified with the genus-1 unknot \eqref{quiver:genus1Unknot}, as claimed. The result confirms the that the infinitely twisted link invariants, when both components have the same colours, correspond to invariants of the genus-1 unknot. If this holds true in a greater generality, then we obtain a systematic way to find block decompositions for links, given that we know their augmentations.

\section{The Linking identity in the full skein}\label{app:full skein linking identity}

In the linking identity \eqref{eq:linking identity ap}, the addition of two auxiliary nodes of framing 0 and 1, which have linking 0 and 1 with the original quiver, acts as multiplication by a certain normalisation factor. This observation admits a generalisation to the full skein of a solid torus. We consider the following geometric setting: let $Z = \sum_\lambda Z_\lambda W_\lambda$ be some partition function in the skein of a solid torus with coefficients $Z_\lambda$. We then act on $Z$ with the operator
\begin{equation}
    U\left(Z_\textnormal{di}\right) P_{1,0}\left(Z_\textnormal{di}^{f=1}\right).
\end{equation}
Here, $Z_\textnormal{di}$ and $Z_\textnormal{di}^{f=1}$ denote the disk partition functions in trivial framing and framing one. In general framing $k$, this partition function is given by
\begin{equation}
    Z_\textnormal{di}^{f=k}(x) = \sum_\lambda \prod_{\square \in \lambda} \frac{\left(-q^{2c(\square)}\right)^k(-q)^{-c(\square)}}{q^{h(\square)} - q^{-h(\square)}}W_\lambda x^{|\lambda|}.
\end{equation}
The symbols $U$ and $P_{1,0}$ denote the two elements in the skein of the surface $T^2$ given by an unknot and a meridian, here via cabling viewed as maps $\textnormal{Sk}(S^1 \times \mathbb{R}^2) \to \textnormal{Sk}(T^2)$. The skein of $T^2$ acts naturally on the skein of the solid torus by inclusion of the boundary, and the action of a \emph{power sum element} $P_{m,0}$ is given by
\begin{equation}
    P_{m,0}(W_\lambda) = \left(\frac{a_L^m-a^{-m}_L}{q^m-q^{-m}} + a^m_L(q^m-q^{-m}) \left(\sum_{\square \in \lambda} q^{2m c(\square)} \right) \right) W_\lambda.
\end{equation}
In order to compute the action of the above operator, we use the following expression of the disk partition function in framing zero and one in terms of the power sum elements
\begin{equation}
    \begin{split}
        Z_\textnormal{di}(x)& = \exp\left(-\sum_{d > 0} \frac{1}{d} \frac{P_d}{q^d-q^{-d}} x^d\right),\\
        Z_\textnormal{di}^{f=1}(x) &= \exp\left(\sum_{d > 0} \frac{1}{d} \frac{P_d}{q^d-q^{-d}} x^d \right).
    \end{split}
\end{equation}
Since the map
\begin{equation}
    U \cdot P_{1,0}: \textnormal {Sk}(S^1 \times \mathbb{R}^2)^{\otimes 2} \to \textnormal{Sk}(T^2)
\end{equation}
is an algebra homomorphism, we can compute
\begin{equation}
    \begin{split}
        \left(U\left(Z_\textnormal{di}(x)\right) P_{1,0}\left(Z_\textnormal{di}^{f=1}(x)\right)\right)(W_\lambda) &= \exp\left( \sum_{d>0} \frac{1}{d} \frac{1}{q^d-q^{-d}} (P_{d,0} - U(P_d)) x^d\right) W_\lambda\\
        &= \exp\left(\sum_{d>0} \frac{1}{d} a^d_L x^d\sum_{\square \in \lambda} q^{2d c(\square)} \right) W_\lambda\\
        &= \frac{1}{\prod_{\square \in \lambda} \left(1- q^{2 c(\square)}a_Lx\right)}W_\lambda,
    \end{split}
\end{equation}
where we used that $U(P_d) = \frac{a_L^d-a_L^{-d}}{q^d-q^{-d}}$.
Note that if $\lambda = (r)$ is a symmetric representation, then 
\begin{equation}
    \prod_{\square \in (r)} \left(1- q^{2 c(\square)}a_Lx^d\right) = (a_L x;q^2)_r.
\end{equation}
If we now set $x= a_L^{-1} a^2$, we recover the normalisation factor of the linking identity in the $U(1)$-skein.

\begin{SCfigure}[1.5][htbp]
    \centering
    \includegraphics[width=0.25\textwidth]
    {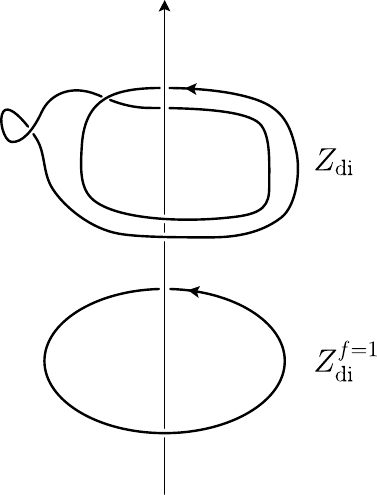}
    \caption{The vertical curve stands for an arbitrary
    element in the skein of a solid torus, where the top
    and bottom parts of the drawing are identified.
    It is linked twice by $Z_\textnormal{di}$, which
    gives the action of $Z_2$, and linked once by
    $Z_\textnormal{di}^{f=1}$.\\\\}
    \label{fig:degree 2 disk linking}
\end{SCfigure}

In the case of strands with parallel orientations in the twist region, we are not able to lift the quiver associated to the $U(1)$-invariants to the full skein. In this case, recall that in the quiver, there is one auxiliary node of framing 0 which links the original quiver once and one node of framing one linking the original quiver twice. For the interest of the reader, we include here an example of two disks partition functions in the full skein which link a given partition function once and twice, respectively:

We act on a partition function $Z$ with $Z^{f=1}_\textnormal{di}$ and the partition function
\begin{equation}
    Z_2 \coloneqq \exp \left( \sum_{d>0} \frac{a_L^d}{2d} \left( P_d^2 - \frac{q^d+q^{-d}}{q^d-q^{-d}} P_{2d}\right) x^d \right)
\end{equation} 
via $P_{1,0}$, i.e.\,we act with $P_{1,0}(Z^{f=1}_\textnormal{di} Z_2)$. 
It can be shown that $Z_2$ is equal to a full twist of $Z_\textnormal{di}$ embedded along a degree 2 curve in the solid torus with one negative crossing, see Figure \ref{fig:degree 2 disk linking}.
As previously, the $W_\lambda$ are eigenvectors for this action, and we can compute the eigenvalues to be
\begin{align}
    \exp\Bigg( &\sum_{d>0}x^d\Bigg[\frac{a_L^d}{2d}\Bigg(\Bigg(\frac{a_L^d-a_L^{-d}}{q^d-q^{-d}} + a_L^d(q^d-q^{-d})\sum_{\square \in \lambda} q^{2dc_\square} \Bigg)^2 \nonumber\\
    &- \frac{q^d+q^{-d}}{q^d-q^{-d}}\left( \frac{a_L^{2d}-a_L^{-2d}}{q^{2d}-q^{-2d}} + a_L^{2d}(q^{2d}-q^{-2d})\sum_{\square \in \lambda} q^{4dc_\square}\right)\Bigg) \\
    &+ \frac{1}{q^d-q^{-d}}\left( \frac{a_L^d-a_L^{-d}}{q^d-q^{-d}} + a_L^d(q^d-q^{-d})\sum_{\square \in \lambda} q^{2dc_\square}\right)\Bigg] \Bigg) \nonumber\\
    &= \exp\Bigg( \sum_{d>0}x^d\frac{a_L^{3d}}{d} \Bigg[ \sum_{\square \in \lambda} q^{2dc_\square} + \frac{1}{2} (q^d-q^{-d})^2 \Bigg(\sum_{\square \in \lambda} q^{2dc_\square}\Bigg)^2 - \frac{1}{2}(q^d+q^{-d})^2 \sum_{\square \in \lambda} q^{4dc_\square}\Bigg]  \Bigg)\,.\nonumber
\end{align}
In the special case that $\lambda = (r)$, this simplifies to
\begin{equation}
\begin{aligned}
    \exp\Bigg(& \sum_{d>0}x^d\frac{a_L^{3d}}{d} \Bigg[ \sum_{i=0}^{r-1} q^{2di} + \frac{1}{2} (q^d-q^{-d})^2 \Bigg(\sum_{i=0}^{r-1} q^{2di}\Bigg)^2 - \frac{1}{2}(q^d+q^{-d})^2 \sum_{i=0}^{r-1} q^{4di}\Bigg]  \Bigg)\\
    &= \exp\Bigg(\sum_{d>0}x^d\frac{a_L^{3d}}{d} \Bigg[ \sum_{i=0}^{r-1} q^{2di} + \frac{1}{2} (q^{d(2r-1}-q^{-d})^2 - \frac{1}{2}(q^{2d} + 2 +q^{-2d}) \sum_{i=0}^{r-1} q^{4di}\Bigg]  \Bigg) \\
    &= \exp\Bigg(-\sum_{d>0}x^d\frac{a_L^{3d}}{d} \Bigg[ \sum_{i=r-1}^{2r-2} q^{2di}\Bigg]  \Bigg) = (a_L^3 x q^{2r-2};q^2)_r\,.
\end{aligned}
\end{equation}
\printbibliography

\end{document}